\documentclass[a4paper,11pt]{article}
\usepackage{jcappub} % for details on the use of the package, please
\usepackage{graphicx,dcolumn,bm,epsfig,cancel,hyperref}
\usepackage{amsmath}
\usepackage{amssymb, times}
\newcommand{\red}[1]{\textcolor{red}{#1}}

\newcommand{\comment}[1]{{}}
\newcommand{\rhogw}[0]{{\rho_{\text{GW}}}}
\newcommand{\aref}[0]{{a_{s}}}
\newcommand{\Href}[0]{{H_{s}}}
\newcommand{\kappam}[0]{{\kappa_{M}}}
\makeatletter
\newcommand{\doublewidetilde}[1]{{%
  \mathpalette\double@widetilde{#1}%
}}
\newcommand{\double@widetilde}[2]{%
  \sbox\z@{$\m@th#1\widetilde{#2}$}%
  \ht\z@=.9\ht\z@
  \widetilde{\box\z@}%
}
\makeatother
\usepackage{color}
\title{
Phase Transitions in an Expanding Universe: Stochastic Gravitational Waves in Standard and Non-Standard Histories
}
\author[a]{Huai-Ke Guo}
\author[a]{Kuver Sinha}
\author[a]{Daniel Vagie}
\author[b]{Graham White}
\affiliation[a]{Department of Physics and Astronomy, University of Oklahoma, Norman, OK 73019, USA}
\affiliation[b]{TRIUMF Theory Group, 4004 Wesbrook Mall, Vancouver, B.C. V6T2A3, Canada}
\emailAdd{ghk@ou.edu}
\emailAdd{kuver.sinha@ou.edu}
\emailAdd{Daniel.d.vagie-1@ou.edu}
\emailAdd{gwhite@triumf.ca}
\begin{document}

%\date{\today}

\abstract
{ We undertake a  careful analysis of stochastic gravitational wave production from cosmological phase transitions in an expanding universe, studying both a standard radiation as well as a matter dominated history. %In particular, we scrutinize the widely adopted conclusion that the effective lifetime of the sound waves is a Hubble time and 
We analyze in detail the dynamics of the phase transition, including the false vacuum fraction, bubble lifetime distribution, bubble number density, mean bubble separation, etc., for an expanding universe. We also study the full set of differential equations governing the evolution of plasma and the scalar field during the phase transition and generalize results obtained in Minkowski spacetime. In particular, we generalize the sound shell model to the expanding universe and determine the velocity field power spectrum. This ultimately provides an accurate calculation of the gravitational wave spectrum seen today for the dominant source of sound waves. For the amplitude of the gravitational wave spectrum visible today, we find a suppression factor arising from the finite lifetime of the sound waves and compare with the commonly used result in the literature, which corresponds to the asymptotic value of our suppression factor. We point out that the asymptotic value is only applicable for a very long lifetime of the sound waves, which is highly unlikely due to the onset of shocks, turbulence and other damping processes. We also point out that features of the gravitational wave spectral form may hold the tantalizing possibility of distinguishing between different expansion histories using phase transitions.
}

%\pacs{11.30.Er, 11.30.Fs, 11.30.Hv, 12.60.Fr, 31.30.jp}
%GW abstract: We undertake a  careful analysis of stochastic gravitational wave production from cosmological phase transitions in an expanding universe, studying both a standard radiation as well as a matter dominated history. In particular, we scrutinize the assumption that the effective lifetime of the sound waves is a Hubble time and analyze in detail the dynamics of the phase transition. This includes the false vacuum fraction, bubble lifetime 
%distribution, bubble number density, etc., for an expanding universe. We also study the full set of hydrodynamic equations governing the field and fluid evolution during the phase
%transition and generalize results obtained in Minkowski space, to determine the velocity field power spectrum. This ultimately  provides an accurate calculation of the gravitational wave spectrum seen today. For the peak amplitude of the gravitational wave spectrum visible today, we find a suppression factor arising from the finite lifetime of the sound waves and compare with the recent results of Ellis et. al. We also point out that changes to the spectral form may hold out the tantalizing possibility of distinguishing between different expansion histories using phase transitions.

\maketitle

\section{Introduction}

Primordial stochastic gravitational waves from first order cosmological phase transitions have become a new cosmic frontier to probe
particle physics beyond the standard model~\cite{Caprini:2015zlo,Weir:2017wfa,Mazumdar:2018dfl,Bertone:2019irm,Caprini:2019egz,Barausse:2020rsu}. Alongside extensive studies on the theory side, direct searches for
stochastic gravitational waves at LIGO and Virgo have also been performed using their O1 and O2 data sets~\cite{TheLIGOScientific:2016dpb,LIGOScientific:2019vic}. 
Perhaps even more significantly, many space-based detectors have been proposed,
such as the Laser Interferometer Space Antenna (LISA)~\cite{amaroseoane2017laser},  
Big Bang Observer (BBO),  DECi-hertz Interferometer Gravitational wave 
Observatory (DECIGO)~\cite{Yagi:2011wg}, Taiji~\cite{Gong:2014mca}, and Tianqin~\cite{Luo:2015ght}.
They will come online within the next decade or so and can probe lower frequencies 
coming from an electroweak scale phase transition~\cite{Grojean:2006bp,Vaskonen:2016yiu,Dorsch:2016nrg,Beniwal:2018hyi,Kang:2017mkl,Delaunay:2007wb,Chala:2018ari,Ellis:2019flb,Alves:2019igs,Alves:2018jsw,Ellis:2019oqb,Morais:2019fnm,Addazi:2018nzm,Wainwright:2011qy,Zhou:2019uzq,Bernon:2017jgv}.\footnote{Note that they are also poised to probe hidden sector transitions \cite{Schwaller:2015tja,Jaeckel:2016jlh,Chala:2016ykx,Addazi:2016fbj,Chao:2017vrq,Baldes:2017rcu,Addazi:2017gpt,Croon:2018erz,Baldes:2018emh,Fairbairn:2019xog,Dunsky:2019upk,Archer-Smith:2019gzq,Hall:2019rld,Bian:2019szo,Wang:2020jrd} and transitions from multi-step GUT breaking \cite{Croon:2018kqn,Greljo:2019xan,Huang:2020bbe,Brdar:2019fur}}

Precise calculations of the gravitational wave power spectrum are required to have any hope of  inferring parameters of the underlying particle physics model. There have been significant advances in this direction in recent years. In particular, it is now generally accepted that the dominant source for gravitational wave production in a
thermal plasma is the sound waves~\cite{Hindmarsh:2013xza}, although a more precise understanding of the onset of the turbulence 
is still needed to settle this issue. For the acoustic production
of gravitational waves, many large scale numerical simulations have been performed~\cite{Hindmarsh:2015qta,Hindmarsh:2017gnf}, 
with the result that standard spectral formulae are now available  for general use. These results have also been understood reasonably well for relatively weak transitions, through the theoretical modeling 
of the hydrodynamics~\cite{Espinosa:2010hh} and with the recently proposed sound shell model~\cite{Hindmarsh:2016lnk,Hindmarsh:2019phv}. 

The first major goal of this paper is to undertake a careful analysis of the gravitational wave power spectrum in a generic expanding universe. This is necessary, 
since the standard result for the spectrum is obtained in  Minkowski spacetime where the effect of the expansion of the universe is neglected. 
In the Minkowski spacetime, the spectrum is proportional to $H_{\ast} \tau_{\text{sw}}$ as derived in Ref.~\cite{Hindmarsh:2015qta}, where
the generalization to the expanding universe with radiation domination 
was also carried out based on rescaling properties of the fluid. It was concluded that the effective lifetime of the sound waves is a Hubble time when comparing 
this spectrum with that derived in the Minkowski spacetime. The reason that this conclusion was reached is due to the absence of the term $H_{\ast} \tau_{\text{sw}}$
in the spectrum for radiation dominated universe and the otherwise very similar form as in Minkowski spacetime (see Appendix~\ref{sec:RDeta} for a re-derivation of this result). 
Later studies suggest that the lifetime generally is smaller than a Hubble time such that $H_{\ast} \tau_{\text{sw}} < 1$~\cite{Ellis:2019oqb,Ellis:2018mja,Ellis:2020awk,Caprini:2019egz}. 
This, when combined with the 
Minkowski result that the spectrum is proportional to $H_{\ast} \tau_{\text{sw}}$, leads to the conclusion that there is a suppression of the spectrum 
when compared with the case when $H_{\ast} \tau_{\text{sw}} = 1$ is used.
We note in retrospect that the spectrum found in above radiation dominated universe is obtained assuming actually an infinite lifetime of the source, i.e., 
$\tau_{\text{sw}} \rightarrow \infty$ and the correct dependence on $\tau_{\text{sw}}$ is a different one. It is the purpose of this paper to provide an
accurate $\tau_{\text{sw}}$ dependence for the spectrum and show its implications.
Moreover the role of the expansion in the process of the phase transition and in the calculation of the spectrum has not been fully revealed. 
We thus present a comprehensive and very careful analysis of the spectrum, clarifying subtle issues when the  calculation is generalized from Minkowski spacetime to an expanding universe, and ultimately providing an accurate spectrum in a standard radiation dominated 
universe and in other expansion scenarios. We also perform a detailed calculation of the nucleation and growth of bubbles in an expanding background, including tracking the shrinking volume available for new bubbles to nucleate in as well as the total area of uncollided walls. Both are needed for an accurate understanding of how the volume fraction and mean bubble separation evolve throughout the phase transition. We then derive and solve the equations governing the evolution of the fluid velocity field in an expanding Universe and then proceed to a derivation of the spectrum for different expansion scenarios. \par

The second major goal of this paper is encapsulated in the title: after  having calculated the gravitational wave spectrum in an expanding universe, we want to explore the extent to which the phase transition can distinguish between \textit{different expansion histories}. In other words, we would like to interrogate how well a phase transition can serve as a \textit{cosmic witness}. This is important, since growing evidence suggests that the standard assumption of radiation domination prior to Big Bang Nucleosynthesis may be too naive~\cite{Kane:2015jia, Allahverdi:2020bys}. An early matter dominated era, for example, is motivated by the cosmological moduli problem \cite{Banks:1995dp, Banks:1995dt, Coughlan:1983ci, Banks:1993en}, hints from dark matter searches \cite{Dutta:2009uf, Allahverdi:2012gk, Acharya:2008bk, Acharya:2009zt, Erickcek:2015bda, Delos:2019dyh, Erickcek:2015jza, Drees:2017iod, Cosme:2020mck}, and perhaps even baryogenesis \cite{Allahverdi:2010im}. Another possibility of a non-standard expansion history is kination, which we do not cover in this paper but can be explored by our methods  \cite{Pallis:2005bb,Lankinen:2019ifa,Nakayama:2010kt,Pallis:2005hm,Grin:2007yg,Dimopoulos:2018wfg, Redmond:2017tja,Bettoni:2018pbl,Bettoni:2019dcw,Bhattacharya:2019bvk}. We note that gravitational waves have been previously employed to investigate early universe cosmology~\cite{Barenboim:2016mjm, Bernal:2019lpc, Caprini:2018mtu, DEramo:2019tit, Geller:2018mwu, Cui:2017ufi}. 

%The study most similar in spirit to ours is \cite{Cui:2017ufi}, who performed the same cosmic ``archaeology" but used cosmic strings. 

%These include an early matter dominated era~\cite{Kane:2015jia, Allahverdi:2020bys, Erickcek:2015bda, Barenboim:2016mjm, Bernal:2019lpc, Caprini:2018mtu, DEramo:2019tit},
% 
%Drees:2017iod,Betancur:2018xtj,Drees:2018dsj,Drees:2018dsj,Bernal:2019lpc,Arias:2019uol,Allahverdi:2019jsc,Cosme:2020mck},
%

%an era dominated by the kinetic energy of some scalar field called the 
%kination \cite{Pallis:2005bb,Lankinen:2019ifa,Nakayama:2010kt,Pallis:2005hm,Grin:2007yg,Cui:2017ufi,Dimopoulos:2018wfg}, a transient inflationary era, or an era dominated by a field with generic equation of state~\cite{Bernal:2019lpc}.

%Reliable applications of this kind demands a faithful calculation of the gravitational waves in the given modified cosmic context.

Our goal is to provide a general theoretical framework to calculate the gravitational wave spectrum 
in different cosmic expansion histories. This includes scrutiny for changes in different aspects.
The dynamics of the phase transition in an expanding universe is studied in Sec.~\ref{sec:dynamics}, 
the velocity field power spectrum is calculated in Sec.~\ref{sec:fluid} and the gravitational wave spectrum in Sec.~\ref{sec:spectrum}.
The main findings of the first two aspects are as follows.
\begin{itemize}
    \item[1.] The mean bubble separation $R_{\ast}$ is related to $\beta$ through a generalized relation for the exponential nucleation (Eq.~\ref{eq:Rsfinal}):
\begin{eqnarray}
R_{\ast}(t) = \frac{a(t)}{a(t_f)} (8\pi)^{1/3} \frac{v_w}{\beta(v_w)} ,
\end{eqnarray}
where $t_f$ is the time when the false vacuum fraction is $1/e$, at which $\beta(v_w)$ is evaluated, and $\beta(v_w)$ can vary by $\sim 20\%$ for different $v_w$.
This relation is also confirmed by numerical calculations and is accurate up to an uncertainty of $2\%$. If one uses the conformal version of $R_{\ast}$ and $\beta$, then
they satisfy the same relation as in Minkowski spacetime (see Eq.~\ref{eq:Rscbetac}).

\item[2.] We derived the bubble lifetime distribution in a generic expanding universe in Eq.~\ref{eq:nbc0}, and the conformal lifetime $\eta_{\text{lt}}$
rather than ordinary lifetime $t_{\text{lt}}$ should be used. It coincides with the distribution $e^{-\tilde{T}}$ found in Minkowski spacetime~\cite{Hindmarsh:2019phv}
for exponential nucleation.

\item[3.] We derived the full set of differential equations in an expanding universe for the fluid and order parameter field model as used in numerical simulations. We find that in the bubble expansion phase the full field equations do not admit rescalings of the quantities that would reduce the expressions to their counterparts in  Minkowski spacetime; this rescaling does, however, work in the bag equation of state model. This implies the velocity profile maintains the same form when appropriate 
rescalings and variable substitutions are used.

\item[4.] We generalized the sound shell model to an expanding universe and calculated the velocity field power spectrum~\cite{Hindmarsh:2016lnk,Hindmarsh:2019phv}.

\end{itemize}

For the gravitational wave energy density spectrum, the main results are:
\begin{itemize}
\item[1.] The peak amplitude of the gravitational wave spectrum visible today has the form (see Eq.~\ref{eq:OmegaFinal})
\begin{eqnarray}
h^2 \Omega_{\text{GW}} =  8.5 \times 10^{-6} \left(\frac{100}{g_{s}(T_e)}\right)^{1/3} \Gamma^2 \bar{U}_f^4  
\left[
\frac{H_s}{\beta(v_w)}
\right]
v_w 
\times
\Upsilon .
\end{eqnarray}
Here $\Gamma \sim 4/3$ is the adiabatic index, $g_s(T_e)$ is the relativistic degrees of freedom for entropy at $T_e$ when the gravitational wave
production ends, $\bar{U}_f$ is the root mean square fluid velocity (see Fig.~\ref{fig:uf}),
$v_w$ is the wall velocity, $H_s$ is the Hubble rate when the source becomes active, and ${\Upsilon}$ 
is the suppression factor arising from the finite lifetime, $\tau_{\text{sw}}$, of the sound waves. 
For radiation domination, it is given by
\begin{equation}
\Upsilon = 1 - \frac{1}{\sqrt{1 + 2 \tau_{\text{sw}} H_s}} ,
\end{equation}
where the standard spectrum generally used corresponds to the asymptotic value $\Upsilon=1$ when $\tau_{\text{sw}} H_s \rightarrow \infty$. 
However the onset of non-linear shocks and turbulence which can disrupt the sound wave source occurs at around
$\tau_{\rm sw} H_s \sim H_s R_{\ast} /\bar{U}_f$. 
This means the asymptotic value will not be reached and there is a suppression to the standard spectrum.
In Fig. \ref{fig:suppression} we compare our result with the suppression factor recently proposed in \cite{Ellis:2019oqb}  (see also \cite{Ellis:2018mja,Ellis:2020awk}).
Similarly, the spectrum for matter domination has also been derived in our work and a similar suppression factor $\Upsilon$ is observed, which has an asymptotic value of $2/3$.
\begin{figure}
    \centering
    \includegraphics[width=0.6\textwidth]{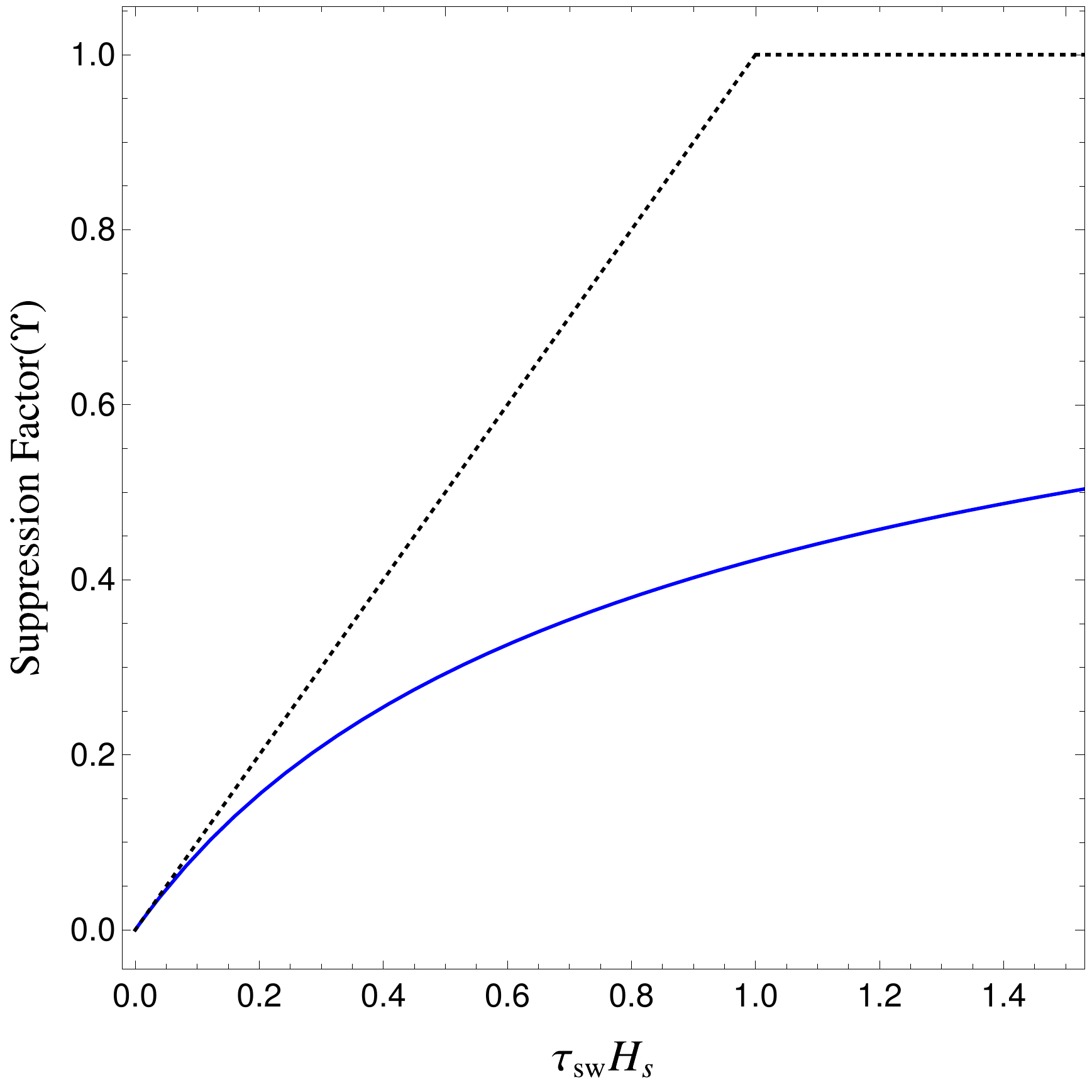}
    \caption{The suppression factor (blue solid line) 
    as a function of the lifetime of the dominant source, the sound waves, in unit of the Hubble time at $t_s$, the time when the source becomes active.
    The black dashed line denotes ${\rm Min}[\tau_{\rm sw} H_s,1]$.
    \label{fig:suppression}
    } 
\end{figure}
 \par 

%Throughout, three different aspects will be scrutinized for changes: 1) the process of phase transition
%and the various parameters characterizing it, such as the nucleation temperature $T_n$, the mean bubble separation $R_{\ast}$, the 
%bubble lifetime distribution, etc.; 2) the set of equations governing the evolution of the fluid and the scalar field and the resulting velocity power spectrum
%of the sound waves; 3) the gravitational wave power spectrum. 

%With the above goals in mind, we lay out the framework for calculations in a generic expanding universe and apply it on two representative expansion scenarios:
%the standard radiation dominated(RD) and an early matter dominated(MD) universe. Our main results are summarized as follows.

\item[2.] We find a change to the spectral form, depending upon whether the phase transition occurs during a period of matter or radiation domination. The change in the form is not leading order, due to the fact that the velocity profiles remain largely unchanged and that 
the autocorrelation time of the source is much smaller than the duration of the transition.
This is in contrast to gravitational waves generated from cosmic strings \cite{Cui:2017ufi}. Even then, the modification of the spectrum presents an enticing possibility that the gravitational waves formed during a phase transition can bear witness to an early matter dominated era. We leave a further detailed exploration of the change of the spectral form for future work. 
\end{itemize}

The remainder of this paper is organized as follows. 
We firstly lay out the theoretical framework for the stochastic gravitational wave calculation in the next Sec.~\ref{sec:framework} and
study the details of the phase transition dynamics in an expanding universe in  Sec.~\ref{sec:dynamics}. After that, we summarize the full set of
fluid equations applicable in an expanding universe and study the velocity profile as well as the velocity power spectrum using the sound shell model in Sec.~\ref{sec:fluid}.
We then analytically calculate  the gravitational waves from sound waves in both radiation dominated and matter dominated scenarios in Sec.~\ref{sec:spectrum}.
We summarize our results in Sec.~\ref{sec:summary}.

\section{Theoretical Framework\label{sec:framework}}

In this section, we set up the framework for calculating the stochastic gravitational waves in the presence of a source, which
also serves to define our notation. The power spectrum of the gravitational waves, as will be discussed, depends on the unequal time 
correlator of the source. Therefore this correlator is of central importance in this work and is discussed in the second subsection. 

\subsection{Gravitational Waves}
The gravitational wave is the transverse traceless part of the perturbed metric. Neglecting the non-relevant scalar and vector perturbations, the 
metric is defined in the FLRW universe as:
\begin{eqnarray}
d s^2 = - d t^2 + a(t)^2 (\delta_{ij} + h_{ij}(\mathbf{x})) d\mathbf{x}^2 ,
\end{eqnarray}
where $h_{ij}$ is only the transverse traceless part of the perturbed $3\times 3$ metric matrix (see, e.g.,~\cite{Weinberg:2008zzc} for a detailed discussion).
It is convenient, most often, to work in Fourier space, with the following convention:
\begin{eqnarray}
&& h_{ij}(t, \mathbf{x}) = \int \frac{d^3 q}{(2\pi)^3} e^{i \mathbf{q}\cdot \mathbf{x}} h_{ij}(t,\mathbf{q}) , 
\end{eqnarray}
where $\bf{q}$ is the comoving wavenumber, in accordance with the comoving coordinate $\bf{x}$. The physical coordinate is $a \bf{x}$ and the physical wavenumber is
${\bf{q}}/a$. The Fourier component $h_{ij}(t,\mathbf{q})$ is thus of dimension $-3$.

Gravitational waves are sourced by the similarly defined transverse traceless part of the perturbed 
energy momentum tensor of the matter content, defined by~\cite{Weinberg:2008zzc}
\begin{eqnarray}
T_{ij} = a^2 \pi^T_{ij} + \cdots ,
\label{eq:piT}
\end{eqnarray}
where ``$\cdots$'' denotes the neglected non-relevant parts. Its Fourier transform is defined by
\begin{eqnarray}
 \pi_{ij}^T(t, \mathbf{x}) = \int \frac{d^3 q}{(2\pi)^3} q e^{i \mathbf{q}\cdot \mathbf{x}} \pi^T_{ij}(t,\mathbf{q}) .
\end{eqnarray}
Since $\pi_{ij}^T$ is of dimension 4, the dimension of its Fourier component $\pi^T_{ij}(t,\mathbf{q})$ is 1. 
The Einstein equation leads to a master equation governing the time evolution of each Fourier component of the gravitational waves, which 
is decoupled from the scalar and vector perturbations,
\begin{eqnarray}
h_{ij}^{\prime \prime}(t,\mathbf{q}) + 2 \frac{a^{\prime}}{a} h^{\prime}_{ij}(t,\mathbf{q}) +q^2 h_{ij}(t,\mathbf{q})= 16 \pi G a^2 \pi_{ij}^T(t,\mathbf{q}) \,\,.
\label{eq:hq}
\end{eqnarray}
Here $\prime \equiv \partial/\partial \eta$, with $\eta$ being the conformal time. 
Derivatives with respect to the coordinate time will be denoted by 
a dot. 
The gravitational wave energy density, as denoted by $\rho_{\text{GW}}$ here, is defined as
\begin{eqnarray}
\rho_{\text{GW}}(t) = \frac{1}{32 \pi G} \langle \dot{h}_{ij}(t,\mathbf{x}) \dot{h}_{ij}(t,\mathbf{x}) \rangle ,
\end{eqnarray}
with the angle brackets, $\langle \cdots \rangle$, denoting both the spatial and ensemble average. Due to the overall spatial homogeneity of the universe, 
we can define the power spectrum of the derivative of the gravitational wave amplitude as:
\begin{eqnarray}
\langle \dot{h}_{ij}(t, \mathbf{q}_1) \dot{h}_{ij}(t, \mathbf{q}_2) \rangle = (2\pi)^{3} \delta^3(\mathbf{q}_1+\mathbf{q}_2) P_{\dot{h}}(q_1,t) .
\label{eq:powerspectrum}
\end{eqnarray}
Then the gravitational wave energy density follows
\begin{eqnarray}
\rhogw(t) = \frac{1}{32 \pi G} \frac{1}{2\pi^2} \int dq\ q^2 P_{\dot{h}}(t, q) ,
\end{eqnarray}
and the gravitational wave energy density spectrum:
\begin{eqnarray}
\frac{d \rhogw(t)}{d \ln q} = \frac{1}{64 \pi^3 G} q^3 P_{\dot{h}}(t, q) .
\label{eq:rho_gw}
\end{eqnarray}
It is conventional to use the dimensionless energy density fraction of the gravitational waves 
$\Omega_{\text{GW}}(t) = \rhogw(t)/\rho_c(t)$ where $\rho_c$ is the critical energy density at time $t$. 
The corresponding dimensionless version of the spectrum is~\footnote{$\mathcal{P}_{\text{GW}}$ is also denoted as $\Omega_{\text{GW}}(t,q)$.}
\begin{eqnarray}
\mathcal{P}_{\text{GW}}(t, q) \equiv \frac{d \Omega_{\text{GW}}(t)}{d \ln q} = \frac{1}{24 \pi^2 H^2} q^3 P_{\dot{h}}(t, q)
=
\frac{1}{24 \pi^2 H^2 a^2} q^3 P_{h^{\prime}}(t, q)
,
\end{eqnarray}
where in the last step $P_{h^{\prime}}(t, q)$ is defined by replacing $\dot{h}$ with $h^{\prime}$ in Eq.~\ref{eq:powerspectrum}.

We thus need to solve for $h_{ij}(\eta,\mathbf{q})$ by solving Eq.~\ref{eq:hq} together with equations governing the evolution of the source. 
We will follow the conventional approach by neglecting the back-reaction of the 
metric on the source and calculate the stress tensor with a modelling of the phase transition process. 
Once $\pi^T_{ij}(t,\mathbf{q})$ is provided in this way, then $h_{ij}(t,\mathbf{q})$ can be solved from Eq.~\ref{eq:hq} with Green's function and with the following boundary conditions
\begin{eqnarray}
G(\tilde\eta \leqslant \tilde \eta_0) = 0, \quad \quad \frac{\partial G(\tilde \eta, \tilde \eta_0)}{\partial \tilde{\eta}}|_{\tilde{\eta} = \tilde \eta_0^+} = 1 ,
\label{eq:boundary}
\end{eqnarray}
where $\tilde{\eta} = q \eta$, which is a dimensionless quantity and $\tilde \eta_0$ is the time when the phase transition starts. 
With the Green's function, the solution of the inhomogeneous Eq.~\ref{eq:hq} is given by
\begin{eqnarray}
h_{ij}(t, \mathbf{q}) = 16 \pi G \int_{\tilde{\eta}_0}^{\tilde{\eta}} d \tilde{\eta}^{\prime} 
G(\tilde{\eta}, \tilde{\eta}^{\prime}) \frac{a^2({\eta}^{\prime}) \pi_{ij}^T(\eta^{\prime}, \mathbf{q})}{q^2} ,
\end{eqnarray}
and its derivative with respect to the conformal time follows simply:
\begin{eqnarray}
h_{ij}^{\prime}(\eta, \mathbf{q}) = 16 \pi G \int_{\tilde{\eta}_0}^{\tilde{\eta}} 
d \tilde{\eta}^\prime
\frac{\partial G(\tilde{\eta}, \tilde{\eta}^{\prime})}{\partial \tilde {\eta}}
\frac{a^2({\eta}^{\prime}) \pi^T_{ij}(\eta^{\prime}, \mathbf{q})}{q} .
\end{eqnarray}
Then we can calculate the 2-point correlation function:
\begin{eqnarray}
\langle h_{ij}^{\prime}(\eta, \mathbf{q}_1) h_{ij}^{\prime}(\eta, \mathbf{q}_2)\rangle
=
(16 \pi G)^2 \int_{\tilde{\eta}_0}^{\tilde{\eta}} d \tilde{\eta}_1 \int_{\tilde{\eta}_0}^{\tilde{\eta}} d \tilde{\eta}_2 
\frac{\partial G(\tilde{\eta}, \tilde{\eta}_1)}{\partial \tilde{\eta}}
\frac{\partial G(\tilde{\eta}, \tilde{\eta}_2)}{\partial \tilde{\eta}}
\nonumber \\
\times \frac{a^2({\eta}_1) a^2({\eta}_2)}{q^2}
\langle \pi_{ij}^T(\eta_1, \mathbf{q}_1) \pi_{ij}^T(\eta_2, \mathbf{q}_2)\rangle .
\label{eq:correh}
\end{eqnarray}
Supposing that the gravitational wave generation finishes at $\tilde{\eta}_f$, the upper limits for the integrals in the expression above will be $\tilde{\eta}_f$. 
Subsequently, the energy density of the gravitational waves for modes inside the horizon will be simply diluted as $1/a^4$.
\comment{
Then the energy density of the previously generated gravitational waves should decay as $1/a^4$. This is indeed the case here, as 
we have already seen $\partial G(\tilde{\eta}, \tilde{\eta}_1)/{\partial \tilde{\eta}} \propto 1/a$ for modes inside
the horizon and the l.h.s will scale as $1/a^2$, which leads to $\mathcal{P}_{\text{GW}} \propto 1/a^4$.
}
We thus see that at the core of the gravitational wave energy density spectrum calculation is the unequal time correlator (UETC) of $\pi_{ij}^T$. It can be parametrized in the following way due to the overall spatial homogeneity of the universe~\cite{Hindmarsh:2015qta}
\begin{eqnarray}
\langle \pi_{ij}^T(\eta_1, \mathbf{q}_1) \pi_{ij}^T(\eta_2, \mathbf{q}_2)\rangle = \Pi^2(q_1, \eta_1, \eta_2) (2 \pi)^{3} \delta^3(\mathbf{q}_1 + \mathbf{q}_2) .
\label{eq:correlator-piT}
\end{eqnarray}
It is obvious that the dimension of $\Pi^2(k, \eta_1, \eta_2)$ is 5.

\subsection{Unequal Time Correlator of the Fluid Stress Energy Tensor}

Let us first write down the energy momentum tensor of the matter content in the universe. Here we keep the dominant contribution from the fluid 
and assume the fluid velocities are non-relativistic following Ref.~\cite{Hindmarsh:2019phv}, then
\begin{eqnarray}
&& T_{ij} = a^2 \left[ p \delta_{ij} + (p + e) \gamma^2 v^i v^j \right],  \nonumber \\
&& T_{i0} = a \left[- (p + e) \gamma^2 v^i \right], \nonumber \\
&& T_{00} = \gamma^2 (e + p v^2) ,
\label{eq:Tij}
\end{eqnarray}
where $e$ is the energy density, $p$ is the pressure and the velocity is defined w.r.t the conformal time $v^i = d x^i/d \eta$.
Then, comparing with Eq.~\ref{eq:piT} and neglecting the non-relevant parts, we have
\begin{eqnarray}
\pi_{ij} = (p + e) \gamma^2 v^i v^j ,
\end{eqnarray}
Here the scale factor dependent $(p+e)$, takes its
homogeneous value (defined with a bar) to leading order $\bar{e} + \bar{p} \equiv \bar{\omega}$ which scales as $1/a^4$, and $\gamma$ is the Lorentz factor.
The calculation of the correlator of $\pi_{ij}^{T}$ parallels that in Minkowski spacetime:

\begin{eqnarray}
&& \langle \pi^T_{ij}(\eta_1, \mathbf{k}) \pi^T_{ij}(\eta_2, \mathbf{q}) \rangle \nonumber \\
&=& \Lambda_{ij,kl}(\hat{\mathbf{k}})
\frac{1}{(2\pi)^6} \int d^3 \mathbf{x} \int d^3 \mathbf{y} e^{-i\mathbf{k}\cdot \mathbf{x}} e^{-i\mathbf{q}\cdot \mathbf{y}}
\langle \pi_{kl}^T(\eta_1, \mathbf{x}) \pi_{ij}^T(\eta_2, \mathbf{y}) \rangle, \nonumber \\
&=& \Lambda_{ij,kl}(\hat{\mathbf{k}}) \bar{\omega}^2
\frac{1}{(2\pi)^6}  \int d^3 \mathbf{x} \int d^3 \mathbf{y} e^{-i\mathbf{k}\cdot \mathbf{x}} e^{-i\mathbf{q}\cdot \mathbf{y}}
\langle v^k(\eta_1, \mathbf{x}) v^l(\eta_1, \mathbf{x}) v^i(\eta_2, \mathbf{y}) v^j(\eta_2, \mathbf{y}) \rangle, \nonumber \\
&=& \bar{\omega}^2 \Lambda_{ij,kl}(\hat{\mathbf{k}}) \frac{1}{(2\pi)^{12}}
\int d^3 \mathbf{q}_1 \int d^3 \mathbf{q}_3 
\underbrace{
\langle 
\tilde{v}_{\mathbf{q}_1}^k(\eta_1) \tilde{v}_{\mathbf{q}_1-\mathbf{k}}^{l \ast}(\eta_1)
\tilde{v}_{\mathbf{q}_3}^i(\eta_2) \tilde{v}_{\mathbf{q}_3-\mathbf{q}}^{j \ast}(\eta_2)
\rangle
}_{\equiv X^{klij}}.
\end{eqnarray}
Here $\Lambda_{ij,kl}$ is the standard projection operator and  $\Lambda_{ij,kl}(\hat{k}) = P_{ik}(\hat{k}) P_{jl}(\hat{k}) -\frac{1}{2} P_{ij}(\hat{k}) P_{kl}(\hat{k})$
with $P_{ij}(\hat{k}) = \delta_{ij} - \hat{k}^k \hat{k}^j$.
$\tilde{v}_{\mathbf{q}}^i$ is the Fourier transform of the velocity field
$v^i(\mathbf{x})$.
Due to the nature of the first order phase transition process and according to the central limit theorem, $\tilde{v}^i_{\mathbf{q}}(\eta)$ follows the Gaussian distribution to a good approximation. 
Also as in Ref.~\cite{Hindmarsh:2019phv}, we neglect the rotational component, then the two point correlator can be defined in the following
way:
\begin{eqnarray}
\langle \tilde{v}_{\mathbf{q}}^i(\eta_1) \tilde{v}_{\mathbf{k}}^{j \ast}(\eta_2)\rangle = \delta^{3}(\mathbf{q}-\mathbf{k}) \hat{q}^i \hat{k}^j G(q, \eta_1, \eta_2) ,
\label{eq:vv}
\end{eqnarray}
and higher order correlators can be reduced to the two point correlator. 
Defining $\tilde{\mathbf{q}}_1 \equiv \mathbf{q}_1 - \mathbf{k}$ and $\tilde{\mathbf{q}}_3 \equiv \mathbf{q}_3 - \mathbf{q}$, then
\begin{eqnarray}
X^{klij} &=&
\langle \tilde{v}_{\mathbf{q}_1}^k(\eta_1) \tilde{v}_{\tilde{\mathbf{q}}_1}^{l \ast}(\eta_1) \rangle
\langle \tilde{v}_{\mathbf{q}_3}^i(\eta_2) \tilde{v}_{\tilde{\mathbf{q}}_3}^{j \ast}(\eta_2) \rangle
+
\langle \tilde{v}_{\mathbf{q}_1}^k(\eta_1) \tilde{v}_{\mathbf{q}_3}^{i}(\eta_2) \rangle
\langle \tilde{v}_{\tilde{\mathbf{q}}_1}^{l \ast}(\eta_1) \tilde{v}_{\tilde{\mathbf{q}}_3}^{j \ast}(\eta_2) \rangle \nonumber \\
&&
+
\langle \tilde{v}_{\mathbf{q}_1}^k(\eta_1) \tilde{v}_{\tilde{\mathbf{q}}_3}^{j \ast}(\eta_2) \rangle
\langle \tilde{v}_{\tilde{\mathbf{q}}_1}^{l \ast}(\eta_1) \tilde{v}_{\mathbf{q}_3}^{i}(\eta_2) \rangle .
\end{eqnarray}
The first term contributes trivially to $\mathbf{k}=0$ and, collecting all other contributions, we have
\begin{eqnarray}
\langle \pi^T_{ij}(\eta_1, \mathbf{k}) \pi^T_{ij}(\eta_2, \mathbf{q}) \rangle
= \delta^3(\mathbf{k} + \mathbf{q}) \bar{\omega}^2 \frac{1}{(2\pi)^6} \int d^3 \mathbf{q}_1 G(q_1, \eta_1, \eta_2) G(\tilde{q}_1, \eta_1, \eta_2)(1-\mu^2)^2
\frac{q_1^2}{\tilde{q}_1^2} .\nonumber \\
\label{eq:correpi}
\end{eqnarray}
Comparing with Eq.~\ref{eq:correlator-piT}, it follows that
\begin{eqnarray}
\Pi^2(k, \eta_1 - \eta_2) = \bar{\omega}^2 \int \frac{d^3 q}{(2 \pi)^3} G(q,\eta_1, \eta_2) G(\tilde{q}, \eta_1, \eta_2) \frac{q^2}{\tilde{q}^2} (1-\mu^2)^2 ,
\label{eq:PI2}
\end{eqnarray}
where $\tilde{q}=|\mathbf{q}-\mathbf{k}|$ and $\mu = \hat{\mathbf{q}} \cdot \hat{\mathbf{k}}$. 
Here $\Pi^2$ depends on $\eta_1 - \eta_2$ rather than on $\eta_1$ and $\eta_2$ separately. This is because the source is 
largely stationary. 

We will later see that the fluid equations maintain the same form as in the Minkowski spacetime once
properly rescaled quantities and previously defined $v^i(\mathbf{x})$ are used (see also Ref.~\cite{Hindmarsh:2015qta}). In particular it means
that we can define a rescaled stress energy tensor ($\tilde{\pi}^T_{ij}$) for the fluid:
\begin{equation}
    \pi_{ij}^T(\mathbf{q},\eta) = \frac{a_s^4}{a^4(\eta)}\tilde{\pi}_{ij}^T(\mathbf{q}, \eta),
\end{equation}
where $a_s$ is a reference scale factor when the source becomes active. 
Similarly we can define a rescaled and dimensionless two point correlator $\tilde{\Pi}$ following Ref.~\cite{Hindmarsh:2015qta} by
\begin{equation}
\Pi^2(q,t_1,t_2) \equiv \frac{a_s^8}{a^4(\eta_1)a^4(\eta_2)} \left[ \left( \bar{\tilde{e}} + \bar{\tilde{p}} \right) 
\bar{U}_f^2\right]^2 L_f^3 \tilde{\Pi}^2(q L_f,q \eta_1,k\eta_2),
\label{eq:PiTilde}
\end{equation}
where $\bar{\tilde{e}}$ and $\bar{\tilde{p}}$ are the rescaled average energy density and pressure, which correspond to the quantities measured
at $t_s$. The quantity $\bar{U}_f$ describes the magnitude of the fluid velocity and is dimensionless. The correlator, $\Pi^2$, on the left hand side of the equation has
dimension 5. Therefore, the additional length factor $L_f^3$ is inserted here to make $\tilde{\Pi}$ dimensionless. Since this length scale is free from
the effect of the expanding universe, it is a comoving length scale. It is found from numerical simulations~\cite{Hindmarsh:2015qta,Hindmarsh:2017gnf} that the typical scale in the gravitational 
wave production is the (comoving) mean bubble separation $R_{\ast c}$. So we will choose $L_f = R_{\ast c}$.

The calculation of the UETC requires us to scrutinize the entire process of the phase transition and the gravitational wave production. 
This task can be separated into two parts. The first part is a study of the bulk parameters characterizing the process of the phase transition, which we will
perform in the next section. The second part is understanding the evolution of the source, which  we go on to perform in Sec.~\ref{sec:fluid}.

\section{Dynamics of the Phase Transition\label{sec:dynamics}}

In this section, we study the changes to the dynamics of the phase transition in an expanding universe. 
This includes parameters characterizing the behavior of the bubble formation, expansion and percolation: the bubble nucleation
rate, the fraction of the false vacuum, the unbroken area of the walls at a certain time, etc. These will eventually be incorporated in the  calculation of the velocity power spectrum in the sound shell model. Another set of important quantities characterize the statistics of the bubbles ever formed: the bubble lifetime distribution, 
as well as the bubble number density. These are also needed in the velocity power spectrum calculation. Moreover, the timing of some important steps in the phase transition are also included, like the nucleation temperature and the percolation temperature. Other changes to the parameters entering the gravitational wave power spectrum calculation are also included, with $\beta/H$ a representative example. We now proceed to a detailed discussion of these quantities.

\subsection{Bubble Nucleation Rate}

The first and most basic ingredient in the analysis of a first order cosmological phase transition is the nucleation rate of the 
bubbles in the meta-stable vacuum at finite temperature~\cite{Linde:1980tt,Linde:1981zj}. 
The number of bubbles nucleated per time per physical volume is given by the following formula:
\begin{eqnarray}
p = p_0 \text{exp}\left[-\frac{S_{3,b}(T)}{T}\right] .
\label{eq:nucrate}
\end{eqnarray}
Here $S_3$ is the Euclidean action of the underlying scalar field $\vec{\phi}$ that minimizes the solution
\begin{equation}
    S_3 (\vec{\phi},T) = 4 \pi \int dr r^2 \left[ \frac{1}{2} \left( \frac{d \vec{\phi}(r)}{dr} \right)^2 + V(\vec{\phi},T)\right] ,
\end{equation}
with the following bounce boundary conditions:
\begin{eqnarray}
  \frac{d \vec{\phi}(r)}{d r}\Big\vert_{r=0} = 0, \quad \quad
  \vec{\phi}(r=\infty) =  \vec{\phi}_{\text{out}} ,
\end{eqnarray}
where $\vec{\phi}_{\text{out}}$ are the components of the vacuum expectation value for the scalar field outside the bubble.
%%%%%%%%%%%%%%%%%%%%%%%%%%%%%%%%%%%%%%%%%%%%%%%%%%%%%%%%%%%%%%%%%%%%%
\begin{figure}
\centering
\includegraphics[width=0.6\textwidth]{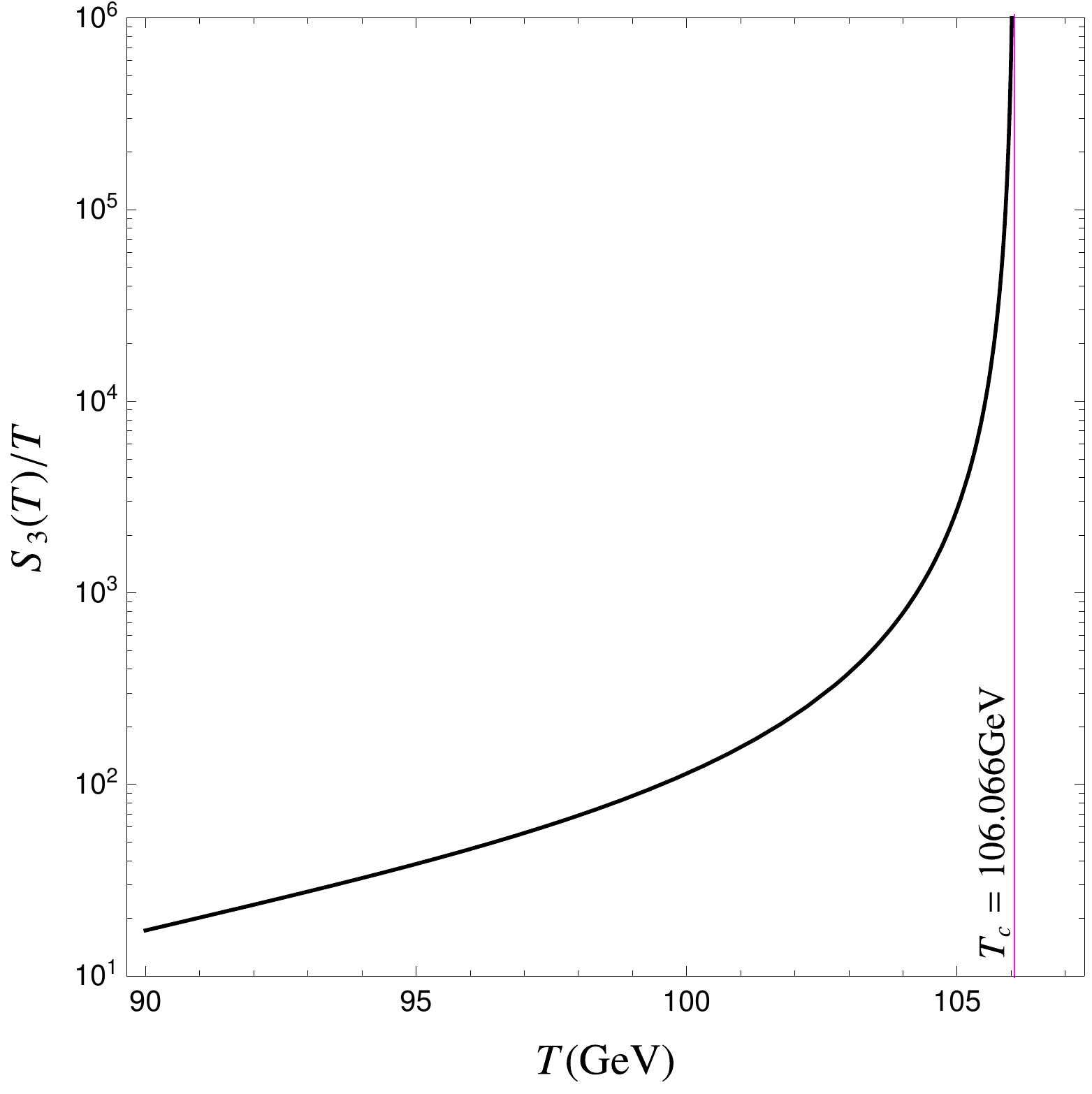}
\caption{\label{fig:S3T}
The representative profile of $S_3(T)/T$ for the example used in Sec.~\ref{sec:dynamics}. 
See Appendix.~\ref{sec:example} for details on how to reproduce this.
}
\end{figure}
%%%%%%%%%%%%%%%%%%%%%%%%%%%%%%%%%%%%%%%%%%%%%%%%%%%%%%%%%%%%%%%%%%%%%
For the pre-factor, we see that $p_0 \propto T^4$ on dimensional grounds, while its precise determination requires integrating out 
fluctuations around the bounce solution (see e.g.,~\cite{Dunne:2005rt,Andreassen:2016cvx} for detailed calculations 
or~\cite{Weinberg:1996kr} for a pedagogical introduction). 

The function $S_3(T)/T$ generally starts from infinity at $T_c$ and drops sharply as temperature decreases,
with a typical profile shown in Fig.~\ref{fig:S3T}. 
Bubbles will be nucleated within a short range of time, say at $t_{\ast}$, when this rate changes slowly, which admits the following Taylor expansion:
\begin{eqnarray}
p(t) = p_0 \text{exp}\left[- S_{\ast} + \beta(t-t_{\ast})\right] ,
\label{eq:pTaylort}
\end{eqnarray}
where $S_{\ast} \equiv S_3(T_{\ast})/T_{\ast}$ and $\beta \equiv d \ln p(t)/dt|_{t=t_{\ast}}$~\footnote{
If there exists a barrier at zero temperature, then $S_3(T)/T$ will reach a minimum, say at $t_{\ast}$. The rate can be expanded around the minimum:
\begin{eqnarray}
p(t) = p_0 \text{exp} \left[-S_{\ast} - \frac{1}{2} \beta_2^2 (t-t_{\ast})^2\right] ,
\end{eqnarray}
with $\beta_2 \equiv S^{\prime \prime}(t_{\ast})$ and the first derivative vanishing. 
The bubble nucleation will happen mostly around $t_{\ast}$, making it look like an instantaneous nucleation~\cite{Hindmarsh:2016lnk}. 
}. More explicitly, we have 
\begin{eqnarray}
\frac{S_3}{T} = \left.\frac{S_3}{T}\right|_{t_{\ast}} + 
\underbrace{\left.\frac{d(S_3/T)}{d T} \frac{d T}{d t}\right|_{t = t_{\ast}}}_{\equiv - \beta} (t - t_{\ast}) ,
\end{eqnarray}
and thus
\begin{eqnarray}
\frac{\beta}{H_{\ast}} = - \frac{1}{H_{\ast}}  \frac{d T}{d t}\left.\frac{d(S_3/T)}{d T}\right|_{t = t_{\ast}} .
\label{eq:betaH}
\end{eqnarray}
We will later see how $t_{\ast}$ should be chosen. For now, we provide a generic expression for $\beta$ during an expanding universe, which
needs the relation between $t$ and $T$.
Suppose the universe is expanding as $a = c_a t^n$ and the radiation sector is expanding adiabatically such that entropy $s_R$ is conserved per comoving volume
for the radiation sector:
\begin{eqnarray}
s_R(T) a^3 = \text{const} .
\end{eqnarray}
Here $s_R \propto T^3$, giving then $T \propto 1/a \propto t^{-n}$. This is the case for a radiation dominated universe, and for 
a matter dominated universe where the non-relativistic matter does not inject entropy to the radiation sector. 
However when the matter decays into radiation, entropy injection into the radiation sector gives a different dependence $T \propto a^{-3/8}$~\cite{Scherrer:1984fd}. 
Generically, we can assume~\footnote{Not to be confused with the Lorentz factor.}
\begin{eqnarray}
T \propto a^{-\gamma}\,\,,
\end{eqnarray}
which then leads to $T = c_T t^{-n \gamma}$, with $c_T$ being another constant.  We thus have
\begin{eqnarray}
\frac{d T}{dt} = - c_T n \gamma\ t^{-n \gamma -1} .
\end{eqnarray}
Moreover $H = \dot{a}/a = n/t$. Then 
\begin{eqnarray}
\frac{1}{H} \frac{d T}{d t} = - c_T \gamma\ t^{-n \gamma} = - \gamma\ T .
\label{eq:dTdt}
\end{eqnarray}
Therefore $\beta/H_{\ast}$ reduces to the following form
\begin{eqnarray}
\frac{\beta}{H_{\ast}} = \gamma \left.T \frac{d(S_3/T)}{d T}\right|_{t=t_{\ast}}  .
\end{eqnarray}
It is obvious from this result that $\beta/H_{\ast}$ does not depend on $n$, i.e., it does not depend on how the scale factor evolves with time but rather 
on how $T$ decreases with the scale factor through $\gamma$. For both the standard radiation dominated universe and an early matter dominated universe wherein the matter is decoupled from the radiation, $\gamma = 1$.
For the matter dominated universe wherein the matter decays into radiation, $\gamma = 3/8$, which gives a smaller $\beta/H_{\ast}$~\cite{Barenboim:2016mjm}.

\subsection{False Vacuum Fraction}

%%%%%%%%%%%%%%%%%%%%%%%%%%%%%%%%%%%%%%%%%%%%%%%%%%%%%%%%%%%%%%%%%%%%%
\begin{figure}[t]
\centering
\includegraphics[width=0.7\textwidth]{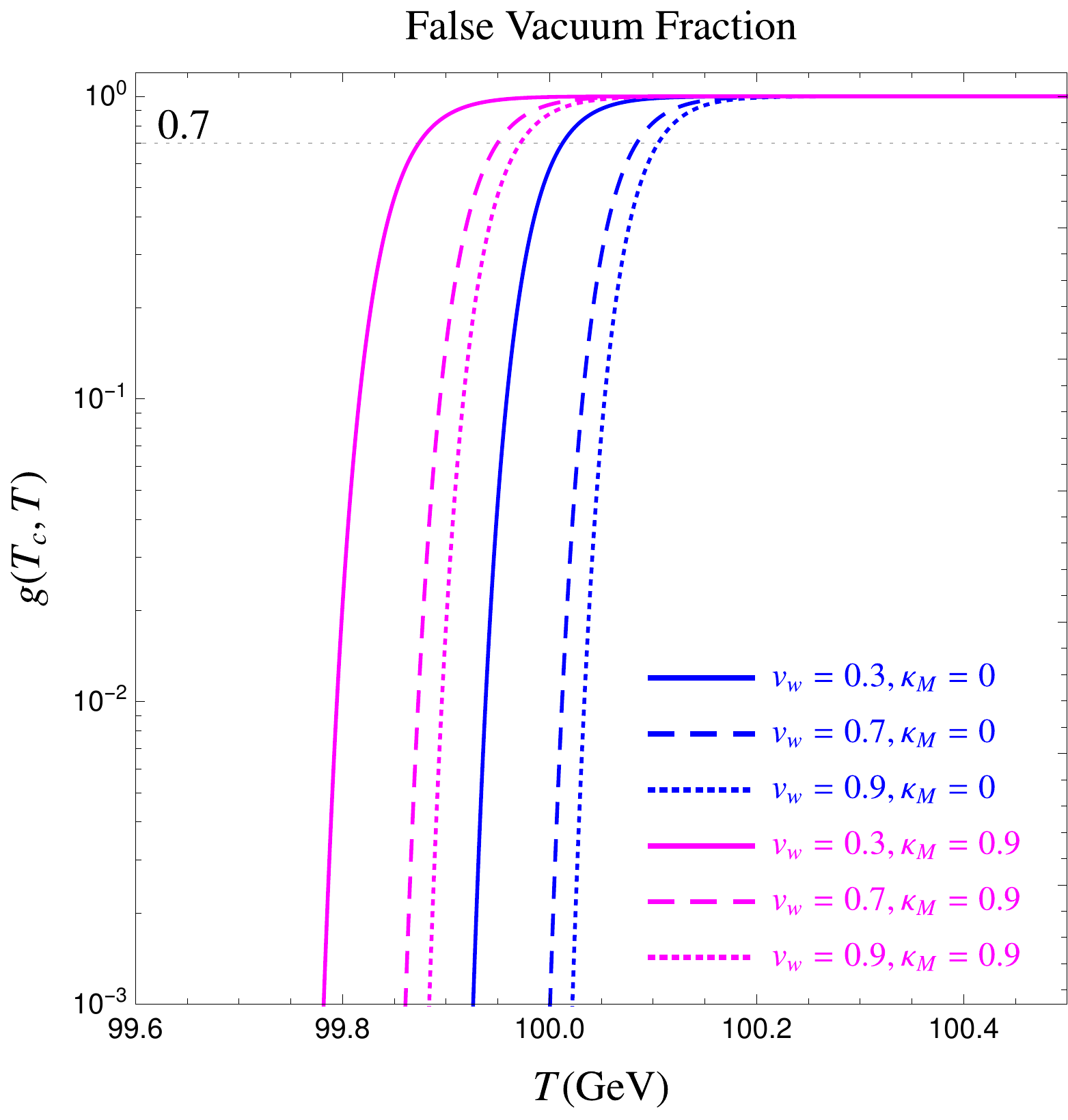}
\caption{\label{fig:g}
The false vacuum fraction as defined Eq.~\ref{eq:fracfalse} for different fractions of matter energy density at $T_c$ ($\kappa_M=0, 0.9$, defined in Eq.~\ref{eq:km}) and for several bubble wall velocities ($v_w=0.3, 0.7, 0.9$). 
The case of $\kappa_M =0$ corresponds to a radiation dominated universe and $\kappa_M = 0.9$ for matter domination. The horizontal line at $g = 0.7$ is
roughly the time when the bubbles percolate.
}
\end{figure}
%%%%%%%%%%%%%%%%%%%%%%%%%%%%%%%%%%%%%%%%%%%%%%%%%%%%%%%%%%%%%%%%%%%%%
The false vacuum fraction $g(t_c, t)$ at $t > t_c$ can be obtained following the derivation in Ref.~\cite{Guth:1981uk}
\begin{eqnarray} 
g(t_c, t) = \text{exp}\left[ - \frac{4 \pi}{3} \int_{t_c}^t d t^{\prime} p(t^{\prime}) a^3(t^{\prime}) r(t^{\prime}, t)^3 \right] \equiv \text{exp}[-I(t)] .
\label{eq:fracfalse}
\end{eqnarray} 
Here $I(t)$ corresponds to the volume of nucleated bubbles per 
comoving volume, double counting the overlapped space between bubbles and virtual bubbles within others. 
$r(t^{\prime}, t)$ is the comoving radius of the bubble nucleated at $t^{\prime}$ and measured at $t$, 
\begin{eqnarray}
r(t^{\prime}, t) = \int_{t^{\prime}}^t d t^{\prime \prime} \frac{v_w}{a(t^{\prime \prime})} .
\end{eqnarray}
For Minkowski spacetime, $r(t^{\prime}, t) = v_w (t-t^{\prime})$. For a FLRW spacetime $r(t^{\prime}, t) = v_w (\eta^{\prime} - \eta)$, which takes the same form
as the Minkowski spacetime, irrespective of the detailed expansion behavior, when conformal time is used.
In obtaining the above results, a constant bubble wall velocity $v_w$ has been assumed and the initial size of the bubble has been neglected. 
This is justified as the initial size is very small.

Eq.~\ref{eq:fracfalse} can be recast in a form that is convenient for calculations, in terms of the temperature. Suppose that the scale factor at the time of
the critical temperature is $a_c$ and that the  scale factor at a later time is related to it by
\begin{eqnarray}
\frac{a}{a_c} \equiv \left(\frac{T_c}{T}\right)^{1/\gamma} .
\end{eqnarray}
%For RD, $\gamma=1$. For MD with a coexisting radiation sector which has no entropy exchange with the matter sector, the temperature is that of
%the radiation and $\gamma=1$. For MD, where the matter content decays and thus injects entropy to radiation, $\gamma=3/8$~\cite{Scherrer:1984fd}.
%Of course this parametrization does not work for all kinds of expansions, and when it happens one can always start with the generic expression in Eq.~\ref{eq:fracfalse}. 
%Assuming that above parametrization is applicable, 
The comoving bubble radius can be conveniently expressed with an integral over temperature:
\begin{eqnarray}
r(T^{\prime}, T) = \frac{v_w}{a_c} \int_T^{T^{\prime}} \frac{d T^{\prime \prime}}{T^{\prime \prime}} \frac{1}{\gamma H(T^{\prime \prime})} 
\left(\frac{T_c}{T^{\prime \prime}}\right)^{-1/\gamma} .
\end{eqnarray}
Accordingly $I(T)$ can be written as
\begin{eqnarray}
I(T) = \frac{4\pi}{3} \int_T^{T_c} \frac{d T^{\prime}}{T^{\prime}}
\frac{1}{\gamma H(T^{\prime})} \bar{p}_0 T^{\prime 4} \text{exp}\left[-\frac{S_3(T^{\prime})}{T^{\prime}}\right]
\left(\frac{T_c}{T^{\prime}}\right)^{3/\gamma} [a_c r(T^{\prime}, T)]^3 .
\label{eq:I}
\end{eqnarray}
Here the factor $\bar{p}_0$ is defined by $p_0 = \bar{p}_0 T^4$ and we choose $\bar{p}_0 = 1$ in the examples of analysis as is usually done in the literature.
A different choice of $\bar{p}_0$ would, of course, affect the resulting false vacuum fraction and thus the relevant temperatures defined~\cite{Moore:1995si}.
Since the focus here is on the changes due to different expansion histories, a fixed choice of $\bar{p}_0$ serves our purpose well.
For the Hubble rate, we need to be more precise with regard to the matter content.
We consider a universe consisting of both radiation and non-relativistic matter and define $\kappa_M$ to be the fraction of 
the total energy density at $T_c$ that is non-relativistic matter:
\begin{eqnarray}
\kappa_M = \left.\frac{\rho_{\text{Matter}}}{\rho_{\text{Total}}}\right|_{T=T_c} .
\label{eq:km}
\end{eqnarray}
We also neglect the vacuum energy for these examples, though it certainly exists during a phase transition.
\begin{eqnarray}
H = H(T_c) \sqrt{\frac{\kappa_M}{y^3} + \frac{1-\kappa_M}{y^4}} , 
\label{eq:HkappaM}
\end{eqnarray}
where $y = a/a(T_c)$. We show in Fig.~\ref{fig:g} the false vacuum fraction during the 
phase transition, for a purely radiation dominated universe with $\kappa_M=0$ and a matter
dominated one with $\kappa_M=0.9$, and for three choices of bubble wall velocities $v_w=0.3,0.7,0.9$. For both choices of $\kappa_M$, it is clear from these figures that 
increasing $v_w$ speeds up the process of phase transition. From $\kappa_M=0$ to $\kappa_M=0.9$,
a larger energy density and thus a larger Hubble rate is obtained, which decreases the function
$r(T^{\prime},T)$ and $I(I)$ and thus slows down the drop of $g(T_c, T)$.

%The horizontal magenta line
%corresponds to the fraction being $0.71$, which is used for the definition of the percolation temperature as the intersection 
%point with the false vacuum fraction curves, to be discussed later.

%%%%%%%%%%%%%%%%%%%%%%%%%%%%%%%%%%%%%%%%%%%%%%%%%%%%%%%%%%%%%%%%%%%%%%
\begin{figure}
\centering
%\includegraphics[width=0.49\textwidth]{Figures/OSA_Area.pdf}
%\includegraphics[width=0.49\textwidth]{Figures/OSA_AreaHubble.pdf}
%\\
\includegraphics[width=0.7\textwidth]{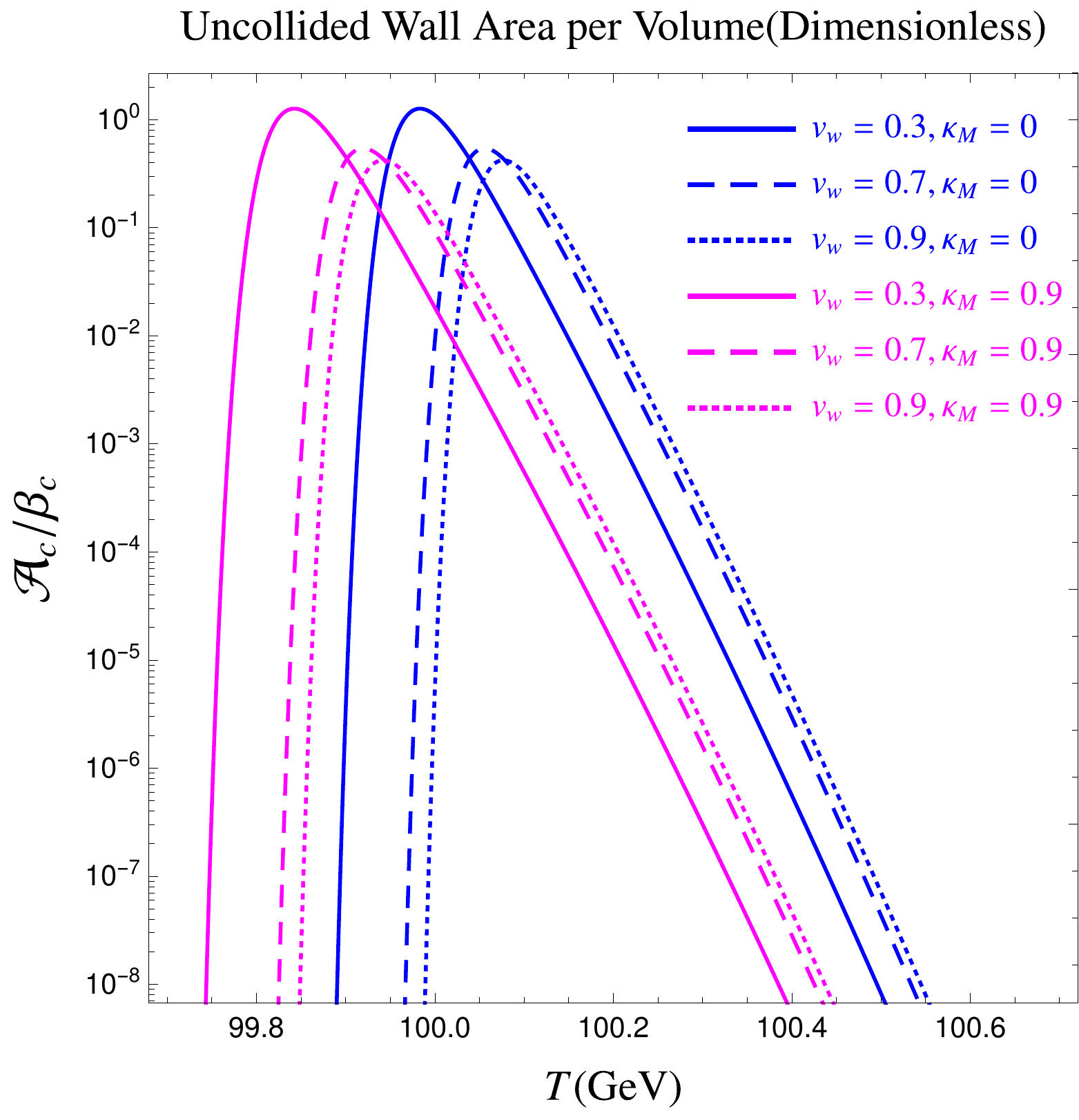}
\caption{
\label{fig:area}
The dimensionless comoving uncollided bubble wall area as defined in 
Eq.~\ref{eq:Ac} and Eq.~\ref{eq:betacdef} for different values of 
$\kappa_M$ (defined in Eq.~\ref{eq:km}) and $v_w$.
}
\end{figure}
%%%%%%%%%%%%%%%%%%%%%%%%%%%%%%%%%%%%%%%%%%%%%%%%%%%%%%%%%%%%%%%%%%%%%%

One often encounters the percolation temperature, which is defined such that the fraction in true vacuum is about $30\%$ of the total volume~\cite{Ellis:2018mja},
i.e., when
\begin{eqnarray}
g(T_c, T_p) \approx 0.7, \quad \text{or} \quad I(T_p) \approx 0.34 ,
\end{eqnarray}
and corresponds to the intersection points of the horizontal line with the curves in Fig.~\ref{fig:g}. Since different choices of $v_w$ 
and $\kappa_M$ lead to different $g(T_c, T)$, the corresponding values of $T_p$ are also different.

\subsection{Unbroken Bubble Wall Area}

With the false vacuum fraction in Eq.~\ref{eq:fracfalse}, the unbroken bubble wall area during the phase transition can be derived~\cite{Hindmarsh:2019phv}
and will be used in the derivation of the bubble
lifetime distribution. Consider a comoving volume of size $V_c$ and a sub-volume occupied by false vacuum $V_{c,\text{False}}$. Then the comoving unbroken 
bubble wall area $\mathcal{A}_c(t)$ at $t$ satisfies the following relation: 
\begin{eqnarray}
d  g(t_0, t) = \frac{d V_{c,\text{False}}}{V_c} = - \mathcal{A}_c(t) \frac{v_w d t}{a(t)} = - \mathcal{A}_c(t) v_w d \eta .
\end{eqnarray}
Then $\mathcal{A}_c$ is given by
\begin{eqnarray}
\mathcal{A}_c(t) = - \frac{1}{v_w} \frac{d g(t_0, t)}{d \eta}   = a(\eta) \frac{H(T) \gamma T}{v_w} \frac{d g(T_c, T)}{d T}  .
\label{eq:Ac}
\end{eqnarray}
One can also define the proper area per proper volume
\begin{eqnarray}
\mathcal{A} = \frac{\text{Proper Area}}{\text{Proper Volume}} = \frac{a^2 \times \text{Comoving Area}}{a^3\times \text{Comoving Volume}} = \frac{1}{a} \mathcal{A}_c .
\end{eqnarray}
Since $\mathcal{A}_c(t)$ and $\mathcal{A}$ are the area per volume, they are of dimension $1$, and can be presented in units of $m^{-1}$ or $\text{GeV}$.
A more meaningful representation can be obtained by comparing it with the typical scale at the corresponding temperature. 
One such quantity is $\beta_c$, to be defined later, which is the comoving version of the $\beta$ parameter and is related to the 
mean bubble separation (also to be defined later).
We show $\mathcal{A}_c/\beta_c$ in Fig.~\ref{fig:area} for different choices $\kappa_M$ and $v_w$, similar to what are used
in Fig.~\ref{fig:g}. We can see the area first increases as more bubbles are formed and expanding. It decreases
as bubbles collide with each other and the remaining false vacuum volume is shrinking to zero. The different behaviors when changing $v_w$
and the amount of non-relativistic matter contents coincide with what we observe in Fig.~\ref{fig:g}.

\begin{figure}
\centering
\includegraphics[width=0.6\textwidth]{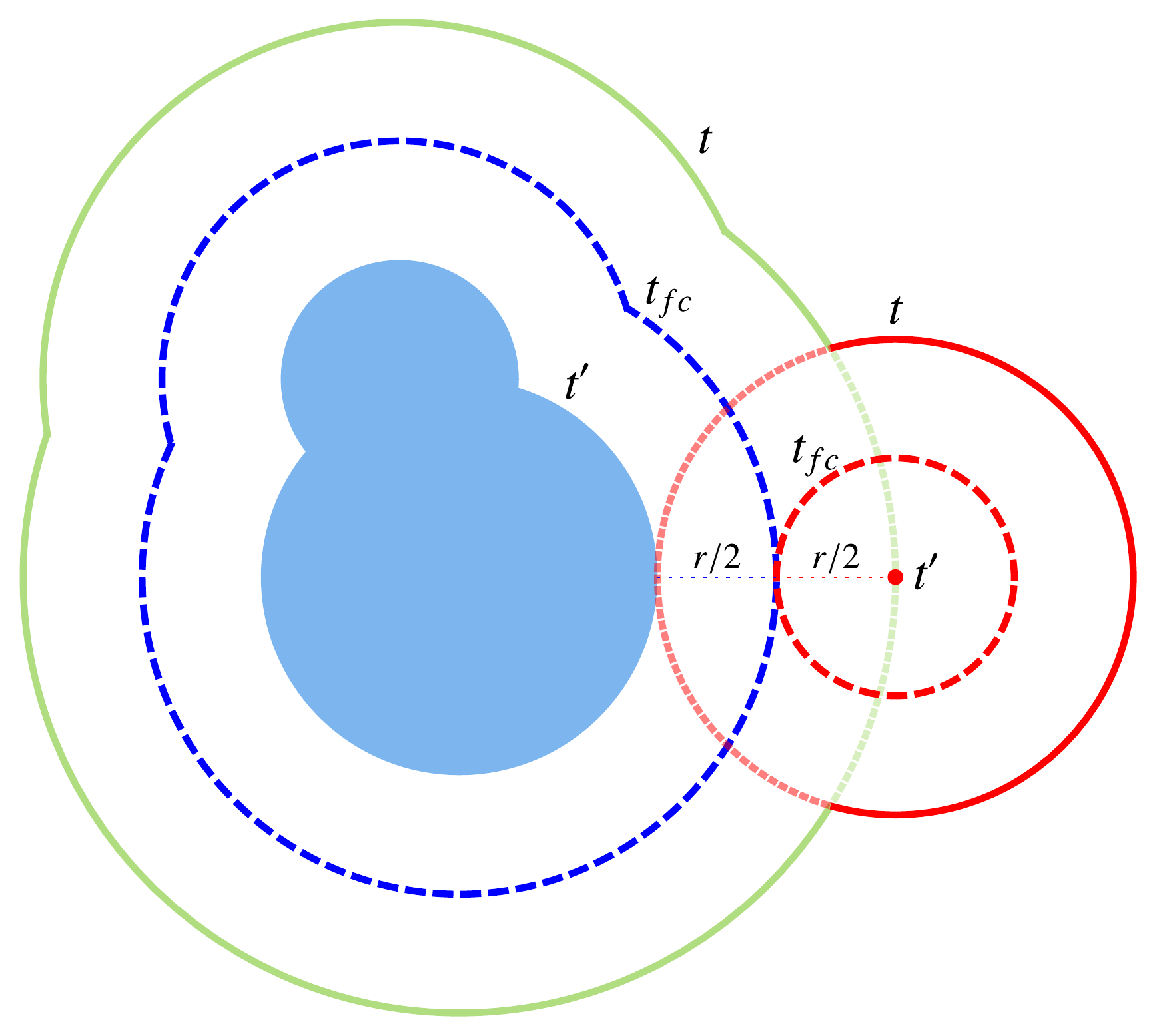}
\caption{
\label{fig:lifetime}
Illustration for the calculation of the bubble lifetime distribution. At $t^{\prime}$, there is a central blue blob composed of two already collided bubbles 
depicting a region of true vacuum space which is expanding into the surrounding false vacuum space, and also a small red nucleus denoting a bubble 
starting to form. At this time, the comoving distance between the red dot and the nearest blue boundary is $r$.
At $t_{fc}$, the walls of the blue blob and the fledged red bubble advance to the place denoted by blue and
red dashed circles respectively, where they make the first contact. At $t$, they reach the place denoted by the solid blue and red circles, where 
half of the red bubble is devoured by the blue one, and the red bubble is defined to be destroyed with a final radius $r$.  
}
\end{figure}

\subsection{Bubble Lifetime Distribution}

The bubble lifetime distribution describes the distribution of bubble lifetime for all the bubbles ever formed and destroyed during the 
entire process of the phase transition. This can be obtained with the help of the unbroken bubble wall area derived earlier, 
by generalizing the result derived in Ref.~\cite{Hindmarsh:2019phv} to the expanding universe. 
We start by considering the number of bubbles that are created at $t^{\prime}$ and are destroyed with comoving radius $r$. 
Here a bubble is defined as destroyed when approximately half of its volume is occupied by the expanding true vacuum space.
These bubbles are therefore at a comoving distance of $r$ at $t^{\prime}$ from the part of the unbroken bubble wall,
assuming constant and universal bubble wall velocity $v_w$. The time $t$ when this set of bubbles is destroyed is connected
with $t^{\prime}$ and $r$ by
\begin{eqnarray}
r = \int_{t^{\prime}}^{t} \frac{v_w d t^{\prime \prime}}{a(t^{\prime \prime})} .
\end{eqnarray}
Since only two quantities out of $(r, t^{\prime}, t)$ are independent, we denote $\mathcal{A}_c(t(t^{\prime},r))$ as $\mathcal{A}_c(t^{\prime}, r)$ and define the number of bubbles per comoving volume as $n_{b,c}$. We then have 
(see an illustration and more details in Fig.~\ref{fig:lifetime}): 
\begin{eqnarray}
d^2 n_{b,c} = p(t^{\prime}) \left[ a^3(t^{\prime}) \mathcal{A}_c(t^{\prime},r)  d r \right] d t^{\prime}  .
\end{eqnarray}
This implies that:
\begin{eqnarray}
d \left(\frac{d n_{b,c}}{d r} \right) \equiv d n_{b,c}(r) = p(t^{\prime}) \left[ a^3(t^{\prime}) \mathcal{A}_c(t^{\prime},r) \right] d t^{\prime}  .
\end{eqnarray}
Now, for fixed $r$, we consider all the bubbles ever formed before a time $t_f$:
\begin{eqnarray}
\left.n_{b,c}(r)\right|_{t_c}^{t_f} =  \int_{t_c}^{t_f} d t^{\prime} p(t^{\prime}) \left[ a^3(t^{\prime}) \mathcal{A}_c(t^{\prime},r) \right] ,
\end{eqnarray}
and $n_{b,c}(r) = 0$ at $t_c$ for all $r$. 
Consider a time when all bubbles have disappeared, when $t_f$ is large enough. Now $n_{b,c}(r)|_{t_f}$ becomes a constant $\tilde{n}_{b,c}(r)$.
We can then relate $r$ with the lifetime of the bubbles. For the bubble nucleated at $t^{\prime}$ and destroyed at $t$, we have
\begin{eqnarray}
r = \int^{t}_{t^{\prime}} dt^{\prime \prime} \frac{v_w d t^{\prime \prime}}{a(t^{\prime \prime})} = v_w \eta_{\text{lt}} ,
\label{eq:retaLT}
\end{eqnarray}
where $\eta_{\text{lt}}$ is the conformal lifetime of the bubble. Thus, $r$ has  
the same relation with the conformal lifetime as its relation with $t_{\text{lt}}$ in Minkowski spacetime.
We can therefore proceed to derive a conformal lifetime distribution for all bubbles ever formed and destroyed:
\begin{eqnarray}
\tilde{n}_{b,c}(\eta_{\text{lt}}) \equiv \frac{d n_{b,c}}{d \eta_{\text{lt}}} = v_w \tilde{n}_{b,c}(r) 
= v_w \int^{t_f}_{t_c} d t^{\prime} p(t^{\prime}) a^3(t^{\prime}) \mathcal{A}_c(t^{\prime}, v_w \eta_{\text{lt}}) .
\label{eq:nbc0}
\end{eqnarray}
Remember $\mathcal{A}_c(t^{\prime}, v_w \eta_{\text{lt}}) = \mathcal{A}_c(t(t^{\prime}, v_w \eta_{\text{lt}}))$ and it is evaluated at $t$,
which should be determined through Eq.~\ref{eq:retaLT} given $t^{\prime}$ and $\eta_{\text{lt}}$. To present a numerically convenient representation
of the above result, we convert coordinate time $t$ to conformal time $\eta$ and then to temperature. For the bubble formed at $t^{\prime}$, the corresponding
conformal time is related to temperature by
\begin{eqnarray}
\eta^{\prime} - \eta_c = \int_{t_c}^{t^{\prime}} \frac{d t^{\prime \prime}}{a(t^{\prime \prime})} = \frac{1}{a_c} \int_{T^{\prime}}^{T_c} \frac{d T^{\prime \prime}}{T^{\prime \prime}} \frac{1}{\gamma H(T^{\prime \prime})}
\left(\frac{T_c}{T^{\prime \prime}}\right)^{-1/\gamma} \equiv \Delta \eta(T^{\prime}, T_c).
\end{eqnarray}
Then for the bubble with conformal lifetime $\eta_{\text{lt}}$, the conformal time for its destruction is given by $\eta_{\text{lt}} + (\eta^{\prime} - \eta_c)$,
with the corresponding temperature $T$ determined through
\begin{eqnarray}
\eta_{\text{lt}} + (\eta^{\prime} - \eta_c)  = \Delta \eta(T, T_c) .
\label{eq:Tlt}
\end{eqnarray}
This temperature, or time, is what should be used in $\mathcal{A}_c$, rather than $T^{\prime}$. 
With the relation between $T$ and $T^{\prime}$ found, it is then straightforward to do the integral in Eq.~\ref{eq:nbc0}, 
which requires only converting $t^{\prime}$ to temperature.

%%%%%%%%%%%%%%%%%%%%%%%%%%%%%%%%%%%%%%%%%%%%%%%%%%%%%%%%%%%%%%%%%%%%%%
\begin{figure}
\centering
\includegraphics[width=0.49\textwidth]{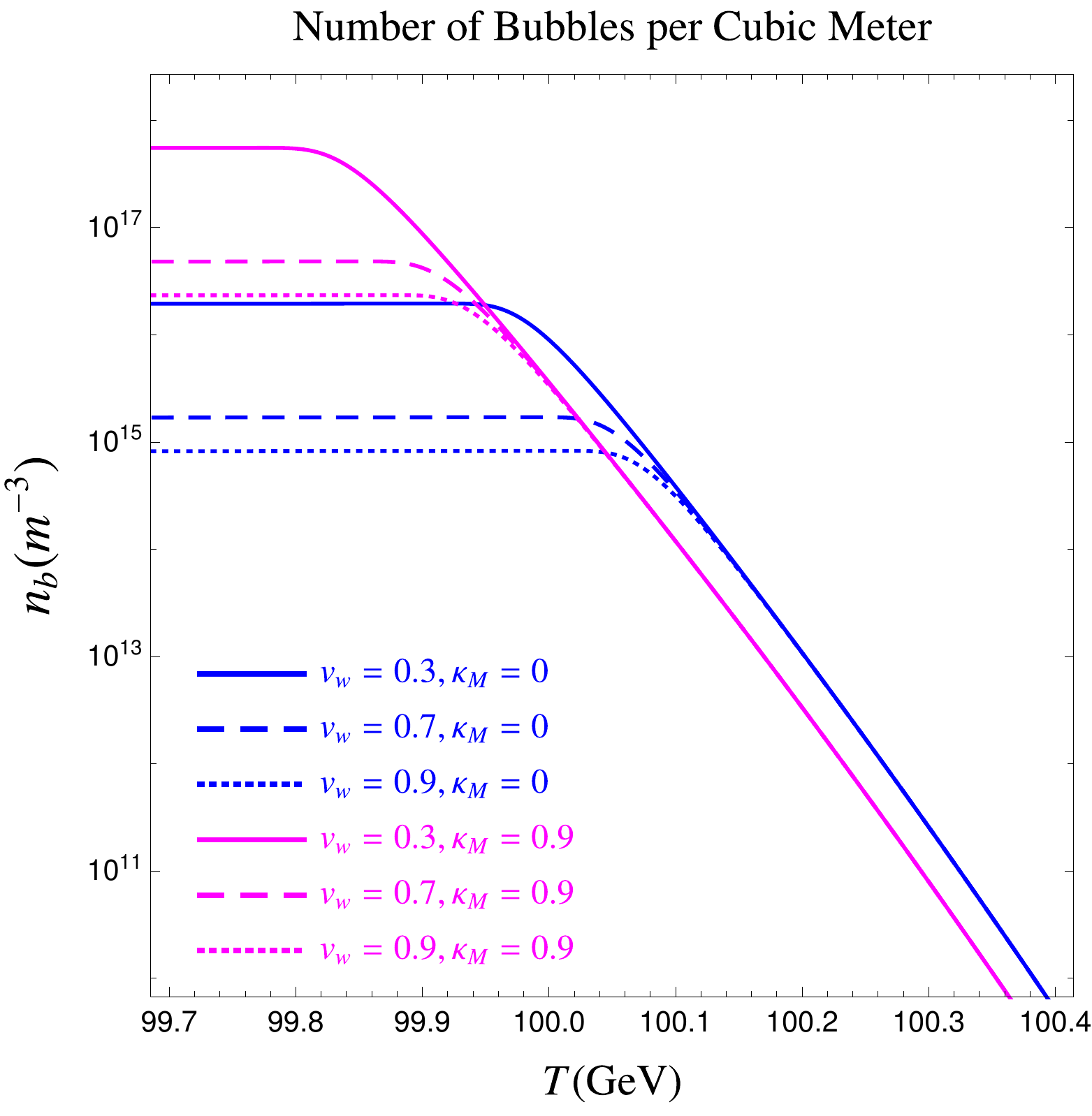}
\includegraphics[width=0.49\textwidth]{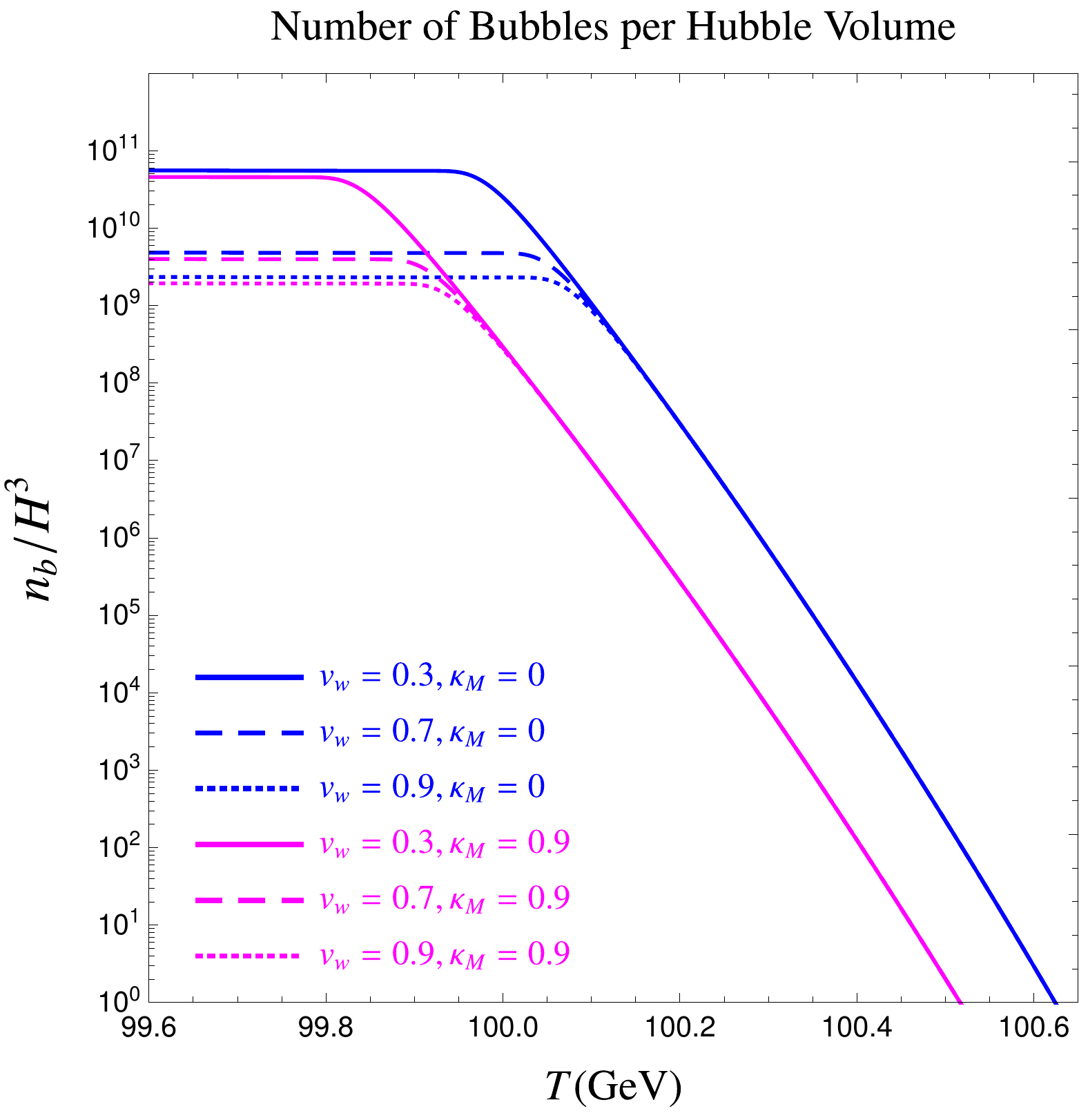}
\caption{
\label{fig:nb}
The number of bubbles (see Eq.~\ref{eq:nbt}) per $m^3$(left) and per Hubble volume(right) as a function of temperature for
difference fractions of non-relativistic matter content at the critical temperature $\kappa_M$ (defined in Eq.~\ref{eq:km}) and for different bubble wall velocities $v_w$.
}
\end{figure}
%%%%%%%%%%%%%%%%%%%%%%%%%%%%%%%%%%%%%%%%%%%%%%%%%%%%%%%%%%%%%%%%%%%%%%
 
\subsection{Bubble Number Density}

\begin{figure}
\centering
\includegraphics[width=0.48\textwidth]{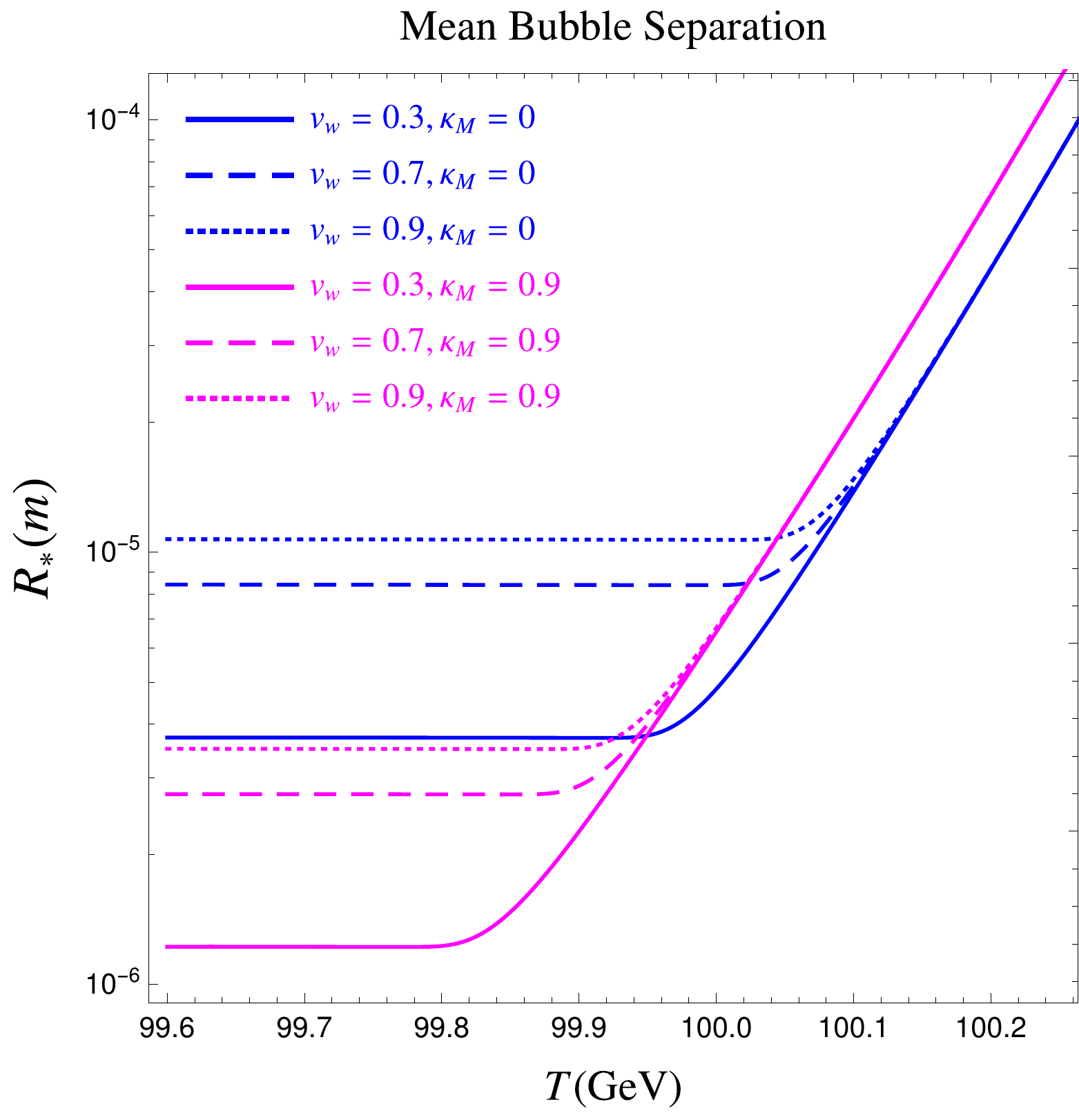}
\includegraphics[width=0.48\textwidth]{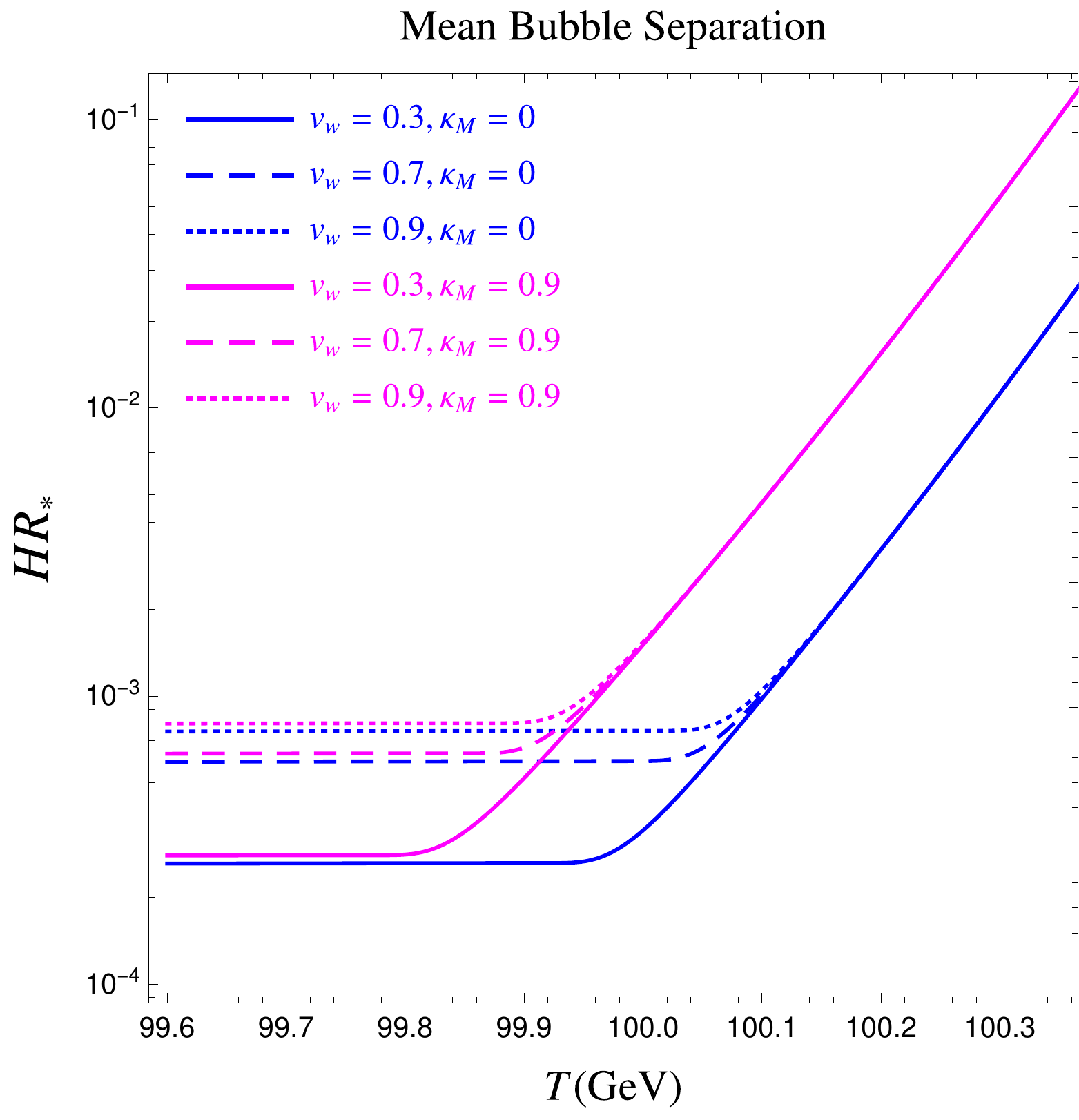}
\caption{
Mean bubble separation $R_{\ast}$ (defined in Eq.~\ref{eq:Rstar}) for different fractions of the non-relativistic matter content at the critical temperature $\kappa_M$ and for different bubble wall velocities $v_w$. The left panel is in unit of meter and the right in unit of Hubble radius.
\label{fig:Rb}
}
\end{figure}

The evolution of the bubble number density per proper volume $n_b = N_b/V$ is governed by the following equation
\begin{eqnarray}
\frac{d [n_b a^3(t)]}{d t} = p(t) g(t_c, t) a^3(t) ,
\label{eq:nbdef}
\end{eqnarray}
which can be integrated to give (noting that $n_b(t_c) = 0$):
\begin{eqnarray}
n_b(t) = \frac{1}{a^3(t) } \int_{t_c}^t d t^{\prime} p(t^{\prime}) g(t_c, t^{\prime}) a^3(t^{\prime}) .
\label{eq:nbt}
\end{eqnarray}
This does not include the decrease of bubble number due to collisions and $n_b$ thus includes all the bubbles ever formed.
The result for $n_b(t)$ can be similarly transformed into a function of temperature.
\begin{eqnarray}
n_b(T) = \left(\frac{T}{T_c}\right)^{3/\gamma} \int_T^{T_c} 
\frac{d T^{\prime}}{T^{\prime}} \frac{1}{\gamma H(T^{\prime})}
\bar{p}_0 T^{\prime 4} \text{exp}\left[- \frac{S_3(T^{\prime})}{T^{\prime}}\right] g(T_c, T^{\prime}) 
\left(\frac{T_c}{T^{\prime}}\right)^{3/\gamma} .
\end{eqnarray}
We show $n_b$ in units of $m^{-3}$ in the left panel of Fig.~\ref{fig:nb} and the total bubble number 
per Hubble volume $n_b/H^3(T)$ in the right panel. We can see that the bubble number density increases for a
delayed false vacuum fraction, which is consistent with physical intuition.
From $n_b$, we can define the mean bubble separation, $R_{\ast}$, as 
\begin{eqnarray}
R_{\ast}(t) = \left[\frac{V(t)}{N_b(t)}\right]^{1/3} = \left[\frac{1}{n_b(t)}\right]^{1/3} .
\label{eq:Rstar}
\end{eqnarray}
This is shown in Fig.~\ref{fig:Rb}. For both $n_b$ and $R_{\ast}$, it appears they both reach an asymptotic value
after the bubbles have disappeared when the curves in these figures become flat. This is misleading as after the time 
the bubbles have disappeared, $n_b$ will be diluted as $1/a^3$ and accordingly $R_{\ast}$ increases as $a$.
The flat curves in the figures are simply due to the very tiny change of temperature plotted.
From numerical simulations~\cite{Hindmarsh:2017gnf,Hindmarsh:2015qta}, it is found
that the peak frequency of the gravitational wave spectrum is related to $R_{\ast}$. Therefore any change on 
$R_{\ast}$ will translate into a shift of the peak frequency of the gravitational waves. Since $R_{\ast}$ is of
particular importance, it is convenient to use the comoving version of it $R_{\ast c} = (V_c/N_b)^{1/3}$, which 
will reach an asymptotic value after the bubble disappearance.

From the right panel of Fig.~\ref{fig:nb}, we can easily read off the nucleation temperature $T_n$, which is defined
such that at this temperature there is about one bubble within a Hubble volume~\cite{Apreda:2001us}. Note $T_n$ obtained
this way differs slightly from the usually used, and a bit crude, criterion:
\begin{eqnarray}
\int_{t_c}^{t_n} d t \frac{p(t)}{H(t)^3}  = 1,
\label{eq:Tndef}
\end{eqnarray}
which for radiation dominated universe where $H^2 = 8 \pi G \rho/3$ and $\rho = \frac{\pi^2}{30} g_{\ast}(T) T^4$ translates into the condition:
\begin{eqnarray}
\int_{T_n}^{T_c} \frac{d T}{T} \left( \frac{90}{8 \pi^3 g_{\ast}} \right)^2 \left(\frac{m_{\text{Pl}}}{T}\right)^4 \text{exp}\left[- \frac{S_3(T)}{T}\right] = 1 .
\end{eqnarray}
Here $m_{\text{Pl}}$ is the Planck mass. A further simplification says that $T_n$ is determined by $S_3(T_n)/T_n = 140$~\cite{Apreda:2001us}. 
These determined $T_n$ differs slightly from the more accurate result obtained by solving directly for $n_b$ with Eq.~\ref{eq:nbdef}.

\subsection{Relation between $\beta$ and Mean Bubble Separation ($R_{\ast}$)}

It was found from numerical simulations that the peak of the gravitational wave power spectrum is located at 
$k R_{\ast} \sim 10$~\cite{Hindmarsh:2017gnf}, where $R_{\ast}$ is the mean bubble separation 
defined earlier. However the standard spectrum people generally use is expressed in terms of $\beta$ (see, e.g., ~\cite{Caprini:2015zlo,Weir:2017wfa}).
So the relation between $\beta$ and $R_{\ast}$ is needed. It can be derived analytically under reasonable assumptions
as was shown in Ref.~\cite{Hindmarsh:2019phv}, which says
\begin{eqnarray}
R_{\ast} = \frac{(8\pi)^{1/3}}{\beta(v_w)} v_w .
\end{eqnarray}
Here we emphasize that $\beta$ varies when $v_w$ is changed. The question is then will this relation still hold in an expanding
universe. We will answer this question by giving a detailed derivation here, which parallels and generalizes the derivation in Ref.~\cite{Hindmarsh:2019phv}.

%%%%%%%%%%%%%%%%%%%%%%%%%%%%%%%%%%%%%%%%%%%%%%%%%%%%%%%%%%%%%%%%%%%%%%
%%%%%%%%%%%%%%%%%%%%%%%%%%%%%%%%%%%%%%%%%%%%%%%%%%%%%%%%%%%%%%%%%%%%%%
We rewrite Eq.~\ref{eq:nbdef} in terms of the conformal time (we still use the same function labels though $t$ is replaced by $\eta$)
\begin{eqnarray}
\frac{d (n_{b,c})}{d \eta} = p(\eta) g(\eta_c, \eta) a^4(\eta) ,
\label{eq:nbeta}
\end{eqnarray}
where $n_{b,c} = n_{b} a^3$ and is the comoving bubble number density.
Here the false vacuum fraction $g$ decreases sharply when its exponent $I(T)$ becomes of order $1$. Since $p(\eta)$
increases exponentially, there is a peak for the r.h.s in above equation, at which time the bubbles are mostly nucleated. 
As $g$ decreases much more sharply than $p$ increases, the rate $p$ only increases slowly during this time duration and it can be Taylor
expanded at around this time. This time can be conveniently chosen to be $\eta_0$ which satisfies $I(\eta_0) = 1$. Then similarly to 
Eq.~\ref{eq:pTaylort}, we define a Taylor expansion but w.r.t the conformal time:
\begin{eqnarray}
p(\eta) = p_0(\eta_0) \text{exp}[-S_0 + \beta_c (\eta - \eta_0)] ,
\end{eqnarray}
where we have neglected the very slow change of $p_0(\eta)$ and defined a comoving version of $\eta$:
\begin{eqnarray}
\beta_c = \left.\frac{d \ln p}{d \eta}\right|_{\eta = \eta_0} .
\label{eq:betacdef}
\end{eqnarray}
Now lets see how $n_{b,c}$ in Eq.~\ref{eq:nbeta} can be solved in terms of $\beta_c$. To do it, lets firstly see how $g$ or its exponent
$I$ can be expressed in terms of $\beta_c$. From Eq.~\ref{eq:fracfalse}, we can write $I$ in the following way:
\begin{eqnarray}
I(\eta) &=& \frac{4 \pi}{3} \int_{\eta_c}^{\eta} d \eta^{\prime} a^4(\eta^{\prime}) p(\eta^{\prime}) r(\eta^{\prime}, \eta)^3 \nonumber \\
&=&  \frac{4 \pi}{3} v_w^3 
\int_{\eta_c}^{\eta} d \eta^{\prime} p_0(\eta_0) e^{-S_0 + \beta_c (\eta^{\prime} - \eta_0)} (\eta - \eta^{\prime})^3  \nonumber \\
&=& 8\pi \frac{v_w^3}{\beta_c^4} p_0(\eta_0) e^{-S_0 + \beta_c (\eta - \eta_0)} .
\end{eqnarray}
Now define a time $\eta_f$ such that 
\begin{eqnarray}
I(\eta_f) = 1 ,
\label{eq:I1}
\end{eqnarray}
then at a later time much simpler expressions can be obtained:
\begin{eqnarray}
I(\eta) = e^{\beta_c (\eta - \eta_f)}, \quad \quad g(\eta_c, \eta) = e^{-I(\eta)} .
\end{eqnarray}
As $I(\eta)$ depends on the bubble wall velocity $v_w$, the resulting $t_f$ and more importantly $\beta_c$ is a function of $v_w$.
Plugging above expressions of $g(\eta_c, \eta)$, $p(\eta)$ into Eq.~\ref{eq:nbeta}, and integrating over $\eta$, we have
\begin{eqnarray}
n_{b,c} = \frac{1}{\beta_c} p_0(\eta_0) e^{-S_0 + \beta_c (\eta_f - \eta_0)} = \frac{\beta_c^3(v_w)}{8\pi v_w^3} .
\end{eqnarray}
Here the second equality comes from the relation in Eq.~\ref{eq:I1}.
As noted in Ref.~\cite{Hindmarsh:2019phv}, the best choice of $t_0$ is $t_f$ so that the Taylor expansion of $p(\eta)$ converges more quickly.
\begin{figure}
    \centering
    \includegraphics[width=0.46\textwidth]{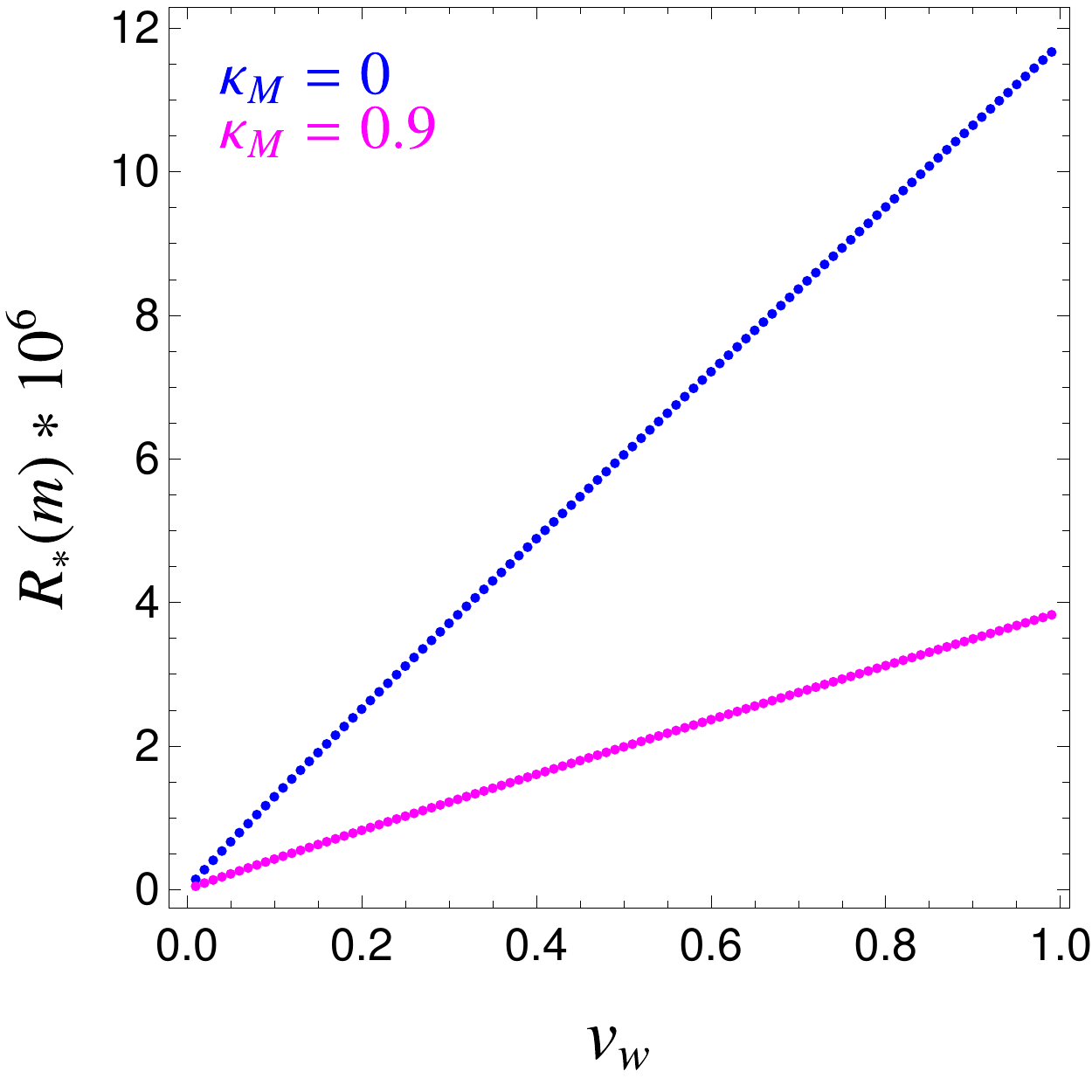}
    \quad
    \includegraphics[width=0.46\textwidth]{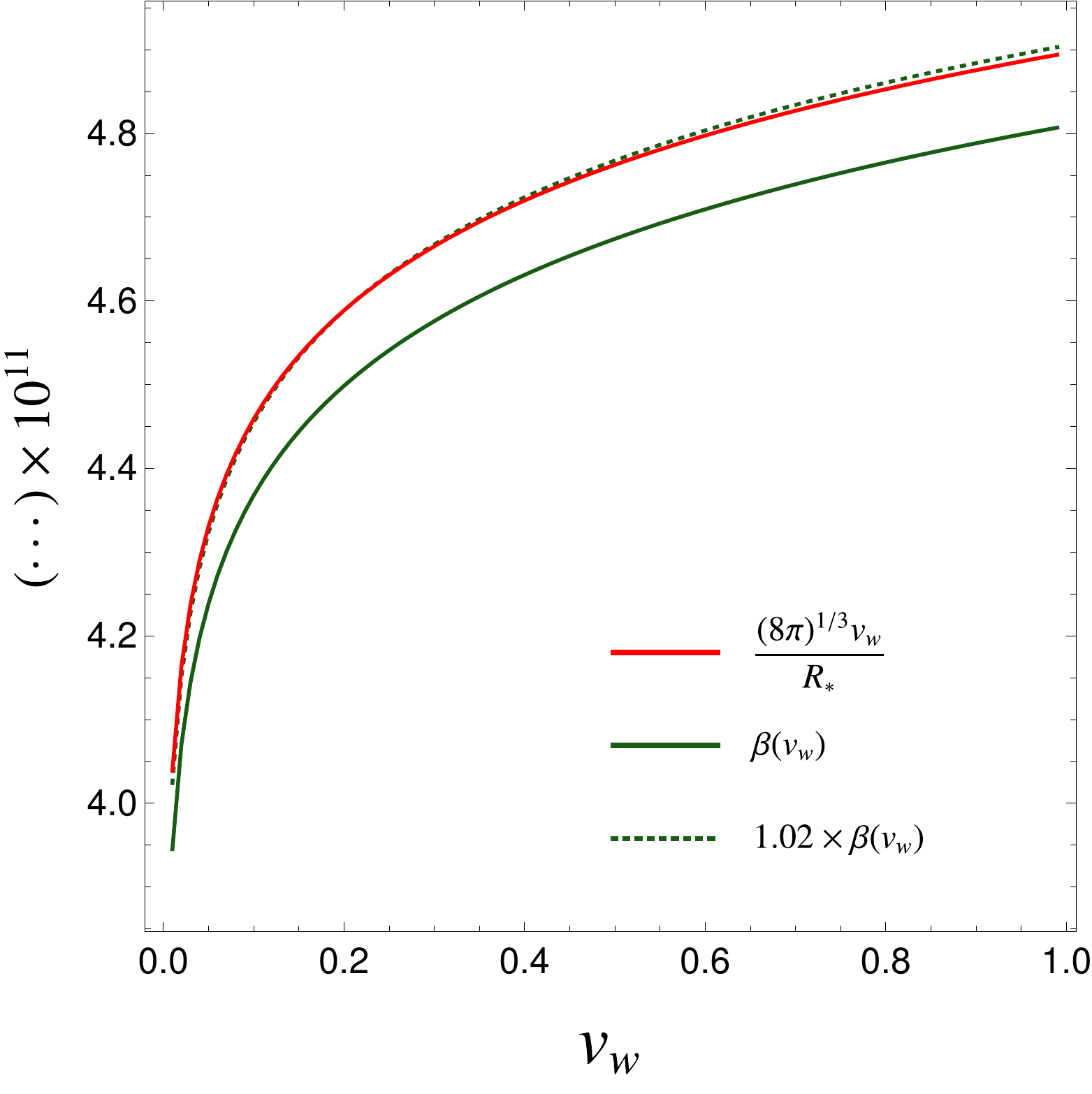}
    \caption{
    The left panel shows the mean bubble separation $R_{\ast}$ immediately after all the bubbles have disappeared versus bubble wall velocity $v_w$ 
    for $\kappa_M=0$ and $\kappa_M=0.9$. 
    The right panel shows $\beta(v_w)$ calculated using Eq.~\ref{eq:betaH} at
    $t_f$, as compared with that calculated from $R_{\ast}$ using Eq.~\ref{eq:Rsfinal} for $\kappa_M=0$. 
    The dotted line shows these differ by roughly $2\%$. For $\kappa_M=0.9$, it shows similar behavior.
    \label{fig:Rbvw}
    }
\end{figure}
This result gives the relation between the comoving mean bubble separation $R_{\ast c}$ and $\eta_c$:
\begin{eqnarray}
R_{\ast c} = (8\pi)^{1/3} \frac{v_w}{\beta_c(v_w)} .
\label{eq:Rscbetac}
\end{eqnarray}
We can also write all results in terms of physical quantities. From Eq.~\ref{eq:betacdef} and enforcing $t_0 = t_f$, we have 
\begin{eqnarray}
\beta_c = a(\eta_f) \beta = a(\eta_f) \left.\left[\gamma H(T) T\frac{d (S_3/T)}{d T}\right]\right|_{T=T_f} ,
\label{eq:betacbeta}
\end{eqnarray}
and thus
\begin{eqnarray}
n_{b,c} = a^{3}(\eta_f) \frac{\beta^3}{8 \pi v_w^3} .
\end{eqnarray}
Note $n_{b,c}$ becomes a constant number as $N_b$ reaches its maximum and the comoving volume is fixed. 
The physical number density after all the bubbles have vanished will be diluted by the expansion. Suppose
we consider the physical number density $n_b$ at time $\eta$, then 
\begin{eqnarray}
n_b(\eta) = \left(\frac{a(\eta_f)}{a(\eta)}\right)^3 \frac{\beta^3}{8 \pi v_w^3} .
\end{eqnarray}
The corresponding physical mean bubble separation would be 
\begin{eqnarray}
R_{\ast}(\eta) = \frac{a(\eta)}{a(\eta_f)} (8\pi)^{1/3} \frac{v_w}{\beta(v_w)} .
\label{eq:Rsfinal}
\end{eqnarray}
Therefore the relation between $R_{\ast}$
and $\beta$ is similar to that derived in Minkowski spacetime and needs only additional attention on the scale factors.
If one uses $R_{\ast c}$ and $\beta_c$, then the relation is exactly the same as in Minkowski spacetime.
We emphasize again that $\beta$ and $\beta_c$ are functions of $v_w$. To see this, we plot $R_{\ast}$ at a time immediately
after all the bubbles have disappeared, as a function of $v_w$, in the left panel of Fig.~\ref{fig:Rbvw}. For each $v_w$, we find the corresponding $\beta(v_w)$
as implied in above equation and compare with $\beta(v_w)$ directly calculated using Eq.~\ref{eq:betacbeta}. This comparison is shown in the right panel and 
the two different determinations differ by at most $2\%$, where the uncertainty can be attributed to the approximations made.

\section{Fluid Velocity Field and Power Spectrum\label{sec:fluid}}

The dominant source of gravitational wave production is the sound waves in a perturbed plasma due to the advancing bubble walls
and their interaction with the surrounding fluid. In the sound shell model~\cite{Hindmarsh:2016lnk,Hindmarsh:2019phv}, the total velocity field is modelled as 
a linear superposition of the individual contribution from each bubble. The first step is then to understand
the velocity profile of the fluid around a single bubble. This topic has been extensively studied several decades ago and is reviewed with
a complete treatment in Ref.~\cite{Espinosa:2010hh}. However the analysis is set in  Minkowski spacetime and it is not clear whether it needs changes
in an expanding universe. Ref.~\cite{Cai:2018teh} studied the velocity profile in an expanding universe and found that there is a 
significant change to the velocity profile and a reduction of energy fraction going into the kinetic energy of the sound waves. But we will see in this section 
the velocity profile actually remains unchanged. 
We first review the full set of fluid and field equations and then analyze the fluid velocity profile around a single bubble. Armed with this information, we then find the total velocity field from a population of bubbles in the sound shell model and calculate the velocity field power spectrum.

\subsection{Fluid and Field Equations}

Numerical simulations that are performed to provide the widely adopted gravitational wave formulae are based on the fluid-order parameter field model~\cite{Hindmarsh:2015qta,Ignatius:1993qn,KurkiSuonio:1995vy} in Minkowski spacetime. Here we generalize the full set of equations used in the simulations to the FLRW universe. Our purpose is to understand whether simulations can be done in Minkowski spacetime and then generalized to an expanding universe by simple rescalings of the physical quantities. This is an important question as it is computationally very expensive to do a numerical simulation.

The universe consists of: (1) the underlying scalar field(s) responsible for the phase transition; (2) the relativistic plasma whose constituent particles
can interact with the scalar field(s); (3) magnetic field produced from the phase transition; (4) other sectors which do not directly 
interact with either the scalar field, the plasma or the magnetic field, though they do interact gravitationally. 
We will neglect (3) by focusing on the dominant source for gravitational wave production, and only 
consider (4) through its effect on the expansion. 
Given our cosmological context, the total energy momentum tensor for (1) and (2) is given by~\cite{Hindmarsh:2015qta}
\begin{eqnarray}
  T^{\mu\nu} = \partial^{\mu} \phi \partial^{\nu} \phi - \frac{1}{2} g^{\mu\nu} \partial_{\mu} \phi \partial^{\mu} \phi +(e + p) U^{\mu} U^{\nu} + g^{\mu\nu} p ,
\end{eqnarray}
where $U^{\mu} = \gamma (1, {\bf v}/a)$ with $\gamma = 1/\sqrt{1-v^2}$ and ${\bf v} = d {\bf x}/d \eta$. The energy and momentum densities are given by
\begin{eqnarray}
&&e = a_B T^4 + V(\phi,T) - T \frac{\partial V}{\partial T} , \nonumber \\
&&p =  \frac{1}{3} a_B T^4 - V(\phi, T),
\end{eqnarray}
where $a_B = g_{\ast} \pi^2/30$ and $g_{\ast}$ is the relativistic degrees of freedom.
It is certainly conserved, i.e., $T^{\mu\nu}_{\ \ \ \ ;\mu} = 0$~\footnote{The subscript ``;'' denotes covariant derivative.}, and it is usually
split into two parts by adding and subtracting a friction term $\delta^{\nu}$~\cite{Ignatius:1993qn}:
\begin{eqnarray}
&& T^{\mu\nu}_{\ \ \ \ ;\mu}|_{\text{field}} = (\partial^2 \phi) \partial^{\nu} \phi 
+ \frac{1}{\sqrt{g}} (\partial_{\mu} \sqrt{g}) (\partial^{\mu} \phi) (\partial^{\nu} \phi)
 - \frac{\partial V}{\partial \phi} \partial^{\nu} \phi = \delta^{\nu} , \nonumber \\
&& T^{\mu\nu}_{\ \ \ \ ;\mu}|_{\text{fluid}} = \partial_{\mu}\left[(e + p) U^{\mu} U^{\nu}\right] 
+ \left[ \frac{1}{\sqrt{g}} (\partial_{\mu} \sqrt{g}) g^{\nu}_{\lambda} 
+ \Gamma^{\nu}_{\mu\lambda}\right] (e + p) U^{\mu} U^{\lambda} 
+ g^{\mu\nu} \partial_{\mu} p + \frac{\partial V}{\partial \phi} \partial^{\nu} \phi = - \delta^{\nu} . \nonumber
\\
\label{eq:Tfriction}
\end{eqnarray}
Note here the appearance of $\partial_{\mu} g$ and $\Gamma^{\nu}_{\mu\lambda}$ as we are using a generic metric.
The friction term $\delta^{\nu}$ is modelled by $\delta^{\nu} = \eta U^{\mu} \partial_{\mu} \phi \partial^{\nu} \phi$.  
For high temperatures it can be chosen as $\eta = \tilde{\eta} \phi^2/T$~\cite{Hindmarsh:2017gnf}, which works well in that case~\cite{Liu:1992tn} 
but may lead to numerical singularities for small temperature. The numerical simulations on sound waves adopted a constant value for the lower temperature case~\cite{Cutting:2019zws}. Note the exact set of equations can also be derived from field theory~\cite{Moore:1995si,Konstandin:2014zta}.

In an FLRW universe, the field energy momentum conservation leads to a scalar equation:
\begin{eqnarray}
- \ddot{\phi} + \frac{1}{a^2} \triangledown^2 \phi - \frac{\partial V}{\partial \phi} - 3 \frac{\dot{a}}{a} \dot{\phi} = \eta \gamma (\dot{\phi} + \frac{1}{a} 
\mathbf{v} \cdot \mathbf{\triangledown} \phi) ,
\end{eqnarray}
which is just the Klein-Gordon equation for the scalar field when the friction term is absent, i.e., when $\eta=0$.
The vector part of the fluid energy-momentum conservation gives:
\begin{eqnarray}
\dot{Z}^i + \frac{1}{a} \triangledown \cdot (\mathbf{v} Z^i) + 5 \frac{\dot{a}}{a} Z^i + \frac{1}{a^2} \partial_i p + 
\frac{1}{a^2} \frac{\partial V}{\partial \phi} \partial_i \phi = - \frac{1}{a^2} \eta \gamma (\dot{\phi} + \frac{1}{a} \mathbf{v}\cdot \triangledown \phi) \partial_i \phi ,
\label{eq:fluid-sclar:vector}
\end{eqnarray}
where $Z^i \equiv \gamma (e + p) U^i = \gamma^2 (e + p) v^i/a$.
The parallel projection along $U_{\nu}$ for the fluid gives another scalar equation:
\begin{eqnarray}
\dot{E} + p [\dot{\gamma} + \frac{1}{a} \triangledown \cdot (\gamma \mathbf{v})] + \frac{1}{a} \triangledown \cdot (E \mathbf{v})
-\gamma \frac{\partial V}{\partial \phi} (\dot{\phi} + \frac{1}{a} \mathbf{v} \cdot \triangledown \phi) + 3 \frac{\dot{a}}{a} \gamma (e + p) \nonumber \\
= \eta \gamma^2 (\dot{\phi} + \frac{1}{a} \mathbf{v} \cdot \triangledown \phi)^2 ,
\label{eq:fluid-scalar:scalar}
\end{eqnarray}
where $E \equiv e \gamma$.
While the above equations form a complete set, the velocity profile is usually derived from a different scalar equation, the perpendicular
projection for the fluid along the direction $\bar{U}_{\nu}$, which is defined by
\begin{eqnarray}
\bar{U}^{\mu} U_{\mu} = 0, \quad \quad \bar{U}^{\mu} \bar{U}_{\mu} = 1,
\end{eqnarray}
and takes the explicit form $\bar{U}^{\mu} = \gamma (v, \hat{v}^i/a)$. This gives 
\begin{eqnarray}
\left[\frac{\dot{a}}{a} v + \gamma^2 \left(\dot{v} + \frac{1}{2 a} \hat{\mathbf{v}} \cdot \triangledown v^2\right) \right]
(e + p) 
+ v \dot{p} + \frac{1}{a} \hat{\mathbf{v}} \cdot \triangledown p
+
\frac{\partial V}{\partial \phi} (v \dot{\phi} + \frac{1}{a} \hat{\mathbf{v}} \cdot \triangledown \phi) \nonumber \\
=
- \eta \gamma (v \dot{\phi} + \frac{1}{a} \hat{\mathbf{v}} \cdot \triangledown \phi)(\dot{\phi} + \frac{1}{a} \mathbf{v} \cdot \triangledown \phi) .
\label{eq:fluid-scalar:scalar-perp}
\end{eqnarray}
These equations are direct generalizations of those in Ref.~\cite{Hindmarsh:2015qta} to an FLRW universe. It is not possible, however, to express the above equations
in a form used in Minkowski spacetime and the problem lies with the scalar field. 
Despite this, the effect on the bubble and fluid motions should be  minor, since the bubble collision process is fast compared with the long duration of the ensuing
sound waves. 

The process of the phase transition can thus be divided into two stages. The first stage is the bubble collision and disappearance of the symmetric phase, and the second is the propagation of sound waves. The difference between them is that the first stage takes a much shorter time, while the second is long-lasting.
This is indeed what is observed from numerical simulations and should well justify simply
neglecting the change of the scale factor during the first stage~\cite{Hindmarsh:2015qta}.
In this sense, the numerical simulations as performed in Ref.~\cite{Hindmarsh:2015qta,Hindmarsh:2017gnf} still give a faithful account of the first step for 
an expanding universe. However we will see in the next subsection that the analytical modelling of this first stage still admits simple rescaling properties
and takes the same form as its Minkowski counterpart.

During the second stage gravitational waves are dominantly produced due to the long-lasting sound waves. Therefore the 
change of the scale factor can not be ignored. The question is: can we still solely perform numerical simulations in Minkowski spacetime.
Fortunately, during this stage, the scalar field plays no dynamical role and we can consider only the fluid.
The corresponding equations can indeed be reduced to the Minkowski form. This is achieved by using the conformal time, 
neglecting the scalar field as well as the friction terms and using $p = e/3$ for the plasma.
Then Eq.~\ref{eq:fluid-sclar:vector}, Eq.~\ref{eq:fluid-scalar:scalar} and Eq.~\ref{eq:fluid-scalar:scalar-perp} reduce to (again, $\prime \equiv \partial/\partial \eta$):
\begin{eqnarray}
&& (a^4 S^i)^\prime + \triangledown \cdot (a^4 S^i \mathbf{v}) + \partial_i(a^4 p) = 0 , \nonumber \\
&& (a^4 e \gamma)^{\prime} + [\gamma^{\prime} + \triangledown \cdot (\gamma \mathbf{v})] (a^4 p) + \triangledown \cdot (a^4 e \gamma \mathbf{v}) = 0 , \nonumber \\
&& \gamma^2 (v^{\prime} + \frac{1}{2} \hat{\mathbf{v}} \cdot \triangledown v^2) [a^4(e + p)] + v (a^4 p)^{\prime}
+ \hat{\mathbf{v}} \cdot \triangledown (a^4 p) = 0 ,
\label{eq:fluid-system:EOM}
\end{eqnarray}
where $S^i = a Z^i = \gamma^2 (e + p) v^i$.
The Minkowski counterpart of these equations can be obtained by setting $a=1$.  
This suggests that we can define rescaled quantities 
$\tilde e = a^4 e/a_s^4$ and $\tilde{p} = a^4 p/a_s^4$, where $a_s$ is the scale factor when the source becomes active.
They are free from the dilution due to the expansion, and that 
the equations governing $\tilde{e}$, $\tilde{p}$ and $v$ take exactly the same form as their Minkowski counterparts, as long as the time $t$ is 
interpreted as the conformal time $\eta$. We will see how these rescaled quantities can be used to derive the modified gravitational wave spectrum in later 
sections.

We note here that these equations were derived earlier in Ref.~\cite{Brandenburg:1996fc,Gailis:1995ohk} when also considering electromagnetism
and it was shown that the above rescaling works not only for the purely fluid system but also for a system containing both fluid and electromagnetism.
Including electromagnetism will add additional terms to the right hand side of the above equations.

\subsection{Velocity Profile around a Single Bubble}

Solving the velocity profile for a single expanding bubble depends on analyzing the behavior of the system consisting of both the fluid and the scalar field.
This is usually done in the so called bag equation of state model, as summarized in Ref.~\cite{Espinosa:2010hh}. 
The energy momentum tensor for the fluid plus scalar field system is assumed to take the following form (``$+$'' for outside the bubble and ``$-$'' for inside):
\begin{equation}
T^{\mu\nu}_{\pm} = p_{\pm} g^{\mu\nu} + (p_{\pm} + \rho_{\pm}) U^{\mu} U^{\nu} ,
\end{equation}
with the bag equation of state:
\begin{eqnarray}
&& p_+ = \frac{1}{3} a_+ T_+^4 - \epsilon, \quad e_+ = a_+ T_+^4 + \epsilon , \nonumber \\
&& p_- = \frac{1}{3} a_- T_-^4, \quad \ \  \quad e_- = a_- T_-^4 ,
\label{eq:bageos}
\end{eqnarray}
where $\epsilon$ is the vacuum energy difference between the false and true vacua. One can also find the enthalpy $\omega = e + p$. Here $v$, $T$ and thus $e,p, \omega$ all vary from the bubble 
center to the region far outside the bubble where there is no perturbation.
The task is to solve for these fields at regions both inside and outside the bubbles and smoothly match these
two sets of solutions through the junction conditions across the bubble wall. 

\subsubsection{Inside the Bubble}

In this region, we drop all terms related to $\phi$ including the vacuum energy from $\epsilon$, and  we also apply the relation 
$p = e/3$~\footnote{Of course, we are assuming a constant value of the speed of sound, i.e., $c_s = 1/\sqrt{3}$. 
Without doing so, the equations cannot be put into the form in Eq.~\ref{eq:fluid-system:EOM}. We also dropped any spatial variation of 
the scalar field and its time variation following the conventional analysis, which amounts to assuming a thin wall.}. 
The resulting equations are already given in Eq.~\ref{eq:fluid-system:EOM} and the equations are exactly the same as the Minkowski 
counterpart when the rescaled quantities are used.
Now, assuming a spherically symmetric profile and denoting
the comoving bubble radius with $r$ and the conformal time elapsed since its nucleation as $\Delta \eta$,  the solution should 
be a self-similar one which  depends solely on the ratio $\xi \equiv r/\Delta \eta$. Then we can obtain the same equations as in Minkowski spacetime:
\begin{eqnarray}
(\xi - v) \partial_{\xi} \tilde{e} &=& \tilde{w} \left[ 2 \frac{v}{\xi} + \gamma^2 (1-\xi v) \partial_{\xi} v\right] , \nonumber \\
(1-v \xi) \partial_{\xi} \tilde{p} &=& \tilde{w} \gamma^2 (\xi - v) \partial_{\xi} v ,
\label{eq:epprofile}
\end{eqnarray}
which can then be combined to give an equation for the velocity field: 
\begin{eqnarray}
2 \frac{v}{\xi} = \gamma^2 (1 - v\xi) \left[\frac{\mu^2}{c_s^2}-1\right] \partial_{\xi} v .
\label{eq:vxi}
\end{eqnarray}
Here $\mu(\xi, v) = (\xi-v)/(1-\xi v)$, which is the Lorentz boost transformation. 
This equation can be directly solved given a boundary condition at the wall, to be specified later. 

\subsubsection{Outside the Bubble}

Outside the bubble, the presence of the constant vacuum energy term $\epsilon$ seemingly does not allow us to reach Eq.~\ref{eq:fluid-system:EOM}
for two possible reasons: (1) we can not apply $p = e/3$ since $p=-e$ for vacuum energy; (2) $\epsilon$ does not scale like 
radiation with the behavior $1/a^4$ and the rescaled quantity $a^4 e$ still contains the expansion effect.
Let us look more closely at the equations.
The parallel projection in Eq.~\ref{eq:fluid-scalar:scalar}, when the friction and scalar gradient terms are neglected, becomes 
\begin{equation}
\left[ (\gamma e)^{\prime} + 3 \frac{a^{\prime}}{a} \gamma (e + p) \right] 
+ p [\gamma^{\prime} + \triangledown \cdot (\gamma {\bf v})] + \triangledown \cdot (\gamma e {\bf v})  = 0 .
\label{eq:parallel_a}
\end{equation}
Correspondingly, the perpendicular projection in Eq.~\ref{eq:fluid-scalar:scalar-perp} reduces to
\begin{eqnarray}
\left[\frac{a^{\prime}}{a} v (e + p) + v p^{\prime} \right]
+
\gamma^2 (v^{\prime} + \frac{1}{2} \hat{{\bf v}} \cdot \triangledown v^2) 
 (e + p) + \hat{{\bf v}} \cdot \triangledown p = 0 .
\label{eq:perp_a}
\end{eqnarray}
In the absence of the vacuum energy inside $e$ and $p$, both of above equations can be put into the form in Eq.~\ref{eq:fluid-system:EOM}, by combining
the terms in $[\cdots]$ and using $e = 3 p$. The resulting equations for the rescaled quantities are the same as in Minkowski spacetime. 
The presence of $\epsilon$ makes this impossible. In Ref.~\cite{Cai:2018teh}, the self-similar velocity profile is assumed anyway. But the existence of 
an explicit time dependence from $a^{\prime}$ makes it impossible to solve, except in corners of the parameter space where it vanishes numerically.
It is also in doubt if there exists a self-similar solution at all for these equations and we refrain from going in that direction.

Despite this dilemma, we can still cast above equations in the form~\ref{eq:fluid-system:EOM} under the assumption that $\epsilon$ is a constant of time
during this very short period of time.
Then the first equation can be reorganized in the following way:
\begin{equation}
\left[\gamma^{\prime} + \triangledown \cdot (\gamma {\bf v}) + 3 \frac{a^{\prime}}{a} \gamma \right] (e + p)
+ \gamma e^{\prime} + \gamma {\bf v} \cdot \triangledown e = 0 .
\end{equation}
Then $\epsilon$ cancels out in $(e+p)$ and drops out in $e^{\prime}$, and of course also in $\triangledown e$. So above $e$ and $p$ can include
only the fluid part. Then one can put it back into the previous form~\ref{eq:parallel_a} and define the rescaled 
quantities: $\tilde{e}$, $\tilde{p}$, which obey exactly the same equation as in the Minkowski spacetime. Therefore we obtain the second equation in 
Eq.~\ref{eq:fluid-system:EOM} and the first in Eq.~\ref{eq:epprofile}.
Similarly for Eq.~\ref{eq:perp_a}, $\epsilon$ drops out in all terms and one can safely define the rescaled quantities, and obtain the
third equation in Eq.~\ref{eq:fluid-system:EOM} and the second in Eq.~\ref{eq:epprofile}. Combining these two equations again gives the same 
Eq.~\ref{eq:vxi} for the velocity field.

\subsubsection{Matching at Bubble Wall}

The equation~\ref{eq:vxi} for both regions needs the junction conditions at the wall to connect them. 
They are derived by integrating the conservation of energy momentum tensor
across the bubble wall, which gives in the wall frame (note $+,-$ denote quantities at positions immediately outside and inside the wall)~\footnote{
Also we follow the conventional procedure by neglecting the time dependence of the various quantities.
}
\begin{eqnarray}
    T_+^{r\eta} = T_-^{r\eta} , \\
    T_+^{rr} = T_-^{rr} ,
\end{eqnarray}
where $v_-$  and $v_+$ are both at wall frame. These two equations imply 
\begin{equation}
    \label{eq:1}
    (e_+ + p_+)v_+ \gamma^2_+  = (e_- + p_-)v_- \gamma^2_-  ,
\end{equation}
\begin{equation}
    \label{eq:2}
    (e_+ + p_+)v_+^2 \gamma^2_+ + p_+= (e_- + p_-)v_-^2 \gamma^2_- + p_-.
\end{equation}
Here both $e_{\pm}$ and $p_{\pm}$ are the ordinary energy density and pressure and include the vacuum energy $\epsilon$. 
The reason is while they can be neglected away from the bubble wall due to the vanishing
spatial gradient, they jump across the bubble wall and give non-negligible contributions to the above equations. 
The junction equations can be solved by making the change of variables 
$v_{\pm} = \tanh(\vartheta{\pm})$ and $\gamma_{\pm}^2 = \cosh^2(\vartheta{\pm})$ which, after simplifying, 
will yield two linear equations in $\cosh^2(\vartheta+)$ and $\cosh^2(\vartheta-)$. The solution will give 
\begin{eqnarray}
    v_+ = \sqrt{\frac{(p_- - p_+)(e_- + p_+)}{(e_- - e_+)(e_+ + p_-)}}, \nonumber \\
    v_- = \sqrt{\frac{(p_+ - p_-)(e_+ + p_-)}{(e_+ - e_-)(e_- + p_+)}}.
\end{eqnarray}
The product and ratio of $v_+$ and $v_-$ can further be found,
\begin{equation}
    v_+ v_- = \frac{p_+ - p_-}{e_+ - e_-}, \quad \frac{v_+}{v_-} = \frac{e_- + p_+}{e_+ + p_-}.
    \label{eq:v+v-}
\end{equation}
Plugging $e_{\pm}, p_{\pm}$ as specified by the bag equation of state in Eq.~\ref{eq:bageos} leads to 
\begin{eqnarray}
    v_+ v_- = \frac{1 - (1- 3 \alpha_+) \sigma}{3 - 3(1 + \alpha_+) \sigma} ,\\
    \frac{v_+}{v_-} = \frac{3 + (1 - 3 \alpha_+) \sigma}{1 + 3(1 + \alpha_+) \sigma} , 
    \label{eq:v+v-withalpha+}
\end{eqnarray}
where $\alpha_+$ and $\sigma$ are defined by 
\begin{equation}
    \alpha_+ = \left.\frac{\epsilon}{a_+ T^4_{+}}\right|_{\text{wall}}, \quad \sigma = \left.\frac{a_+ T^4_+}{a_- T^4_-}\right|_{\text{wall}}. 
\label{eq:alphasigma}
\end{equation}
$\alpha_+$ characterizes the amount of vacuum energy released from the phase transition normalized by the total radiation energy density 
immediately outside the bubble (as denoted by the subscript ``wall''). 
It is not the $\alpha$ usually used in phase transition analyses. Rather, its value should be solved from the requirement that 
far from the bubble where the plasma is not perturbed (denote by $\infty$), the corresponding $\alpha_+$ at $\infty$ matches $\alpha$.
The two equations in Eq.~\ref{eq:v+v-withalpha+} can be solved for both $r$ and $v_+$ to give two branch solutions for the velocity in the symmetric phase,
\begin{equation}
    v_+ = \frac{1}{1+\alpha_+} \left[ \left( \frac{v_-}{2} + \frac{1}{6 v_-} \right) \pm \sqrt{ \left(\frac{v_-}{2} + \frac{1}{6 v_-} \right)^2 + \alpha_+^2 + \frac{2}{3} \alpha_+ - \frac{1}{3}}  \right].
\label{eq:vpm}
\end{equation}
%%%%%%%%%%%%%%%%%%%%%%%%%%%%%%%%%%%%%%%%%%%%%%%%%%%%%%%%%%%%%%%%%%%%%%
\begin{figure}
\centering
\includegraphics[width=0.8\textwidth]{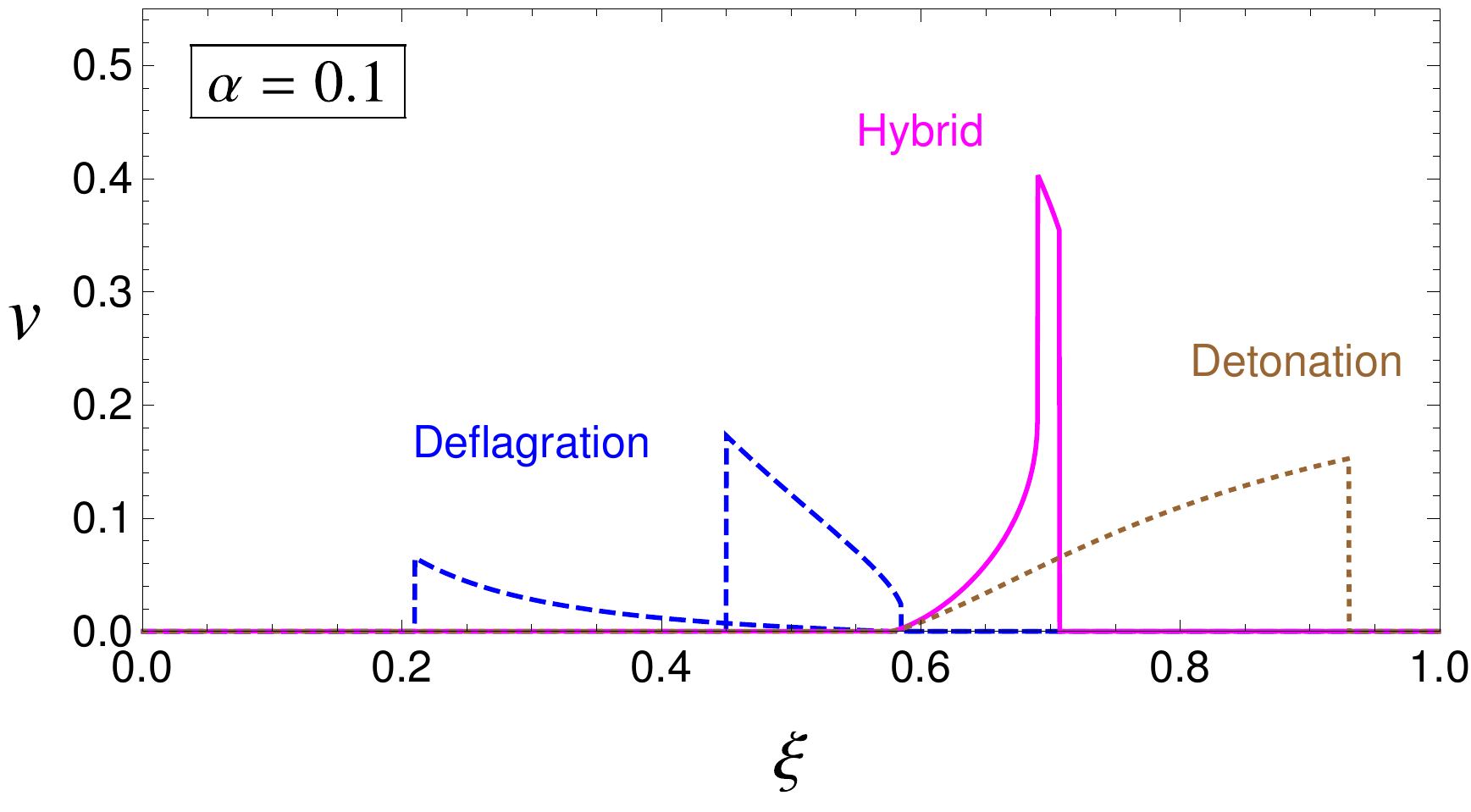}
\caption{
Representative velocity profiles surrounding the bubble walls.
}
\label{fig:velocityprofile}
\end{figure}
%%%%%%%%%%%%%%%%%%%%%%%%%%%%%%%%%%%%%%%%%%%%%%%%%%%%%%%%%%%%%%%%%%%%%%

Up to this point, the results for the velocity profile are exactly the same as in Minkowski spacetime, but with the understanding that
the time $t$ is replaced by the conformal time $\eta$, $v = d {\bf x}/d \eta$ and $(e,p)$ are replaced by $(\tilde{e}, \tilde{p})$.
We will not go into the details of the physics of above results but only summarize the main features of the velocity profile relevant for this study 
and refer the reader to Ref.~\cite{Espinosa:2010hh} for a more detailed analysis.

The fluid admits three modes of motion: deflagration, detonation and supersonic deflagration (also called hybrid)~\cite{KurkiSuonio:1995pp}, with 
representative velocity profiles shown in Fig.~\ref{fig:velocityprofile}. For deflagration, 
the velocity inside the bubble vanishes and is only non-zero outside. Detonation is the opposite, with non-zero velocity inside the bubble.
Supersonic deflagration has non-zero velocity both inside and outside the bubble. Therefore for deflagration, $v_- = v_w$ which 
should be used in Eq.~\ref{eq:vpm} to find $v_+$, choosing a value of $\alpha_+$. This $v_+$ is Lorentz transformed to the plasma
static frame to find $v(v_w)$ immediately outside the wall, which is then used as the boundary condition to solve for $v(\xi)$ outside the wall.
It might not consistently drop to zero, in which case a shock front is encountered and should be determined. Beyond the shock $v(\xi)=0$.
This gives a complete profile, but not yet the correct one, since a specific value of $\alpha_+$ is used in above determination of the profile.
This value needs to be tuned such that $\alpha_+ = \alpha$ far outside the bubble. For detonation, $v_+ = v_w$ and $v_-$ can be determined
from Eq.~\ref{eq:vpm} with $\alpha_+ = \alpha$ as outside the bubble the plasma is not perturbed. Then one can Lorentz transform $v_-$ to $v(v_w)$ immediately
inside the wall and use it as a boundary condition to determine the full profile. No inconsistency or shock front will be encountered in this case.
For supersonic deflagration, the condition $v_- = c_s$ is the boundary condition used. Shock front can exist in this case and should be treated similarly.
We refer the reader for more details in Ref.~\cite{Espinosa:2010hh}.

\subsection{Velocity Field in the Sound Shell Model}

With the velocity profile surrounding a single bubble determined, we can now find the total velocity field, as needed in Eq.~\ref{eq:vv}.
As we have already seen, in an expanding universe the equations of motion of the fluid are exactly the same as those in  non-expanding Minkowski
spacetime. This means that the equation of motion for the sound waves remain the same as its Minkowski counterpart, as long as 
we replace $t$ by $\eta$ and interpret the velocity as obtained by differentiation with respect to the conformal time. So the procedure parallels that in Ref.~\cite{Hindmarsh:2019phv}.

Lets start with the contribution from one bubble. Before it collides with another bubble at $\eta_{fc}$ (see Fig.~\ref{fig:lifetime}), the velocity profile is governed by equations given in previous sections. 
After the collision, the friction vanishes and the velocity field starts freely propagating and becomes sound waves, with the speed of  sound $c_s$.
So we need to match the velocity profile surrounding this bubble with the velocity field at the time when the friction vanishes.
Before collision, we can Fourier decompose the velocity field as
\begin{eqnarray}
v^i(\eta < \eta_{fc}, \mathbf{x}) = \frac{1}{2} \int \frac{d^3 q}{(2\pi)^3} 
\left[
\tilde{v}^i_{\mathbf{q}}(\eta) e^{i \mathbf{q} \cdot \mathbf{x}} 
+
\tilde{v}^{i \ast}_{\mathbf{q}}(\eta) e^{-i \mathbf{q} \cdot \mathbf{x}} 
\right] ,
\label{eq:vqeta}
\end{eqnarray}
with  $\bf{x}$ being the comoving coordinate and $\bf{q}$ the comoving wavenumber.
After collision, the velocity field freely propagates as sound waves and admits the following decomposition:
\begin{eqnarray}
v^i(\eta, \mathbf{x}) = \int \frac{d^3 q}{(2\pi)^3} 
\left[
{v}^i_{\mathbf{q}} e^{- i \omega \eta + i \mathbf{q} \cdot \mathbf{x}} 
+
{v}^{i \ast}_{\mathbf{q}} e^{i \omega \eta -i \mathbf{q} \cdot \mathbf{x}} 
\right] ,
\end{eqnarray}
where $\omega = q c_s$. Since the plasma  consists of relativistic particles, $c_s = 1/\sqrt{3}$.
Here $v_{\mathbf{q}}^i$ is independent of $\eta$, different from $\tilde{v}^i_{\mathbf{q}}(\eta)$. 

The task is then to find the contribution to $v_{\mathbf{q}}^i$ from 
$\tilde{v}^i_{\mathbf{q}}(\eta)$ at $\eta_{fc}$. Since the equation governing the sound waves is of second order,
we need the following initial conditions:  $\tilde{v}^i_{\mathbf{q}}(\eta)$ and $\tilde{v}^{i \prime}_{\mathbf{q}}(\eta)$ at $\eta_{fc}$. 
While one can obtain $\tilde{v}^i_{\mathbf{q}}(\eta)$ directly from the velocity profile in the previous section, one subtlety appears here
for $\tilde{v}^{i \prime}_{\mathbf{q}}(\eta)$. As demonstrated in Ref.~\cite{Hindmarsh:2019phv}, 
the equation governing $\tilde{v}^{i \prime}_{\mathbf{q}}(\eta)$ before the collision 
relies on a force term from the scalar field, which disappears once the collision occurs. So the value $\tilde{v}^{i \prime}_{\mathbf{q}}(\eta)$
calculated with this force (as was previously used in Ref.~\cite{Hindmarsh:2016lnk}) is different from the corresponding value without it. 
It is the latter one that should enter the initial conditions for the sound waves. In this case, rather than calculating $\tilde{v}^{i \prime}_{\mathbf{q}}(\eta)$
from the velocity profile $\tilde{v}^i_{\mathbf{q}}(\eta)$, we need to calculate it directly from the energy fluctuation:
\begin{eqnarray} 
\lambda(x) = \frac{\tilde{e}(x) -\bar{\tilde{e}}}{\bar{\tilde{\omega}}},
\end{eqnarray} 
where a bar denotes averaged quantity and tilde denotes rescaled quantity.
Similarly its Fourier component $\tilde{\lambda}_{\mathbf{q}}$ can be defined in analogy to Eq.~\ref{eq:vqeta}.
%%%%%%%%%%%%%%%%%%%%%%%%%%%%%%%%%%%%%%%%%%%%%%%%%%%%%%%%%%%%%%%%%%%
\begin{figure}
\centering
\includegraphics[width=0.6\textwidth]{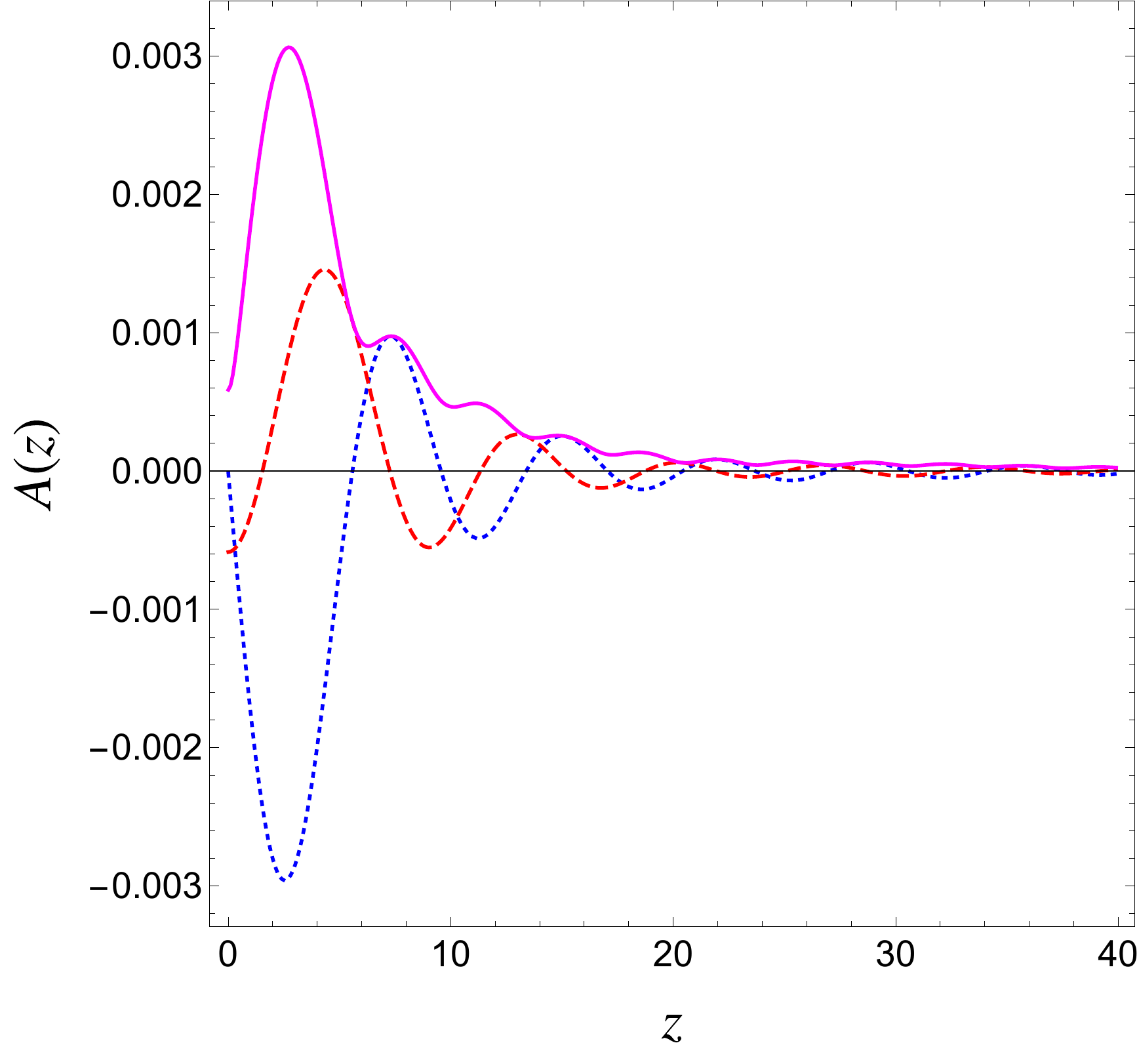}
\caption{
\label{fig:Az}
The real (blue dotted), imaginary (red dashed) parts and absolute value (magenta solid) of $A(z)$ (defined below Eq.~\ref{eq:vqexplicit}) for $v_w=0.92$ and $\alpha=0.0046$.
}
\end{figure}
%%%%%%%%%%%%%%%%%%%%%%%%%%%%%%%%%%%%%%%%%%%%%%%%%%%%%%%%%%%%%%%%%%%
The equations for sound waves then follow:
\begin{eqnarray}
&&\tilde{\lambda}^{\prime}_{\mathbf{q}} + i q^j \tilde{v}^j_{\mathbf{q}} = 0, \nonumber \\
&&\tilde{v}^{j \prime}_{\mathbf{q}} + c_s^2 i q^j \tilde{\lambda}_{\mathbf{q}} = 0 .
\end{eqnarray}
Therefore $\tilde{v}^{j \prime}_{\mathbf{q}} = - c_s^2 i q^j \tilde{\lambda}_{\mathbf{q}}$, and one needs to 
calculate $v^i(\eta, \bf{x})$ and $\lambda(\eta,\bf{x})$ from the self-similar velocity profile for one bubble.
In coordinate space, the velocity profile for the n-th bubble can be written as 
\begin{eqnarray}
\mathbf{v}^{(n)}(\eta, \mathbf{x}) = \hat{\mathbf{R}}(\mathbf{x}) v(\xi),
\end{eqnarray}
where $\mathbf{R}(\mathbf{x}) \equiv \mathbf{x} - \mathbf{x}^{(n)}$, 
$\xi \equiv |\mathbf{R}^{(n)}|/T^{(n)}$ and $T^{(n)}(\eta) \equiv \eta - \eta^{(n)}$, with $\mathbf{x}^{(n)}$ and $\eta^{(n)}$
the coordinate of the bubble center and the conformal time when the bubble is nucleated. Similarly for $\lambda$, as it is a scalar field,
we can define $\lambda(\eta, \mathbf{x}) \equiv \lambda(\xi)$. With the profile specified in  coordinate space, the corresponding Fourier
coefficients can be obtained straightforwardly
\begin{eqnarray}
\tilde{v}^{j (n)}_{\mathbf{q}}(\eta_{fc})   &=& e^{-i \mathbf{q} \cdot \mathbf{x}^{(n)}} (T^{(n)})^3 i \hat{z}^j f^{\prime}(z) |_{\eta = \eta_{fc}} , \nonumber \\
\tilde{\lambda}^{(n)}_{\mathbf{q}}(\eta_{fc}) &=& e^{-i \mathbf{q} \cdot \mathbf{x}^{(n)}} (T^{(n)})^3 l(z)|_{\eta = \eta_{fc}},
\end{eqnarray}
with $\mathbf{z} \equiv \mathbf{q} T^{(n)}$ and the two functions $f(z)$ and $l(z)$ given by
\begin{eqnarray}
f(z) &=& \frac{4 \pi}{z} \int_0^{\infty} d \xi\ v(\xi)\ \sin(z \xi), \nonumber \\
l(z) &=& \frac{4 \pi}{z} \int_0^{\infty} d \xi\ \xi\ \lambda(\xi) \sin(z \xi) .
\end{eqnarray}
Then the n-th bubble's contribution to the Fourier coefficient of the sound waves is
\begin{eqnarray}
v_{\mathbf{q}}^{j(n)} = \frac{1}{2} \left[
\tilde{v}_{\mathbf{q}}^{j(n)}(\eta_{fc}) + c_s \hat{q}^{j} \tilde{\lambda}_{\bf{q}}^{(n)}(\eta_{fc}) 
\right] e^{i \omega \eta_{fc}} ,
\end{eqnarray}
and after using the explicit expression of the bubble profile, 
\begin{eqnarray}
v_{\mathbf{q}}^{j (n)}= i \hat{z}^j (T_{fc}^{(n)})^3 e^{i \omega \eta_{fc} - i \mathbf{q} \cdot \mathbf{x}^{(n)}}A(z_{fc}) ,
\label{eq:vqexplicit}
\end{eqnarray}
where $A(z_{fc}) = [f^{\prime}(z_{fc}) - i c_s l(z_{fc})]/2$, with an example shown in Fig.~\ref{fig:Az}.  Thus we have calculated the contribution to $v_{\bf{q}}^i$ 
from one bubble that is nucleated randomly. The randomness of this bubble is reflected in its formation time, location, collision time and its radius.
Since the radius at collision is fixed once its formation and collision times are given, there are three independent random variables.

The velocity field after all bubbles have disappeared, can be assumed to be the linear addition of the contributions from all bubbles, 
which is the essence of the sound shell model~\cite{Hindmarsh:2016lnk,Hindmarsh:2019phv}.
Suppose the total number of bubbles nucleated within a Hubble volume with comoving size $V_c$ is $N_b$. Then the velocity field can be assumed,
according to the sound shell model, to be given by
\begin{eqnarray}
v_{\bf{q}}^i = \sum_{n=1}^{N_b} v_{\bf{q}}^{i (n)} . 
\end{eqnarray}

\subsection{Velocity Power Spectrum}

As these $N_b$ bubbles are just one realization of the phase transition, the resulting $v_{\bf{q}}^i$ has a random nature with it and
follows a Gaussian distribution to a good approximation according to the central limit theorem~\footnote{
If there is a sufficiently large population of bubbles within this single volume, the summation of these contributions can also remove the randomness,
equivalent to an ensemble average. 
}.
Randomness of this kind can be removed by doing an ensemble average of the product: $\langle v_{\bf{q}}^i v_{\bf{q}}^{j \ast} \rangle$, which is 
all needed for a Gaussian distribution. Now let us see how this is achieved. 

The $N_b$ bubbles can be separated into groups with the bubbles within 
each group sharing a common formation and collision time. Then the only variable that is random across the bubbles of one group, e.g., group $g$ with
$N_{g}$ bubbles, is the spatial locations of the bubbles when they form. Now consider group $g$. Its contribution to the correlator is
\begin{eqnarray}
\langle v_{\bf{q}_1}^i v_{\bf{q}_2}^{j \ast} \rangle_g = 
\hat{q}^i_1 \hat{q}_2^j 
[T_{fc}^{(g)}]^6 A(z_{fc}^{(n)}) A(z_{fc}^{(m)})^{\ast} e^{i (\omega_1 - \omega_2) \eta^{(g)}_{fc}} 
\langle
\sum_{m,n=1}^{N_g} e^{i \bf{q}_2 \cdot \bf{x}^{(m)} - i \bf{q}_1 \cdot \bf{x}^{(n)}}
\rangle .
\end{eqnarray}
Here the order of the ensemble average and the summation can be switched. Since the ensemble average of each of these $N_g$ terms gives
the same result and oscillatory cross terms vanish, we have
\begin{eqnarray}
\langle
\sum_{m,n=1}^{N_g} e^{i \bf{q}_2 \cdot \bf{x}^{(m)} - i \bf{q}_1 \cdot \bf{x}^{(n)}}
\rangle 
&=&
N_g 
\delta_{m n} 
\langle e^{i \bf{q}_2 \cdot \bf{x}^{(m)} - i \bf{q}_1 \cdot \bf{x}^{(n)}} \rangle \nonumber \\
&=& N_g \frac{1}{V_c} \int d^3 {\bf{x}^{(*)}}\  e^{i (\bf{q}_2 - \bf{q}_1) \cdot \bf{x}^{(*)}} \nonumber \\
&=& N_g \frac{1}{V_c} (2 \pi)^3 \delta^3(\bf{q}_1 - \bf{q}_2)
.
\end{eqnarray}
\begin{figure}[t]
\centering
\includegraphics[width=0.6\textwidth]{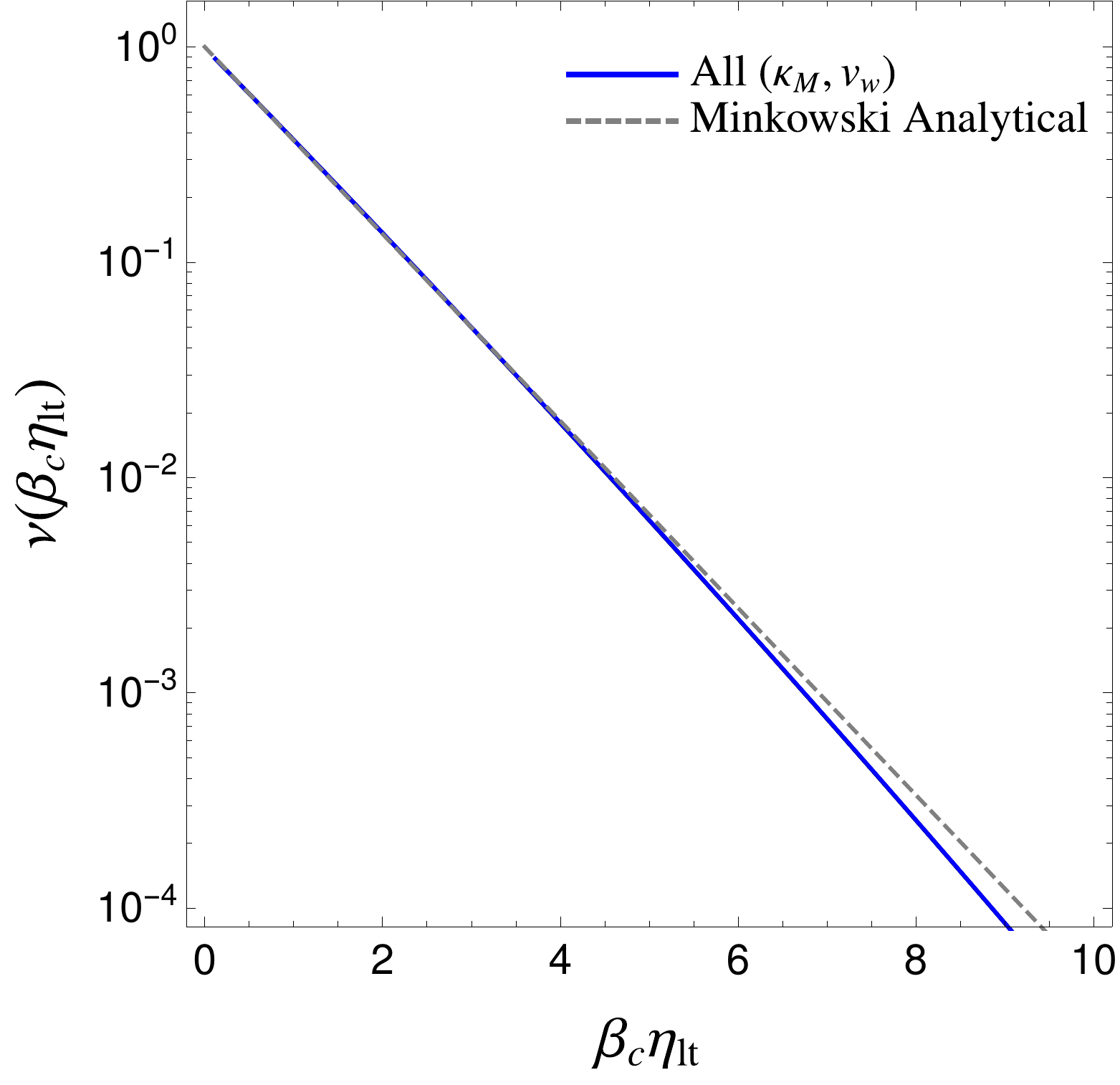}
\caption{
\label{fig:nu}
The dimensionless bubble lifetime distribution $\nu(\beta_c \eta)$ defined in Eq.~\ref{eq:nudef} and more explicitly in Eq.~\ref{eq:nudefexplicit}. 
All previously used choices of $\kappa, v_w$ give the same blue line. The gray dashed line is the analytically derived result $e^{-\beta t_{\text{lt}}}$
in Ref.~\cite{Hindmarsh:2019phv}.
}
\end{figure}
The constraint $\bf{q}_1 = \bf{q}_2$ removes the $\eta_{fc}^{(g)}$ dependence, leading to a result solely dependent on 
the conformal lifetime of the bubble $T_{fc}^{(g)} \equiv \eta_{\text{lt}}$ but not their absolute formation or destruction time:
\begin{eqnarray}
\langle v_{\bf{q}_1}^i v_{\bf{q}_2}^{j \ast} \rangle_g =  
\hat{q}_1^i \hat{q}_2^j 
\eta_{\text{lt}}^6 |A(q \eta_{\text{lt}})|^2 \frac{N_g}{V_c} (2 \pi)^3 \delta^3(\bf{q}_1 - \bf{q}_2) .
\end{eqnarray}
This result means that we can combine groups with the same $\eta_{\text{lt}}$, and of course, different formation time, 
by solely enlarging the value of $N_g$. In the following 
we will simply stick to the group label ``$g$'', though its definition is changed and now includes all bubbles with the same $\eta_{\text{lt}}$.
Restricting to a sufficiently small region centered at $\eta_{\text{lt}}$, the number $N_g$ is \red{an} still an
infinitesimally small fraction of $N_b$ and can be written as
\begin{eqnarray}
N_g = N_b P(\eta_{\text{lt}}) d \eta_{\text{lt}} ,
\end{eqnarray}
where $P(\eta_{\text{lt}})$ is the probability density for bubbles to have conformal lifetime in the range $(\eta_{\text{lt}}, \eta_{\text{lt}} + d \eta_{\text{lt}})$, thus
with dimension $1$ and normalized by
\begin{eqnarray}
\int d \eta_{\text{lt}} P(\eta_{\text{lt}}) = 1.
\end{eqnarray}
Adding the contributions from all the groups and noting that cross terms vanish due to the oscillatory behavior, we have 
\begin{eqnarray}
\langle v_{\bf{q}_1}^i v_{\bf{q}_2}^{j \ast} \rangle
=
\hat{q}_1^i \hat{q}_2^j (2 \pi)^3 \delta^3({\bf q}_1 - {\bf q}_2) 
\int d \eta_{\text{lt}} \left[ P(\eta_{\text{lt}}) \frac{N_b}{V_c} \right] \eta_{\text{lt}}^6 |A(q \eta_{\text{lt}})|^2  .
\end{eqnarray}
One can now identify the quantity in the square bracket as the conformal lifetime distribution defined in Eq.~\ref{eq:nbc0}:
\begin{eqnarray}
P(\eta_{\text{lt}}) \frac{N_b}{V_c} = \tilde{n}_{b,c}(\eta_{\text{lt}}) .
\end{eqnarray}
Since $P(\eta_{\text{lt}})$ is of dimension $1$, it is convenient to define a dimensionless version of it: $\nu$, with 
\begin{eqnarray}
P(\eta_{\text{lt}}) \equiv \beta_c \nu( \beta_c \eta_{\text{lt}}) ,
\end{eqnarray}
and thus
\begin{eqnarray}
\tilde{n}_{b,c}(\eta_{\text{lt}}) = \frac{\beta_c}{R_{\ast c}^3} \nu( \beta_c \eta_{\text{lt}}) ,
\label{eq:nudef}
\end{eqnarray}
where $R_{\ast c}$ is the asymptotic comoving mean bubble separation. Then we have

%%%%%%%%%%%%%%%%%%%%%%%%%%%%%%%%%%%%%%%%%%%%%%%%%%%%%%%%%
\begin{figure}[t]
\centering
\includegraphics[width=0.7\textwidth]{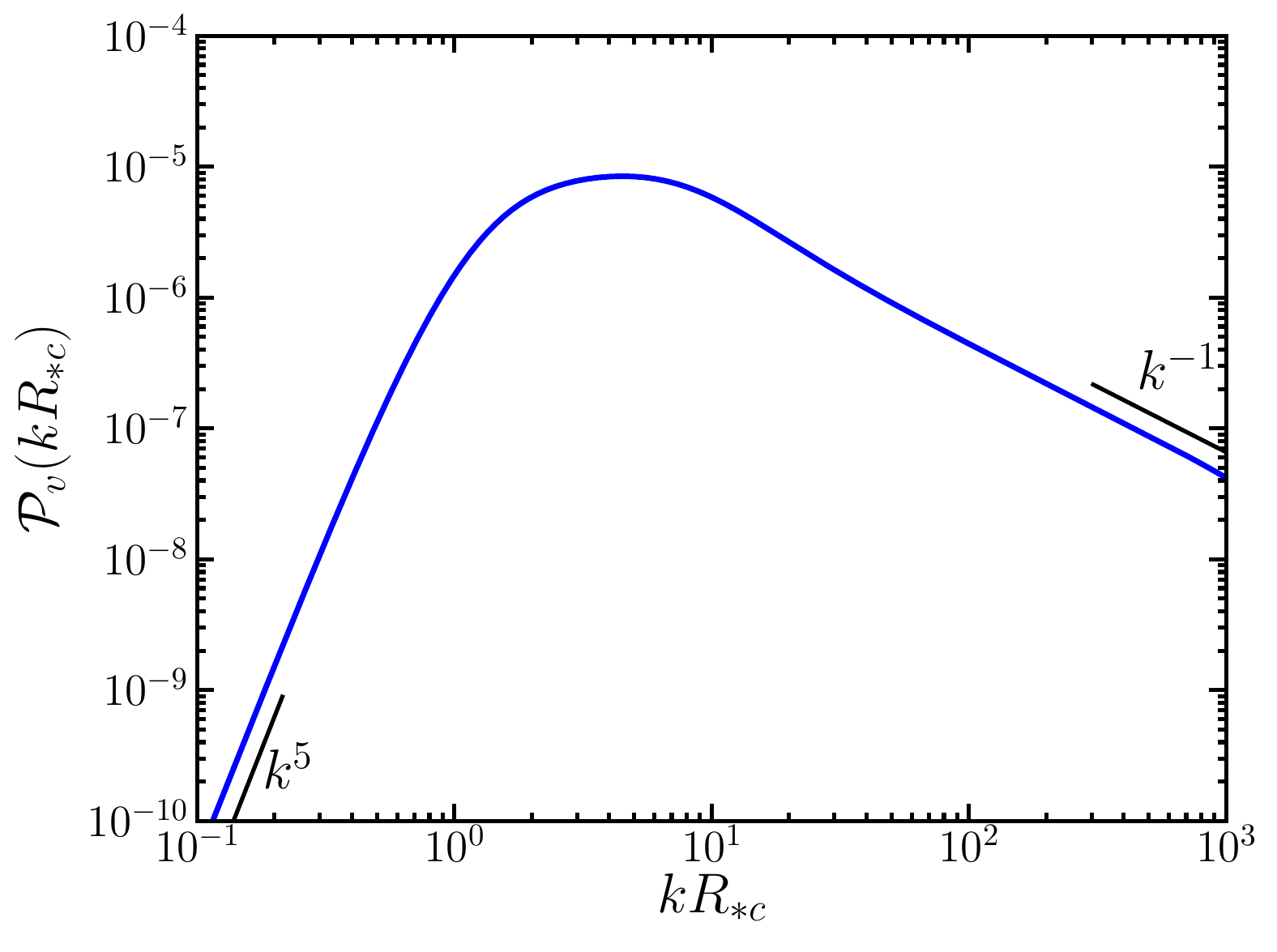}
\caption{\label{fig:Pv}
Representative velocity power spectrum calculated in the sound shell model for a weak phase transition with $\alpha = 0.0046$ and $v_w = 0.92$. 
The bubbles are assumed to nucleate exponentially. The low and high frequency regimes follow the $k^5$ and $k^{-1}$ power law fits respectively (black solid lines).  
See Ref.~\cite{Hindmarsh:2019phv} for more details of its properties.
}
\end{figure}
%%%%%%%%%%%%%%%%%%%%%%%%%%%%%%%%%%%%%%%%%%%%%%%%%%%%%%%%%
\begin{eqnarray}
\langle v_{\bf{q}_1}^i v_{\bf{q}_2}^{j \ast} \rangle
=
\hat{q}_1^i \hat{q}_2^j (2 \pi)^3 \delta^3({\bf q}_1 - {\bf q}_2) 
\underbrace{
\frac{1}{R_{\ast c}^3 \beta_c^6} 
\int d \tilde{T} \tilde{T}^6 \nu(\tilde{T}) |A(\frac{q \tilde{T}}{\beta_c})|^2 ,
}_{\equiv P_v(q)}
\label{eq:Pvdef}
\end{eqnarray}
with here $\tilde{T} = \beta_c \eta_{\text{lt}}$, and we have defined the spectral density $P_v(q)$ for the plane wave amplitude $v_{\bf{q}}^i$. 
Lets write down the explicit expression for $\nu(\tilde{T})$. From Eq.~\ref{eq:nudef} and ~\ref{eq:nbc0}, we have
\begin{eqnarray}
\nu(\tilde{T}) &=& 
v_w R_{\ast c}^3 
\int^{t_f}_{t_c} d t^{\prime} p(t^{\prime}) a^3(t^{\prime})\frac{\mathcal{A}_c(t^{\prime}, v_w \tilde{T}/\beta_c)}{\beta_c} ,
\label{eq:nudefexplicit}
\end{eqnarray}
which can be directly used for numerical calculations once $t^{\prime}$ is transformed to $T^{\prime}$ as demonstrated in previous sections.
The numerically calculated distribution for the examples we have been using is shown in Fig.~\ref{fig:nu}.
For all choices of $\kappa, v_w$, the distributions are almost indistinguishable, shown as the blue curve, and
it coincides with the gray dashed curve which denotes the distribution $e^{-\tilde{T}}$, derived analytically in Ref.~\cite{Hindmarsh:2019phv}. 
With $\nu(\tilde{T})$ obtained, the spectral density $P_v(\tilde{T})$ can be calculated straightforwardly from its definition in Eq.~\ref{eq:Pvdef}.

To calculate the velocity power spectrum, we need to evaluate the correlator
\begin{eqnarray}
\langle \tilde{v}_{\mathbf{q}}^i(\eta_1) \tilde{v}_{\mathbf{k}}^{j \ast}(\eta_2)\rangle| = \delta^{3}(\mathbf{q}-\mathbf{k}) \hat{q}^i \hat{k}^j G(q, \eta_1, \eta_2) ,
\end{eqnarray}
and it can be shown that 
\begin{eqnarray}
G(q, \eta_1, \eta_2) = 2 P_v(q) \cos[\omega (\eta_1 - \eta_2)] .
\label{eq:G}
\end{eqnarray}
Plugging it into Eq.~\ref{eq:correpi} or ~\ref{eq:PI2} gives the stress energy correlator.
Also the velocity field power spectrum $\mathcal{P}_{v}$ follows naturally,
\begin{eqnarray}
\mathcal{P}_v &=& \frac{q^3}{2 \pi^2} [2 P_v(q)]  \nonumber \\
&& = \frac{1}{64 \pi^4 v_w^6}  (q R_{\ast c})^3
\int d \tilde{T} \tilde{T}^6 \nu(\tilde{T}) \left|A\left(\frac{(q R_{\ast c}) \tilde{T}}{(8\pi)^{1/3} v_w}\right)\right|^2 ,
\label{eq:calPv}
\end{eqnarray}
and we have used $\beta_c R_{\ast c} = (8 \pi)^{1/3} v_w$.
It is obvious to see that $\mathcal{P}_v$ is dimensionless, as it is constructed with purely dimensionless quantities.
A representative profile for the velocity power spectrum is shown in Fig.~\ref{fig:Pv} assuming an exponential bubble nucleation rate, 
and more details about its properties can be found in Ref.~\cite{Hindmarsh:2016lnk}.

\section{Gravitational Wave Power Spectrum\label{sec:spectrum}}

We can now go back to Eq.~\ref{eq:correh} and collect all the pieces to calculate the gravitational power spectrum.
It only remains to calculate the Green's function, and it requires to specify an expansion scenario. 
We will as usual focus on the RD and MD scenarios as examples, but the method here is applicable to any expansion history.

\subsection{Solutions in Radiation and Matter Domination}

First, we choose a parameter to measure the time of the cosmic history. 
It can either be the actual time $t$, the conformal time $\eta$, the redshift $z$ or the scale factor $a$. 
To present a result independent of the origin of the time coordinate, we choose the dimensionless
scale factor ratio $y \equiv a/\aref$, giving then $d/dt = \dot{a}/\aref d/dy$. 
Here $\aref$ is the time when the source, the sound waves, becomes active, so that $y$ starts from $1$. 
The Friedmann equation gives the relation between $y$ and the conformal time
\begin{eqnarray}
y = \frac{\kappa_M}{4} (\aref H_s)^2 (\eta - \eta_s)^2 + \aref H_s (\eta - \eta_s) + 1.
\end{eqnarray}
It is obvious that when $\eta=\eta_s$, we have $y=1$. Also it does not matter how the origin of the conformal time is chosen as
it only depends on $\Delta \eta \equiv \eta-\eta_s$. 
For RD, where $\kappa_M \sim 0$, we have $y = \aref H_s (\eta - \eta_s) + 1$. For MD, $\kappa_M \approx 1$ and 
$y = [\frac{1}{2}\aref H_s (\eta - \eta_s) + 1]^2$.
In the literature, it is usually approximated that $a \propto \eta$ deep inside the radiation era or $a \propto \eta^2$ deep inside the matter era.
However we remain agnostic about when the phase transition happens and do not require it to start deep inside the radiation or matter era.
Also the duration of the phase transition is very small compared with the conformal time, which makes 
such approximation quite crude. But our choice using $y$ is free from above limitations and offers a more accurate description of phase transition
process.

With $y$, the Hubble rate, when assuming the existence of both matter and radiation components, takes the following form
\begin{eqnarray}
H = H_s \sqrt{\frac{\kappa_M}{y^3} + \frac{1-\kappa_M}{y^4}} ,
\label{eq:Hy}
\end{eqnarray}
where $\kappa_M$ is the matter fraction of the total energy density at $t_s$. Note this $\kappa_M$ is defined differently from that 
in Eq.~\ref{eq:HkappaM}, which is defined at $T_c$. If the lifetime of the sound waves is sufficiently long, we can neglect this difference.

Switching from the conformal time $\eta$ to $y$ in Eq.~\ref{eq:hq}, the Einstein equation becomes~\footnote{We are using a simplified notation for $h$ and $\pi^T$}:
\begin{eqnarray}
(\kappam y + 1 - \kappam) \frac{d^2 h_q}{d y^2} + \left[\frac{5}{2} \kappam + \frac{2(1-\kappam)}{y}\right] \frac{d h_q}{d y}
+ \doublewidetilde{q}^2 h_q = \frac{16 \pi G a(y)^2 \pi_q^T(y)}{(\aref H_s)^2}.
\end{eqnarray}
Here $\doublewidetilde{q} \equiv q/(\aref \Href)$, and characterizes the number of wavelengths contained within a Hubble radius at $t_s$.
The Green's function can be found by solving the homogeneous version of this equation, together with a slightly modified boundary conditions compared with Eq.~\ref{eq:boundary}:
\begin{eqnarray}
G(y \leqslant y_0) = 0, \quad \quad \frac{\partial G(y, y_0)}{\partial y}|_{\tilde{\eta} = \tilde y_0^+} = \frac{1}{\kappam y_0 + 1 - \kappam} .
\end{eqnarray}
The solution to the homogeneous equation is a linear combination of the hypergeometric function and Bessel functions. For the case of radiation domination $\kappa_M \ll 1$
and matter domination $\kappa_M \approx 1$, the solutions take simpler forms that can be expressed in terms of elementary functions. 
For RD, the equation becomes simpler when expressed using the parameter $\tilde{y}$, defined by
\begin{eqnarray}
\tilde{y} = y \doublewidetilde{q} = q (\eta - \eta_s) + \doublewidetilde{q} = \Delta \tilde{\eta} + \doublewidetilde{q} .
\label{eq:yTildeRD}
\end{eqnarray}
Then the Einstein equation becomes
\begin{eqnarray}
\frac{d^2 h_q}{d \tilde{y}^2} + \frac{2}{\tilde{y}} \frac{d h_q}{d \tilde{y}} + h_q
 = \frac{16 \pi G a(y)^2 \pi_q^T(y)}{q^2} .
\end{eqnarray}
The corresponding Green's function can be easily solved:
\begin{eqnarray}
G(\tilde{y}, \tilde{y}_0) = \frac{\tilde{y}_0 \sin(\tilde{y} - \tilde{y}_0)}{\tilde{y}} .
\end{eqnarray}
For MD, the wave equation can be similarly simplified with 
\begin{eqnarray}
\tilde{y} = y \doublewidetilde{q}^2 = \left[\frac{1}{2} \Delta \tilde{\eta} + \doublewidetilde{q} \right]^2 .
\label{eq:yTildeMD}
\end{eqnarray}
Note this definition is different from that in the radiation dominated case. Then the Einstein equation becomes
\begin{eqnarray}
\tilde{y} \frac{d^2 h_q}{d \tilde{y}^2} + \frac{5}{2} \frac{d h_q}{d \tilde{y}} + h_q
 = \frac{16 \pi G a(\tilde{y})^2 \pi_q^T(\tilde{y})}{q^2} .
\end{eqnarray}
The homogeneous equation for $h_q$ can be transformed into the Bessel equation for a different variable $Z(\lambda)$ defined by $h_q = (\lambda/2)^{-3/2}Z(\lambda)$
with $\lambda = 2 \sqrt{\tilde{y}}$:
\begin{eqnarray}
\lambda^2 Z^{\prime \prime}(\lambda) + \lambda Z^{\prime}(\lambda) + \left[ \lambda^2 - \left(\frac{3}{2}\right)^2 \right] Z(\lambda) = 0.
\end{eqnarray}
The two independent solutions are the first and second kind Bessel functions both with order $3/2$, which can all be expressed in elementary functions. 
Upon using the boundary conditions, the Green's function is found to be~\footnote{
Alternatively, one can express above Green's functions using the conformal time. The corresponding Green's functions are
defined to be zero for $\eta \leqslant \eta_0$ and for $\eta > \eta_0$,
\begin{eqnarray}
G(\tilde{\eta}, \tilde{\eta}_0) = \left\{
\begin{array}{l}
\frac{\tilde \eta_0}{\tilde \eta} \sin(\tilde\eta - \tilde \eta_0) , \hspace{5.1cm} \text{RD} \\
 \frac{\tilde{\eta}_0}{\tilde{\eta}^3}\left[
(\tilde{\eta}_0-\tilde{\eta}) \cos(\tilde{\eta} - \tilde{\eta}_0)
+(\tilde{\eta}_0 \tilde{\eta}+1) \sin(\tilde{\eta} - \tilde{\eta}_0)
\right] .  \hspace{0.43cm} \text{MD}
\end{array}
\right.
\end{eqnarray}
We note that there is a typo in the Green's function for the matter dominated universe given in Ref.~\cite{Barenboim:2016mjm}, where
instead of $(\tilde{\eta}_0 - \tilde{\eta}) \cos(\tilde{\eta} - \tilde{\eta}_0)$, they have $-(\tilde{\eta}_0 - \tilde{\eta}) \cos(\tilde{\eta} - \tilde{\eta}_0)$. 
}
:
\begin{eqnarray}
G(\tilde{y}, \tilde{y}_0) = \frac{(\lambda \lambda_0 + 1)\sin(\lambda - \lambda_0) - (\lambda - \lambda_0)
\cos(\lambda - \lambda_0)}{\lambda^3/2} .
\end{eqnarray}
%Here the paramter $\lambda$ is actually $\tilde{\eta}$
Finally in both cases, the gravitational wave amplitude is given by 
\begin{eqnarray}
h_{ij}(\tilde{y}, {\bf q}) = 
\int_{\tilde{y}_s}^{\tilde{y}} d \tilde{y}^{\prime} G(\tilde{y}, \tilde{y}^{\prime}) \frac{16 \pi G a(\tilde{y}^{\prime})^2 \pi_{ij}^T(\tilde{y}^{\prime}, {\bf q})}{q^2} .
\end{eqnarray}

\subsection{Gravitational Wave Power Spectrum}

The spectral density for $h^{\prime}$, when using $\tilde{y}$ and the dimensionless stress energy tensor correlator $\tilde{\Pi}$ defined in Eq.~\ref{eq:PiTilde}, becomes
\begin{eqnarray}
P_{h^{\prime}}
&=&
[16 \pi G \left( \bar{\tilde{\epsilon}} + \bar{\tilde{p}} \right) \bar{U}_f^2]^2 L_f^3 
\int_{\tilde{y}_s}^{\tilde{y}} d \tilde{y}_1 \int_{\tilde{y}_s}^{\tilde{y}} d \tilde{y}_2 
\left(\frac{\partial \tilde{y}}{\partial \tilde{\eta}}\right)^2 
\frac{\partial G(\tilde{y}, \tilde{y}_1)}{\partial \tilde{y}}
\frac{\partial G(\tilde{y}, \tilde{y}_2)}{\partial \tilde{y}} \nonumber \\
&&\hspace{4cm} \times \frac{a_s^8}{a^2(\tilde{y}_1) a^2(\tilde{y}_2)} \frac{\tilde{\Pi}^2(k L_f, k \eta_1, k \eta_2)}{k^2} .
\end{eqnarray}
From the explicit form of the Green's functions derived earlier, we can see $P_{h^{\prime}}$ has the correct behavior $\propto 1/a(\tilde{y})^2$ for the
mode deep inside the horizon~\footnote{
For modes deep inside the horizon, $\tilde{y} \gg 1$ and $\tilde{y}_0 \gg 1$. 
Then both Green's functions take a universal form $\frac{a_0}{a} \sin(\tilde{\eta} - \tilde{\eta}_0)$. This implies that $h^{\prime} \propto 1/a$,
$P_{h^{\prime}} \propto 1/a^2$ and $\mathcal{P}_{\text{GW}} \propto 1/a^4$, behaving like radiation which is true for massless gravitons. 
}. 
%%%%%%%%%%%%%%%%%%%%%%%%%%%%%%%%%%%%%%%%%%%%%%%%%%%
\begin{figure}
\centering
\includegraphics[width=0.6\textwidth]{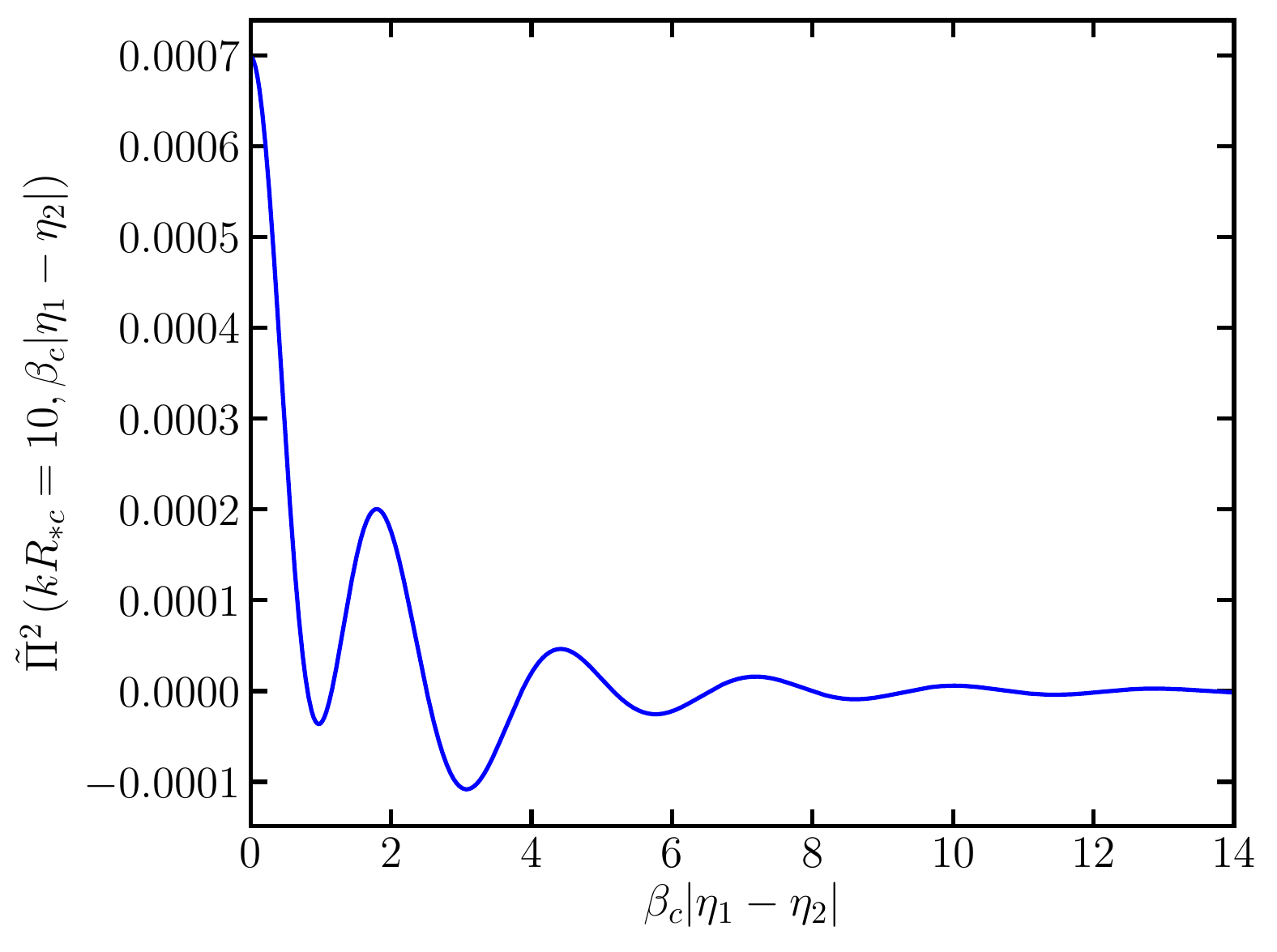}
\caption{
\label{fig:Pi}
Autocorrelation of the source for $k R_{\ast c} = 10$, calculated with the explicit expression in Eq.~\ref{eq:PiExplicit}.
}
\end{figure}
%%%%%%%%%%%%%%%%%%%%%%%%%%%%%%%%%%%%%%%%%%%%%%%%%%%
The dimensionless source correlator can be obtained from Eq.~\ref{eq:PI2}, ~\ref{eq:PiTilde}, ~\ref{eq:G}:
\begin{eqnarray}
\tilde{\Pi}^2 \left( k R_{\ast c}, \beta_c \left|\eta_1 - \eta_2 \right| \right)
&=&
\frac{\pi}{2} \frac{1}{\bar{U}_f^4} 
\int d^3 \tilde{q} \mathcal{P}_v(\tilde{q}) \mathcal{P}_v(\tilde{\bar{q}}) \frac{(1-\mu^2)^2}{\tilde{q} \tilde{\bar{q}}^5} \nonumber \\
&& \hspace{1cm} 
\times
\cos\left[c_s \tilde{q} \frac{\beta_c (\eta_1 - \eta_2)}{\beta_c R_{\ast c}} \right]
\cos\left[c_s \tilde{\bar{q}} \frac{\beta_c(\eta_1 - \eta_2)}{\beta_c R_{\ast c}} \right] .
\label{eq:PiExplicit}
\end{eqnarray}
Here $\tilde{q} = q R_{\ast c}$, a dimensionless quantity, and we use $L_f = R_{\ast c}$. In Fig.~\ref{fig:Pi}, we show this auto-correlator of the
source as a function of $\beta_c|\eta_1 - \eta_2|$. We can see the correlation is quickly lost as $\beta_c |\eta_1 - \eta_2|$ becomes larger than 
$\mathcal{O}(1)$.
Since the source correlator depends only on $\eta_1 - \eta_2$, we can change
the integration variables from $\tilde{y}_{1,2}$ to a quantity proportional to $(\eta_1 - \eta_2)$ and another independent linear combination.
For RD and MD, the relation between $(\eta_1 - \eta_2)$ and $y_{1,2}$ is given from Eq.~\ref{eq:yTildeRD}, ~\ref{eq:yTildeMD}:
\begin{eqnarray}
\frac{\beta_c(\eta_1 - \eta_2)}{\beta_c R_{\ast c}} = 
\frac{1}{R_{\ast c} a_s H_s}
\left\{
\begin{array}{l}
y_1 - y_2 \\
2(\sqrt{y_1} - \sqrt{y_2})
\end{array}
\right. ,
\label{eq:etay}
\end{eqnarray}
where the upper row applies to RD and lower one to MD.
Then for RD, we can make the following change of variables:
\begin{eqnarray}
\left\{
\begin{array}{l}
y_1 \\
y_2
\end{array}
\right.
\quad
\Rightarrow 
\quad
\left\{
\begin{array}{l}
y_1 - y_2 \equiv y_- \ ,\\
\frac{y_1 + y_2}{2} \equiv y_+ \ .
\end{array}
\right.
\end{eqnarray}
The integration range is $1 - \frac{1}{2} y_- \leqslant y_+ \leqslant y + \frac{1}{2} y_-$ when $1-y \leqslant y_- \leqslant 0$, and
$1 + \frac{1}{2} y_- \leqslant y_+ \leqslant y - \frac{1}{2} y_-$ when $0 \leqslant y_- \leqslant y-1$.
Similarly for MD, we can perform the following transformations:
\begin{eqnarray}
\left\{
\begin{array}{l}
y_1 \\
y_2
\end{array}
\right.
\quad
\Rightarrow 
\quad
\left\{
\begin{array}{l}
\lambda_1 - \lambda_2 \equiv y_- \ ,\\
\frac{\lambda_1 + \lambda_2}{2} \equiv y_+ \ ,
\end{array}
\right.
\end{eqnarray}
where $\lambda_i = 2 \sqrt{y_i}$ and the Jacobian is $\sqrt{y_1 y_2}$.
The range of integration is $2 + \frac{1}{2} y_- \leqslant y_+ \leqslant 2 \sqrt{y} - \frac{1}{2} y_-$ when $0 \leqslant y_- \leqslant 2(\sqrt{y} - 1)$
and $2 - \frac{1}{2} y_- \leqslant y_+ \leqslant 2 \sqrt{y} + \frac{1}{2} y_-$ when $2(1-\sqrt{y}) \leqslant y_- \leqslant 0$.

It turns out the relation $y_- \ll y_+$ generally holds, barring special parameter space.
This can be seen from Eq.~\ref{eq:etay} by noting that $\beta_c R_{\ast c} = (8\pi)^{1/3} v_w \approx 3 v_w < 3$, 
$R_{\ast c} a_s H_s \sim \mathcal{O}(10^{-3})$ from Fig.~\ref{fig:Rb}, and thus $y_- \sim \mathcal{O}(10^{-3})/v_w \times \beta_c (\eta_1 - \eta_2)$. 
Except for extremely small $v_w$, which gives highly suppressed gravitational waves, we have $y_- \ll 1$.
On the contrary, $y_+ \sim \mathcal{O}(1)$.
Then we have $y_- \ll y_+$. This means in the integration over $y_+$, we can keep the leading order in $y_-$.

\begin{figure}
\centering
\includegraphics[width=0.65\textwidth]{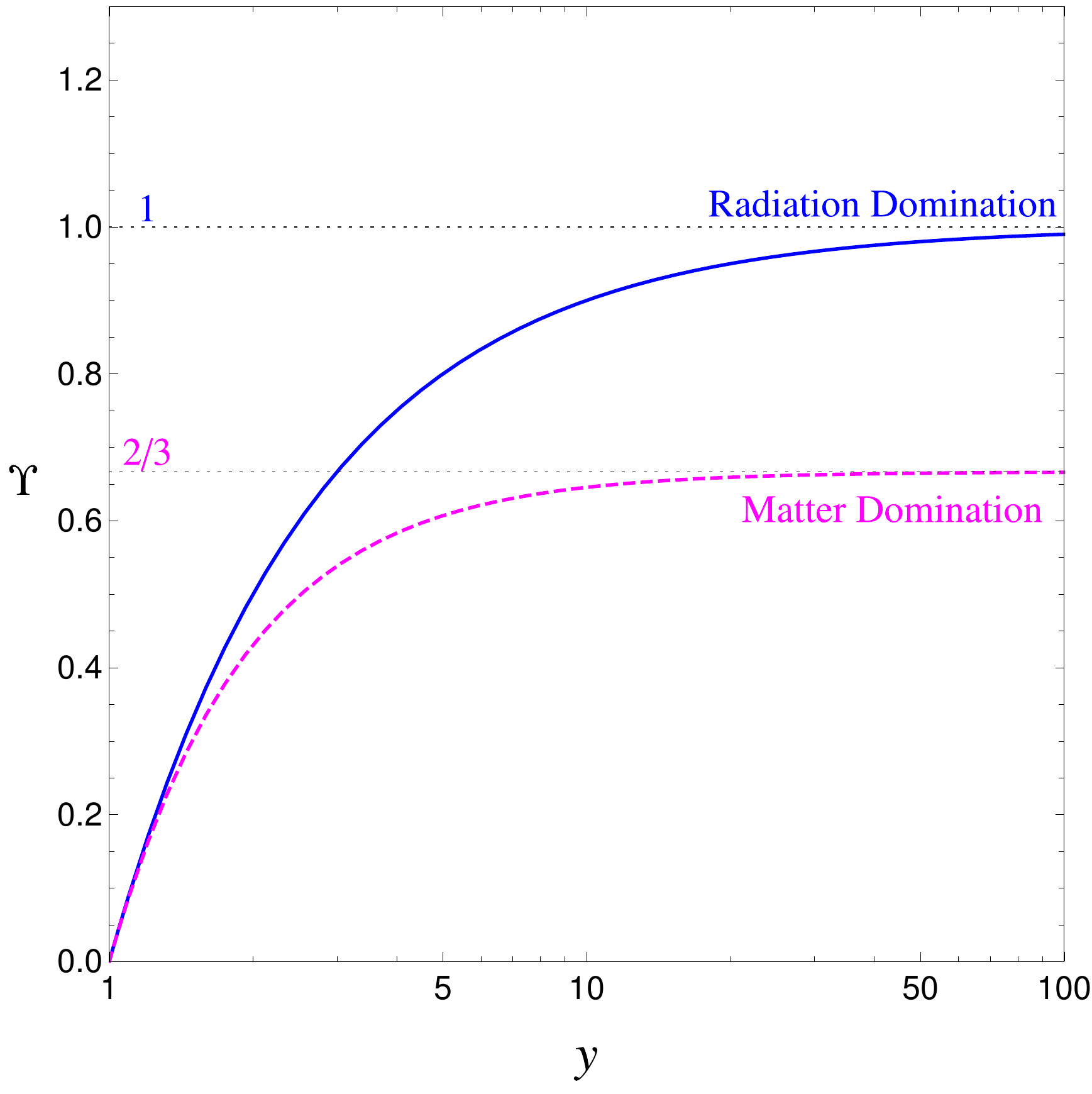}
\caption{
\label{fig:upsilon}
The function $\Upsilon$ for radiation domination(blue solid) and matter domination(magenta dashed).
}
\end{figure}

Now lets look in more detail at the integrand. For RD and MD, the factor containing Green's function can be written as
\begin{eqnarray}
\frac{\partial G(\tilde{y}, \tilde{y}_1)}{\partial \tilde{y}}
\frac{\partial G(\tilde{y}, \tilde{y}_2)}{\partial \tilde{y}} 
=
\left\{
\begin{array}{ll}
\frac{1}{\tilde{y}^2} \left[ c_0^R \tilde{y}^0 + c_{-1}^R \frac{1}{\tilde{y}} + \cdots ]\right] \\
\frac{1}{\tilde{y}^3} \left[ c_0^M \tilde{y}^0 + c_{-1}^M \frac{1}{\tilde{y}} + \cdots ]\right] 
\end{array}
\right.
\equiv
\left\{
\begin{array}{ll}
\frac{1}{\tilde{y}^2} \\
\frac{1}{\tilde{y}^3} 
\end{array}
\right\}
\times
\mathcal{G}_2(\tilde{y}, \tilde{y}_1, \tilde{y}_2) .
\end{eqnarray}
Then 
\begin{eqnarray}
&&\mathcal{P}_{\text{GW}}(y, k R_{\ast c})
= 
\frac{[16 \pi G \left( \bar{\tilde{\epsilon}} + \bar{\tilde{p}} \right) \bar{U}_f^2]^2}{24\pi^2 H^2 H_s^2} \frac{1}{y^4}
(k R_{\ast c})^3
\nonumber \\
&&  \hspace{1.5cm} \times \int d y_- 
\tilde{\Pi}^2\left(k R_{\ast c}, \beta_c |\eta_1 - \eta_2| \right)
\left[
\int d y_+ 
\frac{
\mathcal{G}_2(\tilde{y}, \tilde{y}_1, \tilde{y}_2)}{\tilde{\tilde{k}}^2}
\left\{
\begin{array}{c}
y_1^{-2} y_2^{-2} \\
y_1^{-{3/2}} y_2^{-3/2} \\
\end{array}
\right\} 
\right] .
\label{eq:PgwyFactorized}
\end{eqnarray}
In the square bracket, $y_{1,2}$ are understood to be functions of $y_{\pm}$ (note that 
$\tilde{y}$ is defined differently for matter and radiation cases). The reason we associate a factor of $\tilde{\tilde{k}}^{-2}$
with $\mathcal{G}_2$ is that $\mathcal{G}_2 \propto \tilde{\tilde{k}}^2$ to a good approximation. 
For both RD and MD, the integral over $y_+$ leads to a result in the following form:
\begin{eqnarray}
\left[\int d y_+ \cdots \right]
=
\frac{1}{2} 
\Upsilon(y) \cos \left(\tilde{\tilde{k}} y_-\right).
\end{eqnarray}
The profile in a wide range of $y$ is shown in Fig.~\ref{fig:upsilon}. We can see $\Upsilon$ of RD is slightly larger than MD.
For both cases, $\Upsilon$ approaches an asymptotic value: $1$ for RD and $2/3$ for MD, irrespective of how long the source lasts.
This is due to the dilution of the source over time, which makes the contribution from later time increasingly suppressed. 
To have a better understanding of the behavior of $\Upsilon(y)$, lets see how they can be obtained in a simpler analytical way.

%%%%%%%%%%%%%%%%%%%%%%%%%%%%%%%%%%%%%%%%%%%%%%%%%%%
\begin{figure}
\centering
\includegraphics[width=0.45\textwidth]{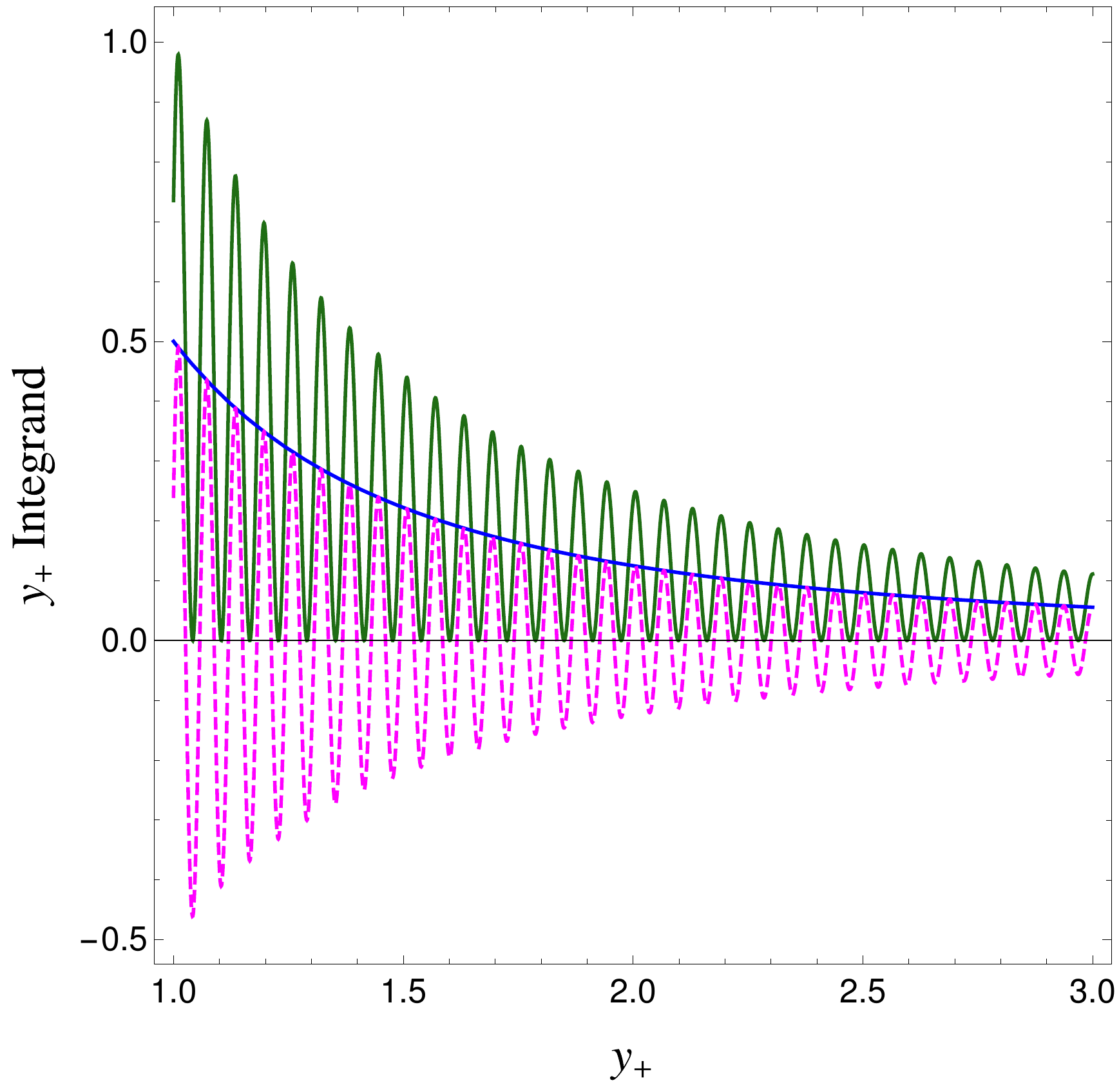}
\includegraphics[width=0.45\textwidth]{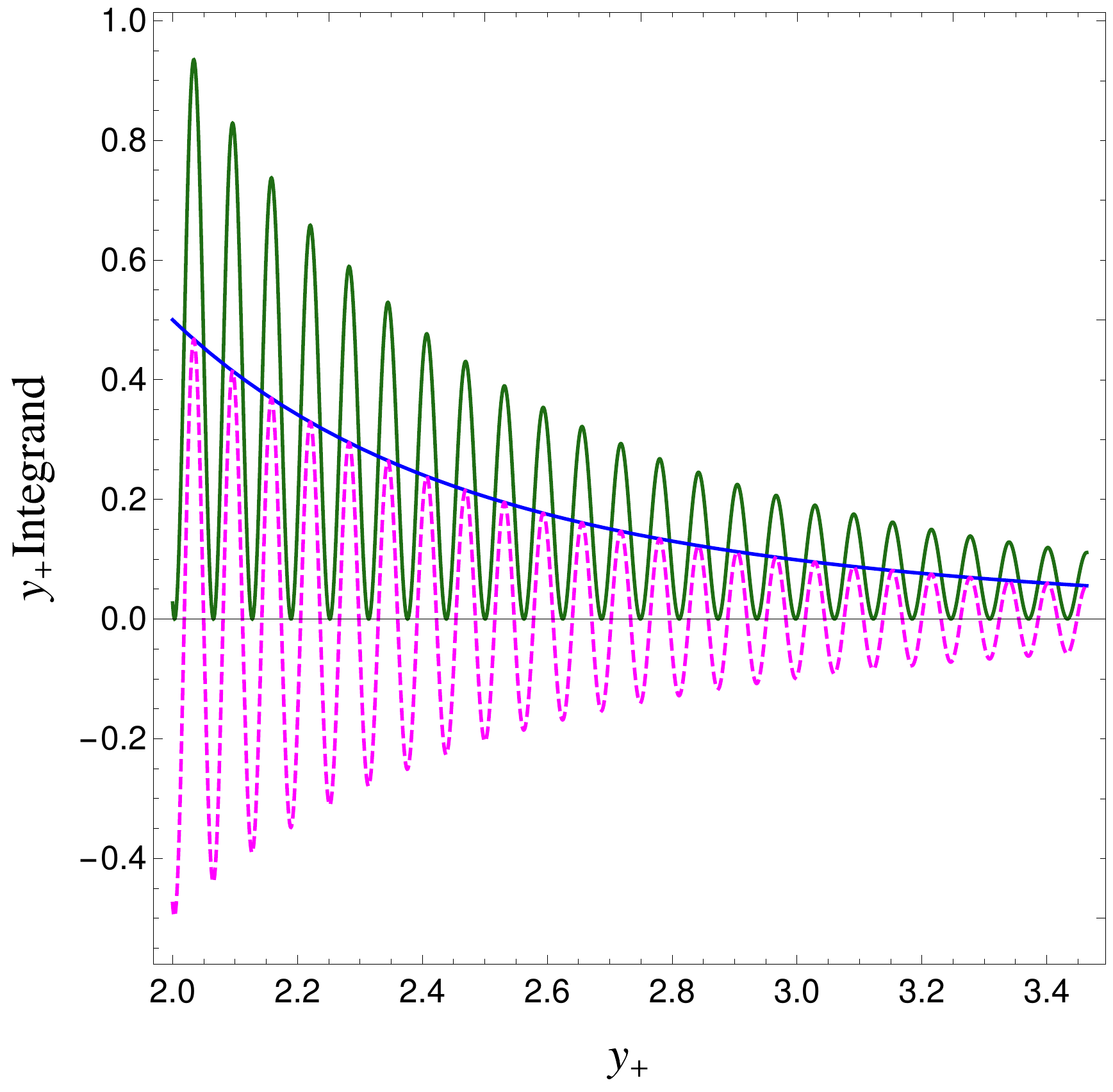}
\caption{
\label{fig:G2}
The integrand of $y_+$ integration, with $y=3$. Left is RD and right is MD. 
The blue is the dominant non-oscillatory part, the magenta dashed is the oscillatory part($k R_{\ast c}$ chosen to be $0.04$) and 
the dark green is the total contribution.
}
\end{figure}
%%%%%%%%%%%%%%%%%%%%%%%%%%%%%%%%%%%%%%%%%%%%%%%%%%%

First for RD, neglecting terms suppressed by $(R_{\ast c} a_s H_s)$ or $y^{-1}$, the dominant contributions to the integrand of the power spectrum are
\begin{eqnarray}
\mathcal{G}_2^{\text{RD}} = \frac{1}{2 y_+^2} \left\{\cos\left[\tilde{\tilde{k}} y_-\right] + \cos\left[2\tilde{\tilde{k}} (y-y_+)\right] \right\} + \cdots .
\end{eqnarray}
The second term is $y_-$ independent and is a highly oscillatory function of $y_+$, 
which averages to zero during the integration over $y_+$ (see Fig.~\ref{fig:G2} for the non-oscillatory and oscillatory contributions). 
On the other hand, the first term, a function of $y_-$, when integrated, gives the dominant contribution:
\begin{eqnarray}
\Upsilon_{\text{RD}} = 1 - \frac{1}{y} .
\label{eq:UpsilonRD}
\end{eqnarray}
For $y \gg 1$, it approaches an asymptotic value of $1$. Since this asymptotic value can only be reached for a long enough source, 
a realistic phase transition might not satisfy this. We will come to this point later.

Similarly for MD, we can perform analogous manipulations and keep only the leading order and also non-oscillatory term: 
\begin{eqnarray}
\mathcal{G}_2^{\text{MD}} = \frac{8}{y_+^4} \cos\left[\tilde{\tilde{k}} y_-\right] + \cdots .
\end{eqnarray}
Upon integration, it gives the dominant contribution:
\begin{eqnarray}
\Upsilon_{\text{MD}} = \frac{2}{3} \left(1 - \frac{1}{y^{3/2}}\right) .
\label{eq:UpsilonMD}
\end{eqnarray}
For $y \gg 1$, it approaches the previously observed asymptotic value of $2/3$. Thus barring other differences for RD and MD, 
the different expansion behaviors lead to a suppression of gravitational wave spectrum for MD, when compared with RD.
%%%%%%%%%%%%%%%%%%%%%%%%%%%%%%%%%%%%%%%%%%%%%%%%%%%
%%%%%%%%%%%%%%%%%%%%%%%%%%%%%%%%%%%%%%%%%%%%%%%%%%%

With $\Upsilon(y)$ obtained, the power spectrum as a function of $y$ can be written in the following form
\begin{eqnarray}
&&\mathcal{P}_{\text{GW}}(y, k R_{\ast c})
= 
\frac{[16 \pi G \left( \bar{\tilde{\epsilon}} + \bar{\tilde{p}} \right) \bar{U}_f^2]^2}{48 \pi^2 H^2 H_s^2} \frac{1}{y^4}
(k R_{\ast c})^3
\nonumber \\
&&  \hspace{1.5cm} \times 
\left[
\int d y_- 
\cos\left(\tilde{\tilde{k}} y_-\right)
\tilde{\Pi}^2\left(k R_{\ast c}, \beta_c |\eta_1 - \eta_2| \right)
\right]
\times \Upsilon(y) .
\end{eqnarray}
Here note that using Eq.~\ref{eq:etay}, we have $\tilde{\tilde{k}} y_- = k (\eta_1 - \eta_2)$. The integral over $y_-$ is obtained
by plugging the explicit expression of $\tilde{\Pi}$, which results in a three-fold integral. The integration of $y_-$ over the three 
trigonometric functions result in a $\delta$ function, and makes the angle integration of $\tilde{q}$ in Eq.~\ref{eq:PiExplicit} trivial. 
We are left eventually with a one fold integral over the magnitude of $\tilde{q}$, and the spectrum can be put in the following standard form:
\begin{eqnarray}
&&\mathcal{P}_{\text{GW}}(y, k R_{\ast c})
= 
3 \Gamma^2\ \bar{U}_f^4 
\frac{H_{R,s}^4}{H^2 H_s} 
(a_s R_{\ast c}) \frac{(k R_{\ast c})^3}{2 \pi^2} \tilde{P}_{\text{gw}}(k R_{\ast}) 
\times 
\frac{1}{y^4}
\Upsilon(y) 
,
\end{eqnarray}
where $\Gamma = \bar{\tilde{w}}/\bar{\tilde{e}} \approx 4/3$, $H_{R,s}$ is defined to contain only the radiation energy density at $t_s$: $H_{R,s} = H_{s} \sqrt{1-\kappa_M}$, and
the integral is hidden inside $\tilde{P}_{\text{gw}}(k R_{\ast})$:
\begin{eqnarray}
\widetilde{P}_{\text{GW}}(k R_{\ast}) = \frac{1}{4 \pi c_s k R_{\ast}} \left(\frac{1-c_s^2}{c_s^2}\right)^2
\int_{z_-}^{z_+} \frac{dz}{z} \frac{(z-z_+)^2(z-z_-)^2}{z_+ + z_--z} \bar{P}_v(z) \bar{P}_v(z_+ + z_- -z) . \quad \quad
\end{eqnarray}
Here $z = q R_{\ast c}$, $z_{\pm} = \frac{1}{2} \frac{k R_{\ast c}}{c_s} (1\pm c_s)$ and 
$\bar{P}_v(z) = \frac{\pi^2}{\bar{U}_f^2} \frac{\mathcal{P}_v(z)}{z^3}$. Using Eq.~\ref{eq:calPv}, the explicit 
expression for $\bar{P}_v$ is 
\begin{eqnarray}
\bar{P}_v(z) = 
\frac{1}{64 \pi^2 v_w^6} \frac{1}{\bar{U}_f^2} 
\int d \tilde{T}\ \tilde{T}^6 \nu(\tilde{T}) \left|A\left(\frac{z \tilde{T}}{(8\pi)^{1/3} v_w}\right)\right|^2 .
\end{eqnarray}
%%%%%%%%%%%%%%%%%%%%%%%%%%%%%%%%%%%%%%%%%%%%%%%%%%%%%%%%%%%
\begin{figure}[t]
\centering
\includegraphics[width=0.7\textwidth]{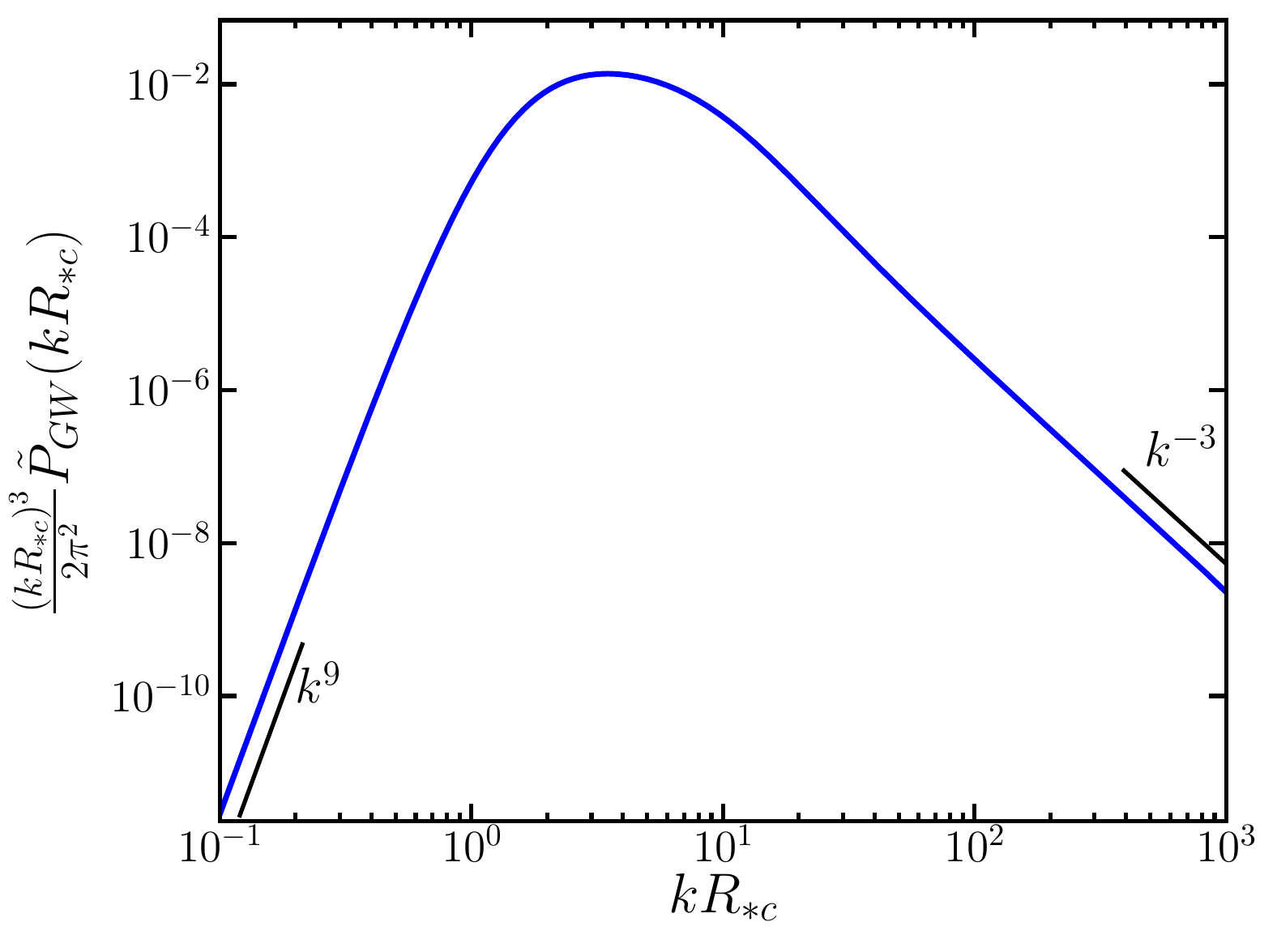}
\caption{
\label{fig:Pgw}
The dimensionless gravitational wave power spectrum computed in the sound shell model. The calculation was performed for a weak phase transition with $\alpha = 0.0046$, $v_w = 0.92$, and exponential bubble nucleation. 
The low and high frequency regimes follow the $k^9$ and $k^{-1}$ power law fits respectively (black solid lines). 
See Ref.~\cite{Hindmarsh:2019phv} for more details of its properties.
}
\end{figure}
%%%%%%%%%%%%%%%%%%%%%%%%%%%%%%%%%%%%%%%%%%%%%%%%%%%%%%%%%%%
Plugging in the explicit expressions of $H$ and $H_{R,s}$, we have
\begin{eqnarray}
&&\mathcal{P}_{\text{GW}}(y, k R_{\ast c})
= 
3 \Gamma^2\ \bar{U}_f^4 
 (H_s a_s R_{\ast c}) \frac{(k R_{\ast c})^3}{2 \pi^2} \tilde{P}_{\text{gw}}(k R_{\ast}) 
\times
\left\{
\begin{array}{c}
1 \\
\frac{(1-\kappa_M)^2}{\kappa_M y + 1 - \kappa_M}
\end{array}
\right\}
\times \Upsilon(y) .  \quad \quad \quad
\label{eq:Pgwfinal}
\end{eqnarray}
For both RD and MD, the shape of the spectra are the same to a good approximation, and are the same as
that derived in the sound shell model and thus the properties of its shape~\cite{Hindmarsh:2019phv} apply here for both cases. 
In particular, the peak frequency of the spectrum is located at around $k R_{\ast c} \approx 10$. This mean a larger
or smaller $R_{\ast}$ can red or blue shift the spectrum respectively. 
For example, as shown in Fig.~\ref{fig:Rb}, increasing $v_w$ reduces $R_{\ast c}$ and thus blue-shift the spectrum.
For MD, it has a larger $R_{\ast}$ and thus red-shift the spectrum.

For RD, we recover the result found in Ref.~\cite{Hindmarsh:2015qta}, as long as $\Upsilon(y) = 1$, which is only true for $y \gg 1$. 
The reason only this asymptotic value is obtained in Ref.~\cite{Hindmarsh:2015qta} is due to the over-simplifying assumptions used (see Appendix~\ref{sec:RDeta}), in
which case the second terms in both Eq.~\ref{eq:UpsilonRD} and ~\ref{eq:UpsilonMD} are missing. Whether or not the asymptotic values 
can be reached depends on how long the source remains active, and we continue in the next section on this question.

%\noindent\blue{To express the peak frequency in terms of $\beta$, we need the relation between $\beta$ and $R_{\ast}$. The relation
%$R_{\ast} \beta = (8\pi)^{1/3} v_w$ probably wont hold in an expanding universe. 
%But we can express the peak frequency in terms of $H_n R_{\ast}$ as done by Ref.~\cite{Hindmarsh:2015qta}(Eq.43).}

\subsection{Lifetime of the Source}

%%%%%%%%%%%%%%%%%%%%%%%%%%%%%%%%%%%%%%%%%%%%%%%%%%%%%%%%%%%%%%%%%%%%%
\begin{figure}[t]
\centering
\includegraphics[width=0.6\textwidth]{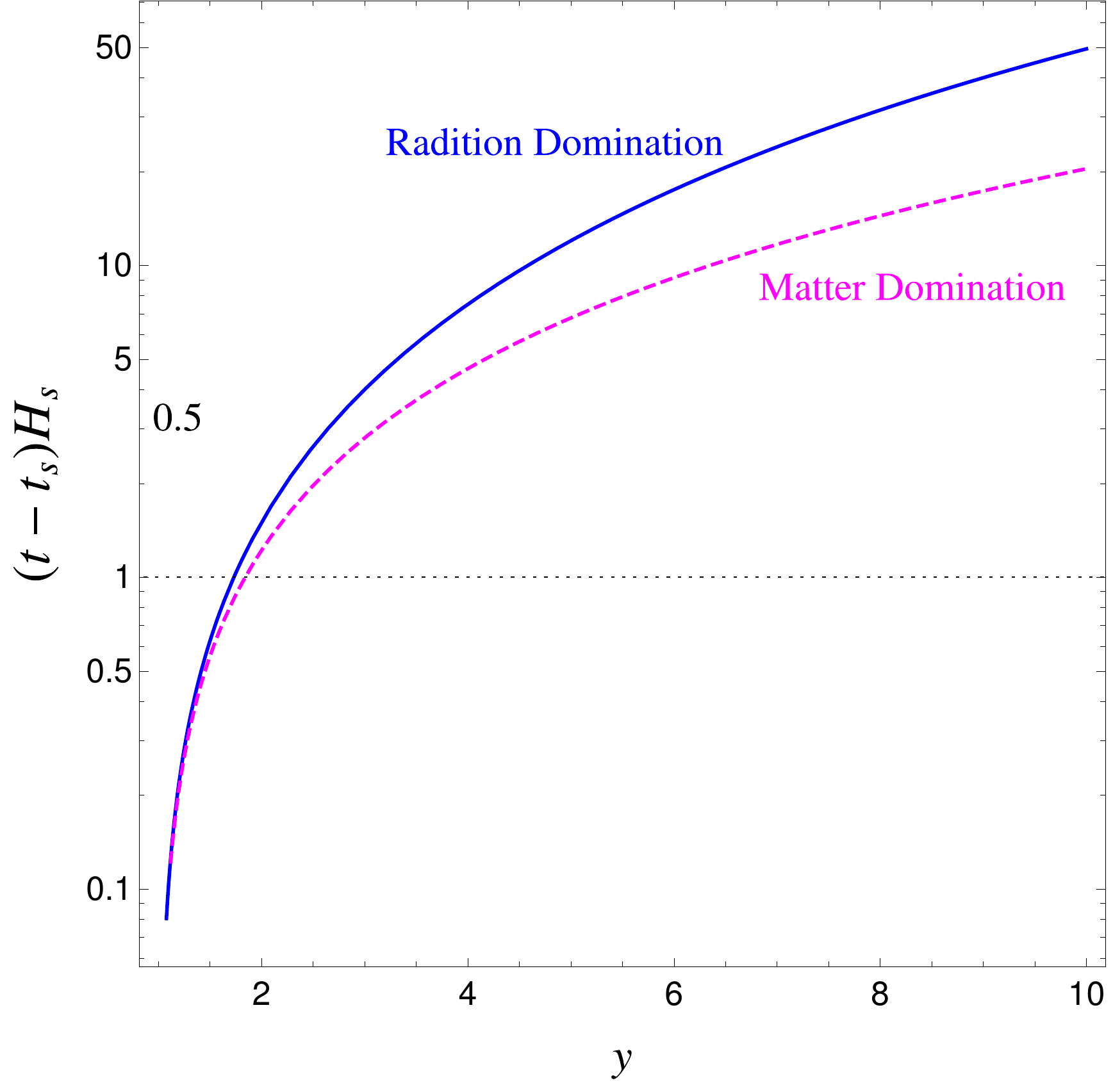}
\caption{\label{fig:srclifetime}
Time elapsed since $t_s$ in unit of Hubble time $H_{s}^{-1}$ at $t_{\ast}$.
}
\end{figure}
%%%%%%%%%%%%%%%%%%%%%%%%%%%%%%%%%%%%%%%%%%%%%%%%%%%%%%%%%%%%%%%%%%%%%
As we saw earlier, the presence of an asymptotic value for $\Upsilon$ for large $y$ in both cases is due to the dilution of the source energy density.
This asymptotic value was used in Ref.~\cite{Hindmarsh:2015qta} to reach the conclusion that for RD the effective lifetime of the source is a Hubble time $H_s^{-1}$
for RD, i.e., $\tau_{\text{sw}} = 1/H_s$, which as we have seen is only true if $\Upsilon=1$ for $y \gg 1$. 
The question is, however, whether this asymptotic value can be reached in a realistic time frame. In Fig.~\ref{fig:srclifetime}, we show the time elapsed since the reference time $t_s$, in unit of the Hubble time $H_s^{-1}$.
For RD, 
\begin{eqnarray}
\frac{t - t_s}{1/H_s} = \frac{y^2-1}{2} ,
\end{eqnarray}
and for MD
\begin{eqnarray}
\frac{t - t_s}{1/H_s} = \frac{2}{3}(y^{3/2}-1) .
\end{eqnarray}
At about a Hubble time, $\Upsilon \approx 0.4$ for both RD and MD, which is less than a half of the asymptotic value for RD and $60\%$ for MD.
We need many Hubble times for $\Upsilon$ to approach the asymptotic value. The problem is certain physical processes might prohibit the sound waves from
being active for such a long time, and thus the asymptotic value might never be reached.
One such process is the possible formation of shocks and turbulence. Another is the existence of possible dissipative processes, whose presence damps the sound waves. 
If either of these processes quenches the sound waves, the asymptotic value will not be achieved. In this case, the effective lifetime is shorter 
than the Hubble time for RD, and the result obtained with an effective lifetime of a Hubble time overestimates the gravitational wave production.
The time scale for turbulence is roughly~\cite{Pen:2015qta,Hindmarsh:2017gnf}
\begin{eqnarray}
\tau_{\text{sw}} \sim \frac{L_f}{\bar{U}_f} \sim \frac{R_{\ast}}{\bar{U}_f} .
\end{eqnarray}
Therefore 
\begin{eqnarray}
\frac{\tau_{\text{sw}}}{1/H_s} \sim \frac{H_s R_{\ast}}{\bar{U}_f} .
\end{eqnarray}
As we have seen in Fig.~\ref{fig:Rb}, $H_s R_s \sim 10^{-3}$ and different expansion histories lead to larger or smaller
values. To delay the appearance of turbulence and thus approach the asymptotic value of $\Upsilon$ thus requires smaller fluid velocity
$\bar{U}_f$ or larger bubble separation. While $H_s R_{\ast}$ depends on specific expansion behavior adopted, the value of $\bar{U}_f$ is 
more or less universal, and its value is shown in Fig.~\ref{fig:uf} on the plane of $(v_w, \alpha)$. 
We show here two versions of it obtained using two different methods: one by solving the velocity profile around a single bubble and the other by 
integrating over the velocity power spectrum (see Ref.~\cite{Hindmarsh:2019phv} for details).
Thus whether or not above ratio becomes large enough depends on the details of the 
phase transition in a given cosmological context. Even in cases where the turbulence is delayed or not present, i.e., for sufficiently 
strong or weak phase transitions respectively, the damping of the sound waves caused by some weak processes could still shorten the lifetime in the form of 
shear viscosity~\cite{Hindmarsh:2015qta}. It seem unlikely for any scenario to be very close to the asymptotic value.

\subsection{Spectrum Today}
%%%%%%%%%%%%%%%%%%%%%%%%%%%%%%%%%%%%%%%%%%%%%%%%%%%%%%%%%%%%%%%%%%%%%
\begin{figure}[t]
\centering
\includegraphics[width=0.49\textwidth]{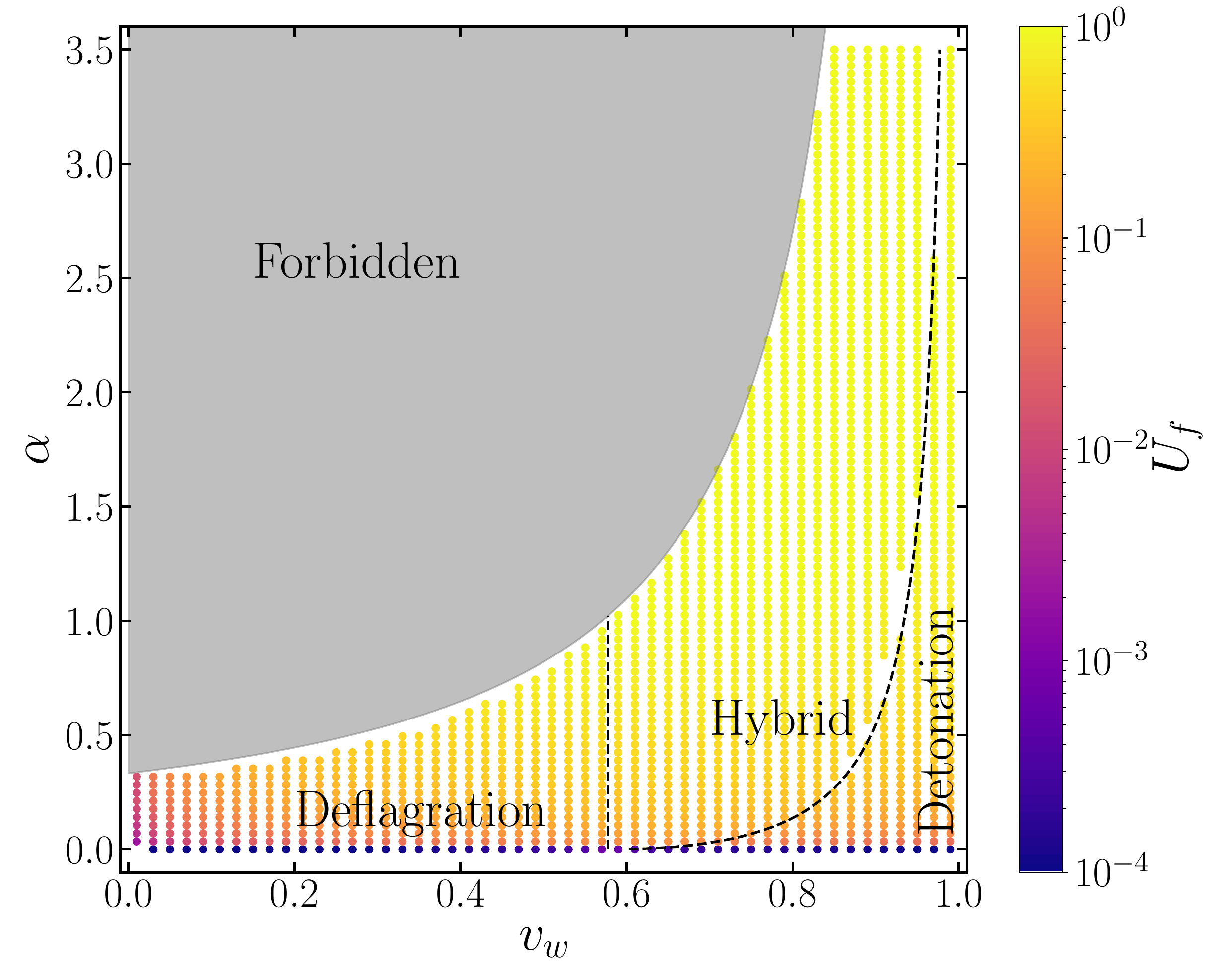}
\includegraphics[width=0.49\textwidth]{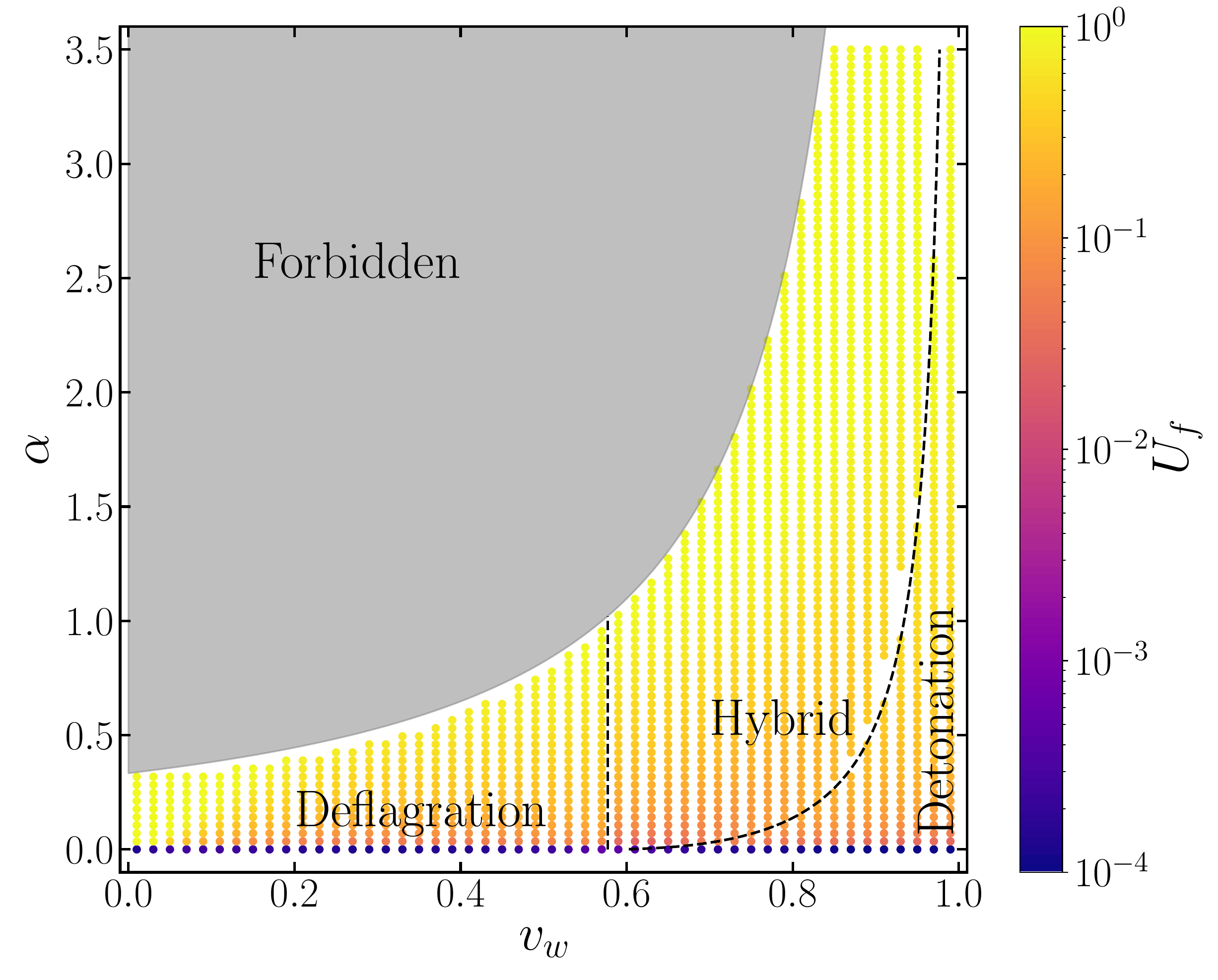}
\caption{
\label{fig:uf}
$\bar{U}_f$ on the plane of $(v_w, \alpha)$. The left figure is $\bar{U}_f$ of the fluid around a single bubble.  The right figure is $\bar{U}_f$ of the fluid calculated from the velocity power spectrum. 
}
\end{figure}
%%%%%%%%%%%%%%%%%%%%%%%%%%%%%%%%%%%%%%%%%%%%%%%%%%%%%%%%%%%%%%%%%%%%%

We will mainly consider the case of RD as it is the most frequently encountered scenario. 
Denote the temperature after the gravitational wave production as $T_e$ with the scale factor being $a_e$. 
The amount of redshifting is described by the scale factor ratio $a_e/a_0$. For radiation in thermal equilibrium
and in adiabatic expansion, the relation between $a_e$ and $a_0$ is governed by entropy conservation:
\begin{eqnarray}
g_{s}(T_e) a_e^3 T_e^3 = g_s(T_{\gamma 0}) a_0^3 T_{\gamma 0}^3 , 
\end{eqnarray}
where $g_s$ is the relativistic degrees of freedom for entropy; $T_{\gamma 0}$ is the temperature of the CMB photon with $T_{\gamma 0} \approx 2.73 K$.
At the present time, the relativistic species includes photons and decoupled neutrinos, thus $g_s = 2 + \frac{7}{8} \times 2 N_{\text{eff}} (\frac{4}{11})^{3/3} \approx 3.94$ for $N_{\text{eff}} = 3.046$. Using
these, the ratio of the scale factor can be put into the following form:
\begin{eqnarray}
\frac{a_e}{a_0} = 1.65 \times 10^{-5} \left(\frac{g_{s}(T_e)}{100}\right)^{1/6} \left(\frac{T_e}{100 \text{GeV}}\right) \left(\frac{1 \text{Hz}}{H_e}\right) .
\end{eqnarray}
For the peak frequency at $k R_{\ast} = z_p$~\footnote{
We use a notation where $k$ in $k R_{\ast}$ is physical wavenumber, and $k$ in $k R_{\ast c}$ is a comoving wavenumber.
} 
where $z_p \approx 10$~\cite{Hindmarsh:2017gnf}, the frequency at $t_e$ is
\begin{eqnarray}
f_p = \frac{z_p}{2 \pi R_{\ast}(t_e)} ,
\end{eqnarray}
where $R_{\ast}(t_e)$ is evaluated at the end of the gravitational wave production and note all previously generated gravitational waves at higer frequencies
at $k R_{\ast c} = z_p$ have all redshifted to the frequency produced at $t_e$.
Then the corresponding frequency today is 
\begin{eqnarray}
f_{\text{SW}} = 2.65 \times 10^{-5} \text{Hz}
\left(\frac{g_{s}(T_e)}{100}\right)^{1/6} \left(\frac{T_e}{100 \text{GeV}}\right) \left(\frac{z_p}{10}\right) \left(\frac{1}{H_e R_{\ast}(t_e)}\right) .
\end{eqnarray}
We can express $R_{\ast}$ by $\beta(v_w)$ using Eq.~\ref{eq:Rsfinal}, so that,
\begin{eqnarray}
\frac{1}{H_e R_{\ast}(t_e)} = 
(8\pi)^{-1/3} \frac{a(t_f)}{a(t_e)} \frac{1}{v_w}\frac{\beta(v_w)}{H_e} 
=
(8\pi)^{-1/3} \frac{1}{v_w}\frac{\beta(v_w)}{H_e} \times \frac{1}{y}.
\label{eq:HRe}
\end{eqnarray}
Here we neglect the very small difference between $t_f$, the time when all the bubbles have disappeared and $t_s$,
and we have shown explicitly the dependence of $\beta$ on $v_w$. 
Also note $\beta$ is evaluated at $t_f$ when $I(t_f) = 1$. 
The factor $y^{-1}$ is significant when the lifetime of the source is long.
Then the present peak frequency becomes
\begin{eqnarray}
f_{\text{SW}} = 8.97 \times 10^{-6} \text{Hz} \frac{1}{v_w}
\left(\frac{g_{s}(T_e)}{100}\right)^{1/6} \left(\frac{T_e}{100 \text{GeV}}\right) \left(\frac{z_p}{10}\right) \left[\frac{\beta(v_w)/y}{H_e}\right] .
\label{eq:fp0}
\end{eqnarray}
%%%%%%%%%%%%%%%%%%%%%%%%%%%%%%%%%%%%%%%%%%%%%%%%%%%%%%%%%%%%%%%%%%%%%
\begin{figure}
\centering
\includegraphics[width=0.9\textwidth]{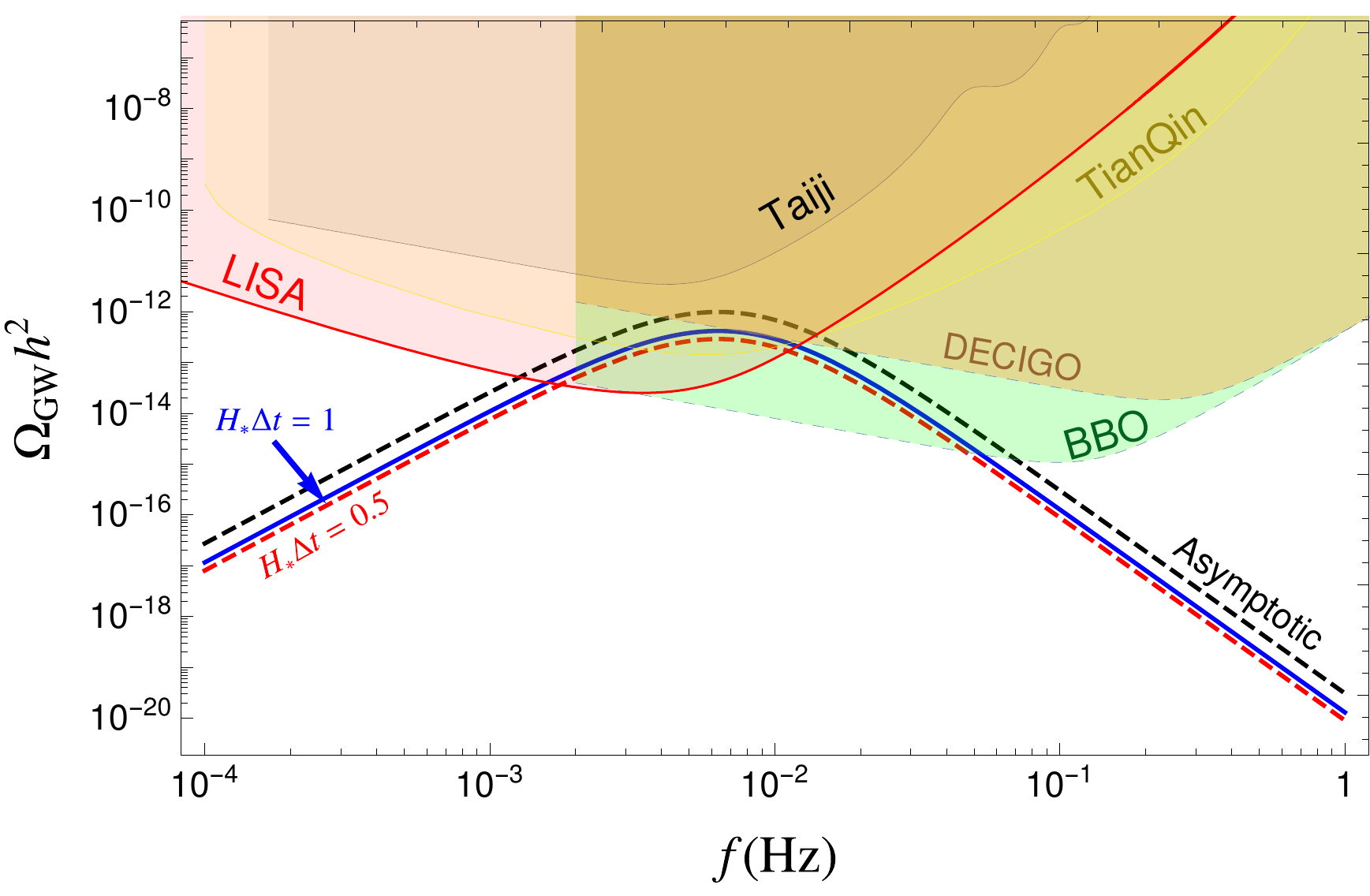}
\caption{
\label{fig:gwtoday}
The present day gravitational wave energy density spectra for $H_{\ast} \Delta t = 0.5, 1$ and for $H_{\ast} \Delta t \gg 1$ when it takes the asymptotic form. Here $\Delta t = t-t_s$ and is the time elapsed since $t_s$, the time when the source becomes active.
In all three cases, $v_w=0.3$, $\alpha = 0.1$, $T_e = 100\text{GeV}$ and $\beta/(y H_{\ast}) = 100$.
The shaded regions at the top are experimental sensitive regions for several proposed space-based detectors.
}
\end{figure}
%%%%%%%%%%%%%%%%%%%%%%%%%%%%%%%%%%%%%%%%%%%%%%%%%%%%%%%%%%%%%%%%%%%%%
For the energy fraction of gravitational waves, the dilution of gravitational waves leads to the following connection:
\begin{eqnarray}
h^2 \Omega_{\text{GW}}(t_0, f) &=& h^2 \left(\frac{a_e}{a_0}\right)^4 \left(\frac{H_e}{H_0}\right)^2  \Omega_{\text{GW}}(t_e, a_0 f/a_e) , \nonumber 
\\
&=& 1.66 \times 10^{-5} \left(\frac{100}{g_{s}(T_e)}\right)^{1/3} \Omega_{\text{GW}}(t_e, a_0 f/a_e) .
\end{eqnarray}
Here $h \approx 0.673$, the Hubble parameter today in unit of $100 \text{km}/s/\text{Mpc}$. 
Then plugging the explicit expression for $\mathcal{P}_{\text{GW}}$ in Eq.~\ref{eq:Pgwfinal}, we have
\begin{eqnarray}
h^2 \Omega_{\text{GW}}(f) = 4.98 \times 10^{-5} \left(\frac{100}{g_{s}(T_e)}\right)^{1/3} \Gamma^2 \bar{U}_f^4 \left[H_s R_{\ast}(t_s)\right] 
A\ \mathcal{S}_{\text{SW}}(f)
\Upsilon(y) .
\end{eqnarray}
Here we have defined $A \mathcal{S}_{\text{SW}}(f)$ to be $(k R_{\ast})^3 \tilde{P}_{\text{gw}}(k R_{\ast})/(2\pi^2)$ with appropriate redshifting factors included.
One can either use the prediction from the sound shell model to determine $A \mathcal{S}_{\text{SW}}(f)$, 
or use result from numerical simulations~\cite{Hindmarsh:2017gnf}. We choose the latter as it should give a more accurate result, in which case
$A \approx 0.058$ and~\cite{Weir:2017wfa}
\begin{eqnarray}
\mathcal{S}_{\text{SW}}(f) = \left(\frac{f}{f_{\text{SW}}}\right)^3 \left[\frac{7}{4 + 3 (f/f_{\text{SW}})^2}\right]^{7/2}.
\end{eqnarray}
For the term $H_s R_{\ast}(t_s)$, similar to Eq.~\ref{eq:HRe}, we can write 
\begin{eqnarray}
H_s R_{\ast}(t_s) = (8\pi)^{1/3} v_w \frac{H_s}{\beta(v_w)} .
\end{eqnarray}
Therefore the final spectrum is~\footnote{
Note current simulations only probe relatively weak transitions and this spectrum might not be applicable for strong transitions $\alpha \sim 1$.
As shown in a recent simulation~\cite{Cutting:2019zws}, a deficit in the gravitational wave production has been found for such strong transitions.
This reduction is more severe for small $v_w$, and of course a large $\alpha$, and would require extremely strong couplings to the plasma
which might be a rare case. We also note that a large $\alpha$, such as the region when $\alpha > 1$ in Fig.~\ref{fig:uf}, might leads to a temporary 
inflationary stage with exponential expansion (see e.g.,~\cite{Ellis:2018mja}) and contradicts the assumed radiation domination for this spectrum. 
In this case, one should use the corresponding Green's function and follow previous steps in deriving this spectrum.
}
\begin{eqnarray}
h^2 \Omega_{\text{GW}}(f) =  8.5 \times 10^{-6} \left(\frac{100}{g_{s}(T_e)}\right)^{1/3} \Gamma^2 \bar{U}_f^4  
\left[
\frac{H_s}{\beta(v_w)}
\right]
v_w \mathcal{S}_{\text{SW}}(f)
\times
\Upsilon(y) .
\label{eq:OmegaFinal}
\end{eqnarray}
For a long lifetime of the source, the main changes are the suppression factor $\Upsilon(y)$. 
In Fig.~\ref{fig:gwtoday}, we show the spectra for several choices of $H_{\ast} \Delta t$, with $z_p=10$ (see caption for more details).

For MD, apparently the extra dominant matter content will decay to radiation at some time 
later, which will inject entropy to the standard radiation sector. This can be studied using two methods. In the first method, 
one can assume a very quick and thus instantaneous decay of the matter, which then allows to use energy conservation to get the
new heated radiation temperature. In the second method, a more precise account of the matter decay is provided, with the conclusion
that there is no heating up of the radiation but one gets a slower cooling of the radiation, as was firstly pointed out in Ref.~\cite{Scherrer:1984fd}. 
Therefore one needs to follow more closely the entropy evolution by taking into account finite matter decay width, following the procedure 
of Ref.~\cite{Scherrer:1984fd} or a more closely related example studied in Ref.~\cite{Bernal:2019lpc}. This however introduces extra
model dependent varieties and is beyond the scope of this work.

\section{Summary\label{sec:summary}}

We studied in detail the cosmological first order phase transition and the calculation of resulting stochastic gravitational waves
in an expanding universe, with radiation and matter dominated universe as two representative examples. Firstly we studied the changes to 
process of bubble formation and collision, including important observables such as the mean bubble separation and its relation with $\beta$.
We also derived the unbroken bubble wall area, the bubble conformal lifetime distribution which are needed for the calculation of the 
gravitational wave spectrum. We then derived the full set of differential equations as used in numerical simulations in an expanding universe.
We found that simple rescalings work such that the equations governing the velocity profile around a single bubble maintains the same form
as in Minkowski spacetime in the bag model and that the velocity profile remains the same when appropriate substitution of variables are used. We then 
generalized the sound shell model to the expanding universe and derived the velocity power spectrum.  This result is used to derive
analytically the gravitational wave power spectrum from the sound waves, the dominant source. We found that the standard formula of the 
spectrum needs to include an additional suppression factor $\Upsilon$, which is a function of the lifetime of the source. 
For radiation domination, the asymptotic value of $\Upsilon$ is $1$ when the lifetime of the source is very long, and corresponds to the 
usually adopted spectrum in the literature. This asymptotic value however can not be reached as the onset of shocks and turbulence may
disrupt the sound waves and possible dissipative processes may further damp it. Therefore an additional suppression factor needs to be 
taken into account when using the gravitational wave spectrum from sound waves and we provided simple analytical expression for $\Upsilon$.

\section{Acknowledgments}

HG and KS are supported by DOE Grant desc0009956. TRIUMF receives federal funding via a contribution agreement with the National Research Council of Canada. 
We thank Mark Hindmarsh for helpful communications. We acknowledge Elizabeth Loggia for her early involvement in this project. May our field become more accommodating to mothers. 

\appendix

\section{\label{sec:example}The Example Effective Potential}

Here we provide details of the example effective potential used in Sec.~\ref{sec:dynamics}, so that those results can be reproduced more easily. 
The effective potential was originally used as a high temperature approximation for the standard model (see, e.g., Ref.~\cite{Dine:1992wr}), given by

\begin{eqnarray}
V(\phi, T) = D (T^2 - T_0^2) \phi^2 -E T \phi^3 + \frac{\lambda}{4} \phi^4 .
\end{eqnarray}
Here $D >0$, $E>0$, $\lambda>0$ and $\lambda$ has a weak dependence on $T$. The first term has a positive coefficient 
when $T> T_0$ to restore the symmetry. The third, the cubic term, when is sufficiently smaller, helps create
a barrier together with the first term, and creates another minimum.
Since this example is only used to provide a simple benchmark 
effective potential to show the effects of the expansion of the universe, we will take these parameters to be $T$
independent. It should be noted that an effective potential of this form can characterize features of a wide class of beyond the standard 
model scenarios in the high temperature approximation. We will use this effective potential to calculate bounce 
solutions and corresponding parameters relevant for the phase transition. 

%%%%%%%%%%%%%%%%%%%%%%%%%%%%%%%%%%%%%%%%%%%%%%%%%%%%%%%%%%%%%%%%%%%%%
\begin{figure}
\centering
\includegraphics[width=0.5\textwidth]{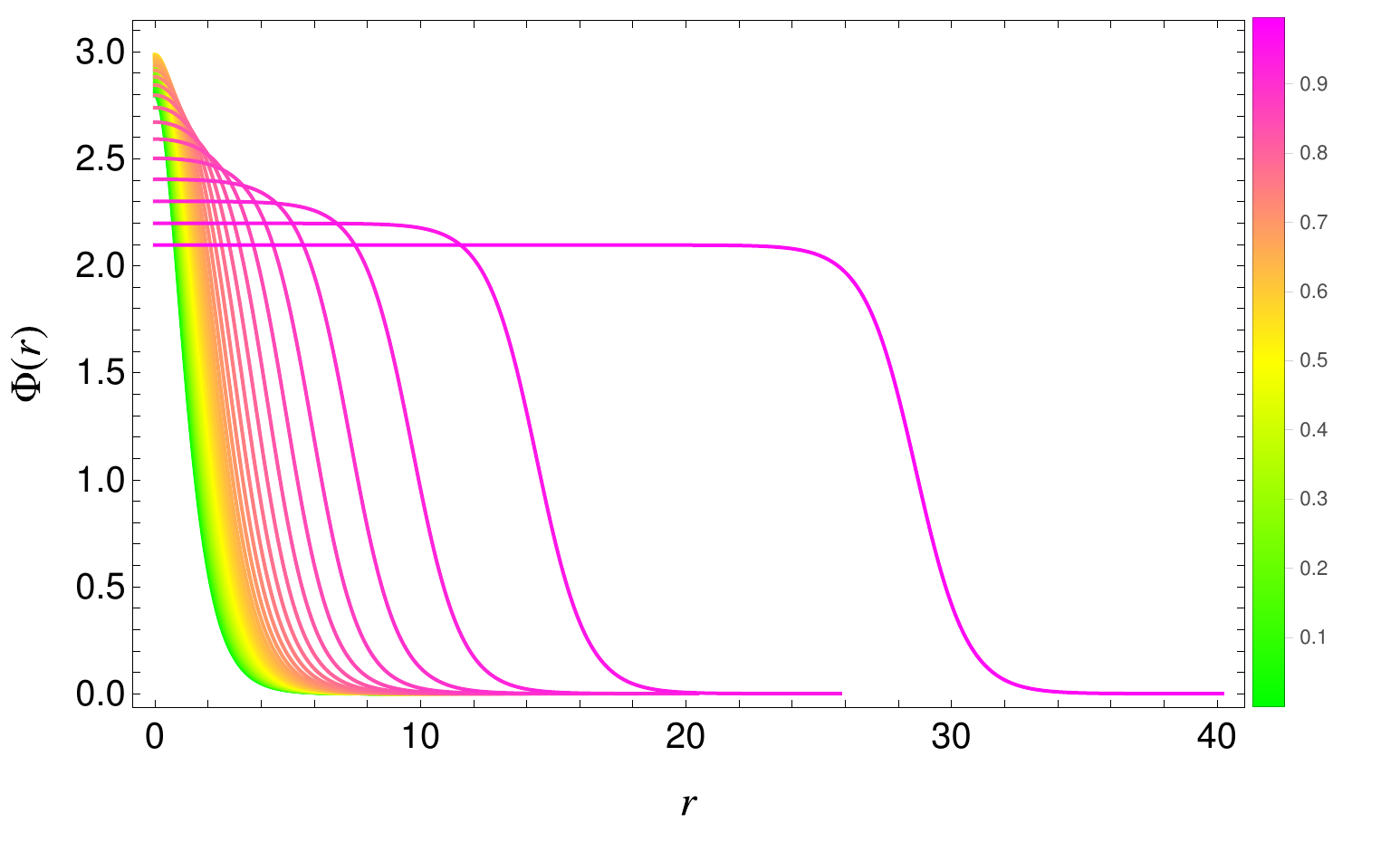}
\includegraphics[width=0.45\textwidth]{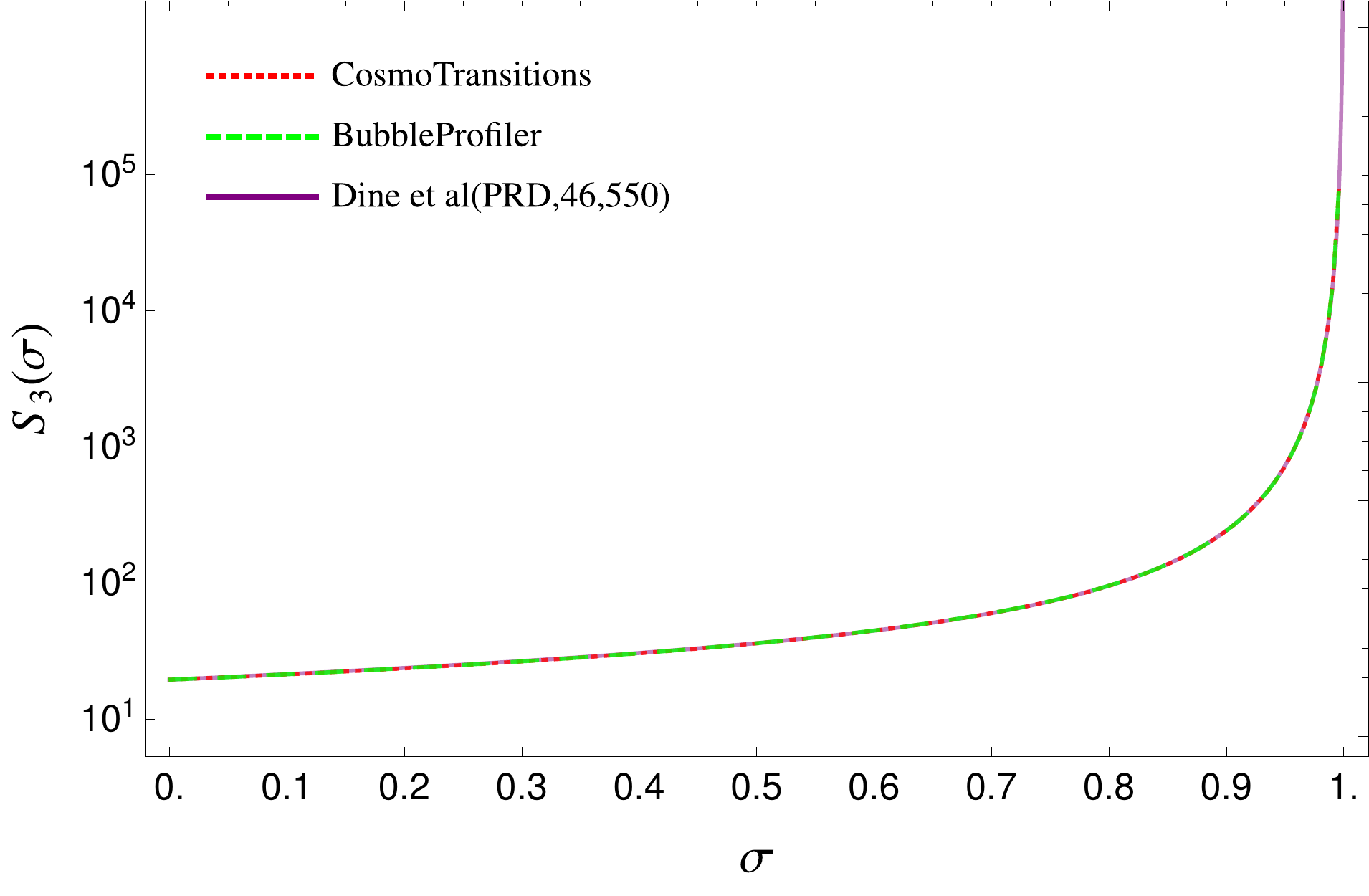}
\caption{
\label{fig:bounce}
Left panel: the bounce solutions for the example effective potential with rescaled fields and coordinates used in this work for different choices of $\sigma$, 
with the color-map denoting values of $\sigma$. Right panel: comparison of the corresponding $S_3(T)/T$ obtained with different packages and the analytical
fit provided in Ref.~\cite{Dine:1992wr}.
}
\end{figure}
%%%%%%%%%%%%%%%%%%%%%%%%%%%%%%%%%%%%%%%%%%%%%%%%%%%%%%%%%%%%%%%%%%%%%

Though there are four free parameters 
for this simple effective potential, a rescaling of both the coordinates and the scalar fields allows to reduce to 
only one dynamical parameter~\cite{Dine:1992wr}. 
The rescaled fields and coordinates are defined as $\Phi = 2 E T \phi/M^2$ and $X = M x$. The Lagrangian then becomes
\begin{eqnarray}
\mathcal{L} = \frac{M^6}{4 E^2 T^2} \left[ 
\frac{1}{2} (\partial_X \Phi)^2 - \frac{1}{2} \Phi^2 + \frac{1}{2} \Phi^3 - \frac{1}{8} \sigma \Phi^4 
\right] ,
\end{eqnarray}
where $\sigma \equiv \lambda M^2/(2 E^2 T^2)$~\footnote{This is of course different from the $\sigma$ defined in Eq.~\ref{eq:alphasigma}}. The behavior of the effective potential for the rescaled fields
during the phase transition is solely controlled by $\sigma$.
When $\sigma < 9/8$, a second minimum develops at the temperature
\begin{eqnarray}
T = \sqrt{\frac{T_0^2}{1- \frac{9 E^2}{8 \lambda D}}} .
\end{eqnarray}
When $\sigma = 1$, this minimum is degenerate with the one at 
the origin, which corresponds to a critical temperature of 
\begin{eqnarray}
T_c = \sqrt{\frac{T_0^2}{1- \frac{E^2}{\lambda D}}} .
\end{eqnarray}
Therefore for the rescaled field $\Phi$ and coordinate $X$, 
there is essentially one parameter $\sigma$ that determines the shape of the potential. Calculating the bounce solution
and $S_3$ for all choices of $\sigma$ is sufficient to cover the full parameter space of the original four parameters.
Define the $S_3$ action for the rescaled fields and coordinates
as $\tilde{S}_3(\sigma)$, then the action $S_3(T)$ for the original four parameter theory can be obtained directly as
\begin{eqnarray}
\frac{S_3(T)}{T} = \frac{M^3}{4 E^2 T^3} \tilde{S}_3(\sigma).
\end{eqnarray}
The bounce solutions for various choices of $\sigma$ are shown in the left panel of Fig.~\ref{fig:bounce} 
and the corresponding $S_3(\sigma)$ shown as red dotted and green dashed lines for solutions solved from 
CosmoTransitions~\cite{Wainwright:2011kj} and BubbleProfiler~\cite{Athron:2019nbd} respectively. In this plot,
there is also a purple curve, corresponding to the analytical fit in Ref.~\cite{Dine:1992wr}:
\begin{eqnarray}
\tilde{S}_3(\sigma) = 4 \times 4.85 \times \left\{
1 + \frac{\sigma}{4} \left[ 1 + \frac{2.4}{1-\sigma} + \frac{0.26}{(1-\sigma)^2} \right] 
\right\} .
\end{eqnarray}
We can see in the whole region plotted, the three results agree very well with each other. So our results in previous sections can be 
followed by simply choosing above analytical fit. For the example used in Sec.~\ref{sec:dynamics}, $T_0 = 75\text{GeV}$, $E=D=0.1$ and $\lambda=0.2$, which
gives $T_c = 106.066 \text{GeV}$.

\section{\label{sec:RDeta}The Previously Derived Effective Lifetime of the Source}
Here we revisit the deviation that led to the conclusion that the effective lifetime of the source is one Hubble time
in a radiation dominated universe, as was originally obtained in Ref.~\cite{Hindmarsh:2015qta}. We will follow closely their notations, 
using the conformal time $\eta$ as variable instead of $y$, and using $a_{\ast}$ rather than $a_s$. Also we study both RD and MD, though
only RD is studied in Ref.~\cite{Hindmarsh:2015qta}.

We start with Eq.~\ref{eq:correh} and do the integrals over $\tilde{\eta}_1$ and $\tilde{\eta}_2$. 
We can keep only the leading contribution by neglecting the highly oscillatory part in the Green's functions. 
This means for the trigonometric function, we keep only the parts with argument $(\tilde{\eta_1} - \tilde{\eta}_2)$ and find
\begin{eqnarray}
&&\frac{\partial G(\tilde{\eta}, \tilde{\eta}_1)}{\partial \tilde{\eta}}
\frac{\partial G(\tilde{\eta}, \tilde{\eta}_2)}{\partial \tilde{\eta}} \nonumber \\
&&=
\frac{\tilde{\eta}_1 \tilde{\eta}_2}{2}
\times
\left\{
\begin{array}{l}
\tilde{\eta}^{-2}(1 + \tilde{\eta}^{-2}) \cos(\tilde{\eta}_1 - \tilde{\eta}_2) , \\
\tilde{\eta}^{-4} (1 + 3 \tilde{\eta}^{-2} + 9 \tilde{\eta}^{-4}) [ (\tilde{\eta}_1 - \tilde{\eta}_2) \sin(\tilde{\eta}_1 - \tilde{\eta}_2) + (1 + \tilde{\eta}_1 \tilde{\eta}_2) \cos(\tilde{\eta}_1 - \tilde{\eta}_2) ] , 
\end{array}
\right.
\end{eqnarray}
where the upper and lower row applies to radiation and matter dominated universe respectively.
Now switch integration variables from $\tilde{\eta}_1$ and $\tilde{\eta}_2$ to $x \equiv (\tilde{\eta}_1 + \tilde{\eta}_2)/2$ and $z=\tilde{\eta}_1 - \tilde{\eta}_2$.
This results in the relation $\tilde{\eta}_1 \tilde{\eta}_2 = x^2 - \frac{z^2}{4}$. Under these manipulations, the power spectral density of $h^{\prime}$ becomes:
\begin{eqnarray}
P_{h^{\prime}}
&=&
[16 \pi G \left( \bar{\tilde{\epsilon}} + \bar{\tilde{p}} \right) \bar{U}_f^2]^2 L_f^3 
\left\{
\begin{array}{l}
\tilde{\eta}^{-2} (1 + \tilde{\eta}^{-2})
\\
\tilde{\eta}^{-4} (1 + 3 \tilde{\eta}^{-2} + 9 \tilde{\eta}^{-4})
\end{array}
\right\}
\int dx \int dz
 \frac{1}{k^2} \frac{\tilde{\eta}_1 \tilde{\eta}_2 a_{\ast}^8}{a^2(\eta_1) a^2(\eta_2)} \nonumber \\
&& \hspace{1.3cm} \times \frac{1}{2}
\left\{
\begin{array}{l}
\cos z \\
z \sin z + (1 + x^2 - \frac{z^2}{4}) \cos z
\end{array}
\right\}
\tilde{\Pi}^2(\tilde{L}_f,\tilde{\eta}_1, \tilde{\eta}_2) .
\end{eqnarray}
Here $\tilde{L}_f \equiv k L_f$. The expression can be reorganized to show the correct dependence on $a(\eta)$ and we have for the correlator of $\dot{h}$:
\begin{eqnarray}
P_{\dot{h}}
&=&
\frac{a_{\ast}^6}{a^4(\eta)}
\frac{1}{k^2}
[16 \pi G \left( \bar{\tilde{\epsilon}} + \bar{\tilde{p}} \right) \bar{U}_f^2]^2 L_f^3 
\left\{
\begin{array}{l}
1 + \tilde{\eta}^{-2}
\\
1 + 3 \tilde{\eta}^{-2} + 9 \tilde{\eta}^{-4}
\end{array}
\right\}
\int_{\tilde{\eta}_{\ast}}^{\tilde{\eta}} dx \int dz \nonumber \\
&& \hspace{1.3cm} \times \frac{1}{2}
\left\{
\begin{array}{l}
\frac{\tilde{\eta}_{\ast}^2}{x^2 - z^2/4}  \\
\frac{\tilde{\eta}_{\ast}^4}{(x^2 - z^2/4)^3}  
\end{array}
\right\}
\left\{
\begin{array}{l}
\cos z \\
z \sin z + (1 + x^2 - \frac{z^2}{4}) \cos z
\end{array}
\right\}
\tilde{\Pi}^2(\tilde{L}_f,\tilde{\eta}_1, \tilde{\eta}_2) .
\end{eqnarray}
As we have seen the source is largely stationary, that is, the correlator 
$\tilde{\Pi}^2(\tilde{L}_f,\tilde{\eta}_1, \tilde{\eta}_2)$ depends only on $z$ but not on $x$. Then it 
can be written as $\tilde{\Pi}^2(\tilde{L}_f,z)$. Also the autocorrelation time $z$ is very small compared with the Hubble time, 
so we can neglect the $z$ dependence on the denominators in the first curly bracket and keep only the $x^2$ term for MD in the second curly bracket, 
which then allows the integration over $x$, giving 
\begin{eqnarray}
\int_{\tilde{\eta}_{\ast}}^{\tilde{\eta}} dx \frac{1}{x^2} = 
\frac{1}{\tilde{\eta}_{\ast}} - \frac{1}{\tilde{\eta}} ,
\quad \quad
\int_{\tilde{\eta}_{\ast}}^{\tilde{\eta}} dx \frac{1}{x^4} =  
\frac{1}{3}(\frac{1}{\tilde{\eta}_{\ast}^3} - \frac{1}{\tilde{\eta}^3}) .
\label{eq:etaint}
\end{eqnarray}
Here is where things become subtle. The second term for RD is neglected in Ref.~\cite{Hindmarsh:2015qta}. This leads to a result that corresponds to the 
asymptotic value $\Upsilon=1$ for RD, and as we have seen the short duration of the source does not allow to neglect this term.
Lets continue to reproduce the result of Ref.~\cite{Hindmarsh:2015qta} by keeping only the first term. This gives
\begin{eqnarray}
P_{\dot{h}}
&=&
\frac{a_{\ast}^6}{a^4(\eta)}
\frac{1}{k^2}
[16 \pi G \left( \bar{\tilde{\epsilon}} + \bar{\tilde{p}} \right) \bar{U}_f^2]^2 L_f^3 
\left\{
\begin{array}{l}
1 + \tilde{\eta}^{-2}
\\
(1 + 3 \tilde{\eta}^{-2} + 9 \tilde{\eta}^{-4})/3
\end{array}
\right\}
\tilde{\eta}_{\ast}
 \nonumber \\
&& \hspace{1.3cm} \times \int dz \frac{\cos(z)}{2}
\tilde{\Pi}^2(\tilde{L}_f,z) \nonumber \\
&=&
\frac{a_{\ast}^4}{a^4(\eta)}
[16 \pi G \left( \bar{\tilde{\epsilon}} + \bar{\tilde{p}} \right) \bar{U}_f^2]^2 L_f^3 
\left\{
\begin{array}{l}
1 + \tilde{\eta}^{-2}
\\
(1 + 3 \tilde{\eta}^{-2} + 9 \tilde{\eta}^{-4})/3
\end{array}
\right\}
{(a_{\ast}{\eta}_{\ast})(a_{\ast} L_f)}  
\widetilde{P}_{\text{GW}}(k L_f). \nonumber \\
\end{eqnarray}
In the second line, the following definition is used:
\begin{eqnarray}
\widetilde{P}_{\text{GW}}(k L_f)
= \frac{1}{k L_f}
\int dz 
\frac{\cos z}{2}
\tilde{\Pi}^2(\tilde{L}_f,z) .
\end{eqnarray}
The variables appearing in above equations can further be reorganized so that we have a result similar to Eq.(A10) in Ref.~\cite{Hindmarsh:2015qta}:
\begin{eqnarray}
\mathcal{P}_{\text{GW}}(t,k) = %\frac{d \Omega_{\text{GW}}(t)}{d \ln(k)}  =
3 \Gamma^2 \bar{U}_f^4
\left(\frac{a_{\ast}^4}{a^4} \frac{H_{\ast R}^4}{H^2 H_{\ast}^2}\right)
\left\{
\begin{array}{l}
1 + \tilde{\eta}^{-2}
\\
(1 + 3 \tilde{\eta}^{-2} + 9 \tilde{\eta}^{-4})/3
\end{array}
\right\} \nonumber \\
\times 
(H_{\ast} a_{\ast} \eta_{\ast}) (H_{\ast} a_{\ast} L_f) \frac{(k L_f)^3}{2 \pi^2} \widetilde{\mathcal{P}}_{\text{GW}}(k L_f) .
\end{eqnarray}
For RD, $H_{\ast} a_{\ast} \eta_{\ast} = 1$ and $a_{\ast} L_f$ is the physical length scale ($L_f^{\ast}$ in Ref.~\cite{Hindmarsh:2015qta}).
If we also neglect the variation of the Hubble rate from $H_{\ast}$ to $H$, and since in this case $H_{\ast R} = H_{\ast}$, and also neglect
the terms suppressed by $1/\tilde{\eta}$ in the curly bracket due to the assumed relation $\tilde{\eta} \gg \tilde{\eta}_{\ast}$, then the result 
for RD reduces to Eq.(A11) in Ref.~\cite{Hindmarsh:2015qta}. 
Because $H_{\ast} a_{\ast} \eta_{\ast} = 1$ and also because the power spectrum in Minkowski 
spacetime is proportional to $H_{\ast} \tau_{\text{sw}}$, it is concluded in Ref.~\cite{Hindmarsh:2015qta} that the effective lifetime is a Hubble time.
This is true if indeed $\tilde{\eta} \gg \tilde{\eta}_{\ast}$, but as we have seen it requires many Hubble times for the asymptotic value to be 
reached. The sound wave, however, is likely to be disrupted by the onset of shocks or turbulence or damped by other dissipative processes, 
which certainly do not allow the sound wave to remain active that long for the asymptotic value to be reached.
So the main point is we can not assume $\tilde{\eta} \gg \tilde{\eta}_{\ast}$ and 
neglect the second term in the first equation of Eq.~\ref{eq:etaint}. 

While non-relevant here for MD, we can still compare its asymptotic value with what we already find in previous sections. From above equation we
can see the quantity in the curly bracket is $1/3$ for MD and $1$ for RD. But for MD, $H_{\ast} a_{\ast} \eta{\ast} = 2$, then the asymptotic
value of $\Upsilon$ is $2/3$ for MD, which is consistent with our previous result.

\end{document}